# KINECT SENSOR BASED GESTURE RECOGNITION FOR SURVEILLANCE APPLICATION

*A Thesis*
*submitted in partial fulfillment of the requirement for the Degree of*
*Master of Technology in Intelligent Automation and Robotics*

Jadavpur University
May 2016

*By*

**Biswarup Ganguly**
Registration No: 125150 of 2013-14
Examination Roll No: M6IAR1604

*Under the Guidance of*

**Prof. Amit Konar**
Department of Electronics & Telecommunication Engineering
Jadavpur University, Kolkata-700032
India

FACULTY OF ENGINEERING AND TECHNOLOGY
JADAVPUR UNIVERSITY

CERTIFICATE

This is to certify that the dissertation entitled **"Kinect Sensor Based Gesture Recognition For Surveillance Application"** has been carried out by BISWARUP GANGULY (University Registration No.: 125150 of 2013-14) under my guidance and supervision and be accepted in partial fulfillment of the requirement for the Degree of Master of Technology in Intelligent Automation and Robotics. The research results presented in the thesis have not been included in any other paper submitted for the award of any degree to any other University or Institute.

……………………………………..
Prof. Amit Konar
Thesis Supervisor
Dept. of Electronics & Telecommunication Engineering
Jadavpur University

……………………………………..

Dr. P. Venkateswaran
Head of the Department
Dept. of Electronics & Telecommunication Engineering
Jadavpur University

……………………………………..

Prof. (Dr.)Sivaji Bandyopadhyay
Dean
Faculty of Engineering and Technology
Jadavpur University

FACULTY OF ENGINEERING AND TECHNOLOGY

JADAVPUR UNIVERSITY

CERTIFICATE OF APPROVAL*

The forgoing thesis is hereby approved as a creditable study of an engineering subject and presented in a manner satisfactory to warrant acceptance as prerequisite to the degree for which it has been submitted. It is understood that by this approval the undersigned do not necessarily endorse or approve any statement made, opinion expressed or conclusion drawn there in but approve the thesis only for which it is submitted.

**Committee on final examination for the evaluation of the thesis**

---------------------------------------
Signature of the Examiner

---------------------------------------
Signature of the Supervisor

*Only in the case the thesis is approved

FACULTY OF ENGINEERING AND TECHNOLOGY
JADAVPUR UNIVERSITY

## DECLARATION OF ORIGINALITY AND COMPLIANCE OF ACADEMIC THESIS

I hereby declare that this thesis entitled **“Kinect Sensor Based Gesture Recognition For Surveillance Application”** contains literature survey and original research work by the undersigned candidate, as part of his Degree of Master of Technology in Intelligent Automation and Robotics. All information have been obtained and presented in accordance with academic rules and ethical conduct. I also declare that, as required by these rules and conduct, I have fully cited and referenced all materials and results that are not original to this work.

Candidate Name: **Biswarup Ganguly**

Examination Roll No.: **M6IAR1604**

Thesis Title: **Kinect Sensor Based Gesture Recognition For Surveillance Application**

-------------------------------------------
Signature of the candidate

# PREFACE

The introduction of gesture recognition has led to a better Human-Computer Interaction (HCI). Gesture recognition is an interesting topic in computer vision and pattern recognition (CVPR) technology which deals with the mathematical interpretation of human gestures via a computing device. It enables human being to interact with machines without any mechanical contact.

Hand gesture recognition has been granted as one of the emerging fields in research today providing a natural way of communication between man and a machine. Gestures are some forms of body motions which a person expresses when doing a work or giving a reply. Human body tracking is a well-studied topic in today's era of Human Computer Interaction and it can be formed by the virtue of human skeleton structures. These skeleton structures have been detected successfully due to the smart progress of some devices, used to measure depth (e.g. Sony PlayStation, Kinect sensor etc.). Human body movements have been viewed using these depth sensors which can provide sufficient accuracy while tracking full body in real-time mode with low cost. In reality action and reaction activities are hardly periodic in a multi-person perspective situation. Also, recognizing their complex, a-periodic gestures are highly challenging for detection in surveillance system.

In a nutshell, both the algorithms proposed for hand gesture recognition may be considered as a relatively unexplored application area. The proposed work is an attempt to recognize the single person as well as multi person gestures and it produces reasonable accuracy (more that eighty five percent) and has scopes for future research.

## ACKNOWLEDGEMENTS

I am availing this opportunity to express my gratitude to everyone who has supported me throughout the course of my M-Tech tenure. I am thankful for their aspiring guidance, invaluably constructive criticism and friendly advice during this duration. I am sincerely grateful to them for their truthful and illuminating views on a number of issues related to this Master Degree Research.

First and foremost, I am gratified to my supervisor Prof. Amit Konar for their continuous support of my M-Tech study and research. I have been amazingly fortunate to have such advisors who gave me the freedom to explore on my own, and at the same time provide guidance when my steps faltered. Their guidance has helped me in all the time of research and writing of this dissertation. I am also thankful to them for morally supporting me during my time of crisis. I am thankful to Dr. P. Venkateswaran who have acted as Head of the Department of Electronics and Telecommunication Engineering during the tenure of my studentship. I would also like to show my gratitude to the respected professors of the Department of Electronics and Telecommunication Engineering for their constant guidance and valuable advices.

The realization of this research work would not have been possible without the whole-hearted cooperation from fellow researchers and students of the Artificial Intelligence Lab under the Department of Electronics and Telecommunication Engineering for making my journey memorable. I am grateful to my seniors Sriparna Saha, Pratyusha Rakshit, Saugat Bhattacharyya, Anwesha Khasnobish, Tanmoy Dasgupta, Lidia Ghosh for their constant support and encouragement. In this regard, I am indebted to my classmates, friends and seniors for their constant support and good wishes with a special mention to Subho Paul and Saurabh Bhattacharya who supported me in a number of ways.

Last but not the least, I want to pay my sincere respect to my father Mr. Utpal Ganguly and mother Mrs. Aparna Ganguly, who have always endowed me with their love, affection, care and trust. I owe everything to my parents who gave me the strength to achieve my dreams.

Date :

Place:

Biswarup Ganguly
Dept. of Electronics & Telecommunication Engineering
Examination Roll No.: M6IAR1604
Jadavpur University

# List of Publications

1. Sriparna Saha, **Biswarup Ganguly**, Amit Konar, "Gesture Based Improved Human-Computer Interaction Using Microsoft's Kinect Sensor", *IEEE International Conference on Microelectronics, Computing and Communication* 23-25 January, 2016, (Accepted, to be published).

2. Sriparna Saha, **Biswarup Ganguly**, Amit Konar, "Gesture Recognition from Two-Person Interactions Using Ensemble Decision Tree" 4$^{th}$ *International Conference on Advanced Computing, Networking, and Informatics* 22-24 September, 2016, (Accepted, to be published).

*“Gestures, in love, are incomparably more attractive, effective and valuable than words.’’*

Francois Rabelais

# CONTENTS

## Chapter 1: Introduction

## Chapter 2: Gesture Based Improved Human-Computer Interaction Using Microsoft's Kinect Sensor

## List of Figures

## List of Tables

# CHAPTER 1

# INTRODUCTION

*The chapter gives an overview of gestures and its recognizing procedures. Section 1.1 introduces the concepts of gestures followed by the introduction of gesture recognition, gesture recognition tools and applications in section 1.2. Section 1.3 describes the Kinect sensor, its hardware and software tools along with performance. Also gesture recognition through Kinect has been illustrated. Section 1.4 deals with some literature surveys regarding Microsoft's Kinect sensor and its application areas. The scope and the organization of the thesis are detailed in section 1.5 and section 1.6 respectively.*

## 1.1 Gesture

Gestures [1] are a type of nonverbal communication where visible body movements take place and deliver certain messages along with speech. Gestures can be expressed by suitable body movements of face, hands, head which can be used for controlling specific devices or transferring significant knowledge to the environment. Human gestures can originate from any body movements.

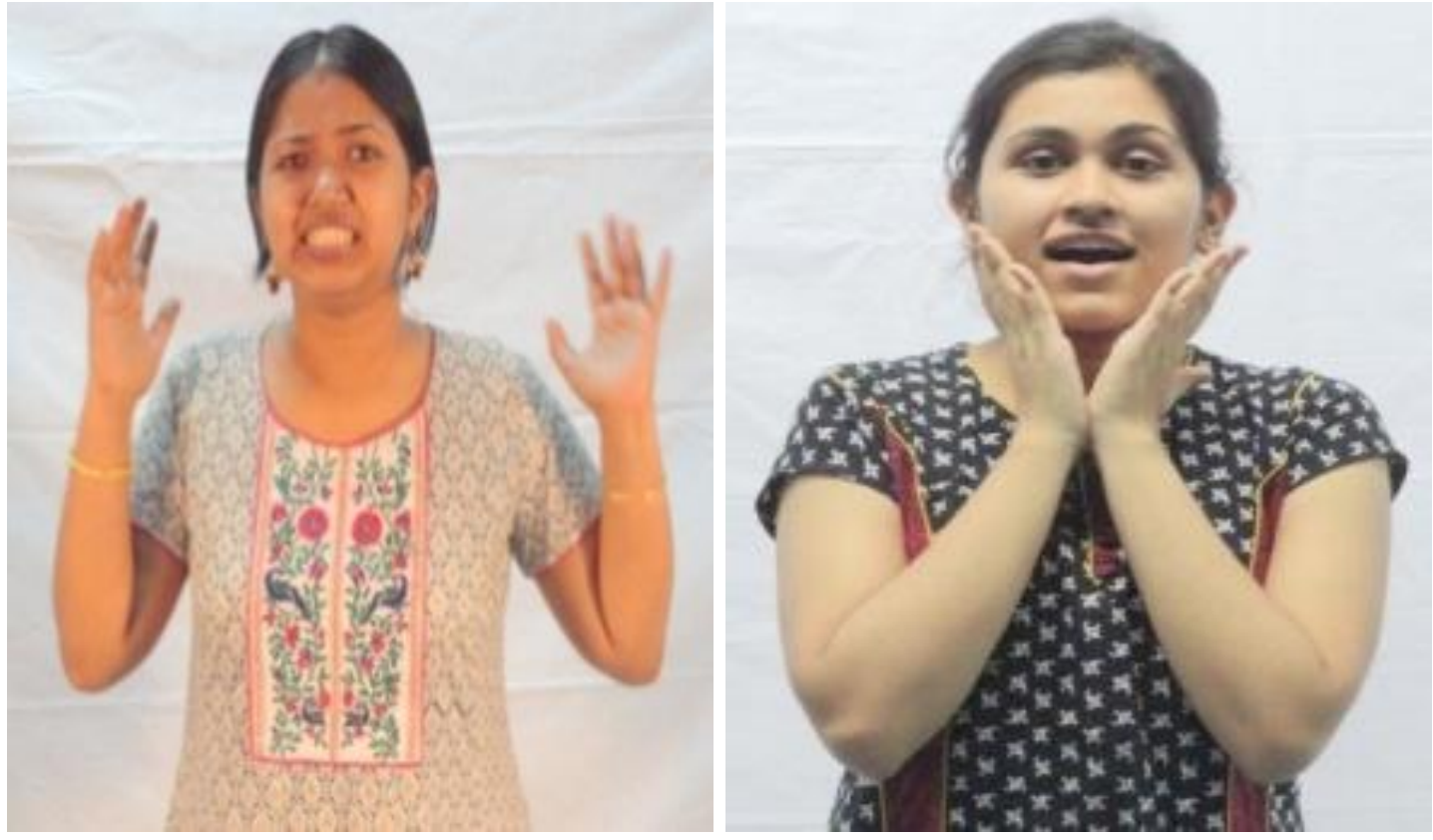

(a) (b)

Fig 1.1 (a) ‘Disguist’ gesture (b) ‘Surprise’ gesture

In the above figures, gestures like disguist and surprise are shown, where the hands and the faces are moved in such a way to express their feelings.

### 1.1.1 Classification of Gestures

Gestures are classified as;

a. Static type : definite pose or configuration, or still body posture.
b. Dynamic type: movements of different body parts.
c. Both static and dynamic type: Sign language representation.

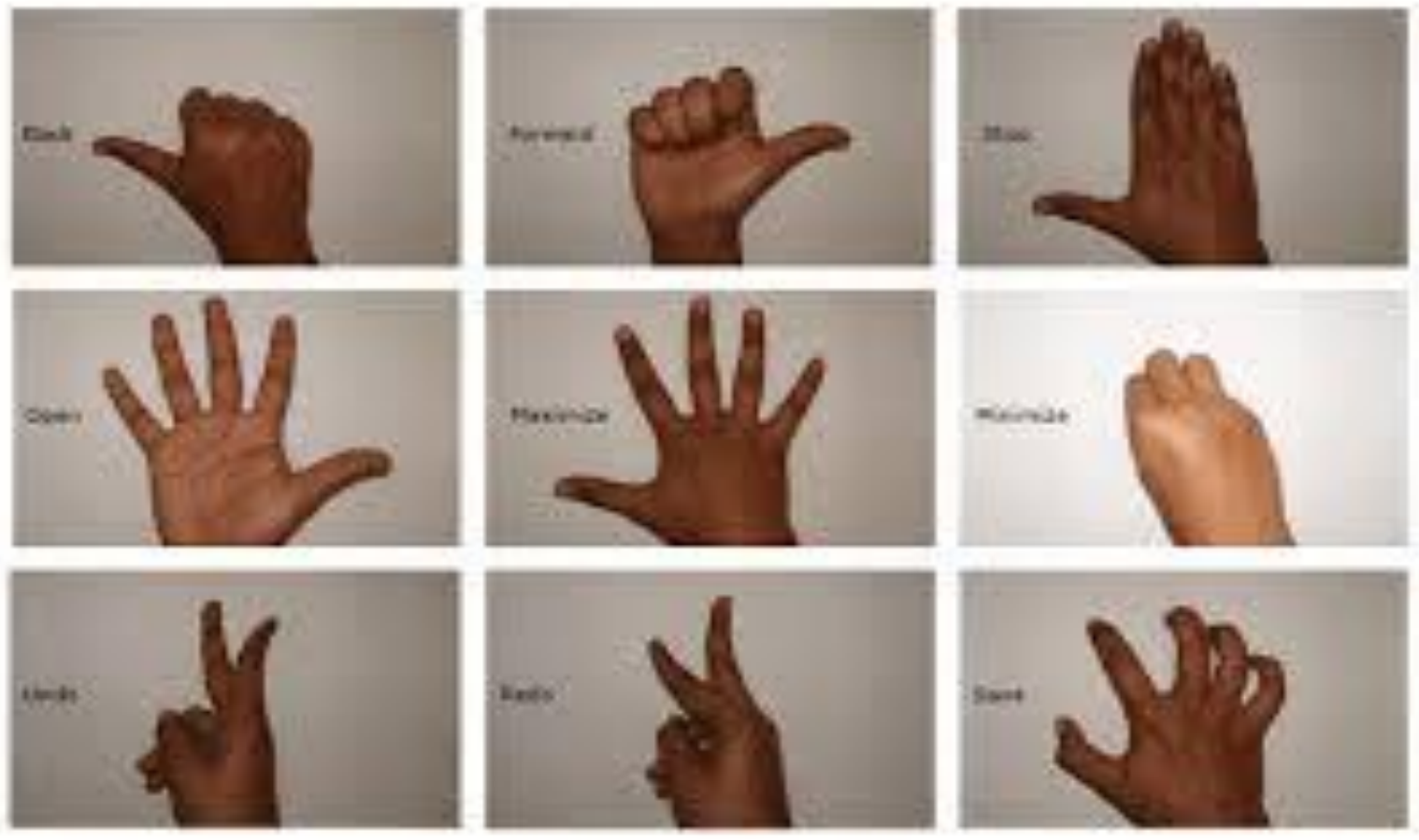

Fig 1.2 Static and dynamic gestures for sign language recognition

Moreover, gestures can be broadly classified into the following types;

1. ***Hand and arm gestures*****:** identification of hand poses, sign languages, and some recreation applications.

2. ***Head and face gestures:*** shaking of head, raising eyebrows, opening mouth to speak, flaring the nostrils, directing the eyes, looks of surprise, happiness, disgust, anger, relax, etc.

3. ***Body gestures:*** tracking movements of single as well as two persons interaction, recognizing movements of a dancer, identifying human gaits and postures.

Gestures are dependent on the following factors;

1. Spatial information : where the gesture takes place.
2. Pathic information : the path it follows.
3. Symbolic information : the sign it generates.
4. Affective information : the excitement it produces.

## 1.2 Gesture Recognition

We, the human being, communicate with the machines via direct contact mechanism. Now-a-days, instead of pressing switches, touching monitor screens, twisting knobs, raising voices, work can be simply done by pointing fingers, waving hands, movements of bodies and so on.

Gesture recognition [2] concerns about recognizing meaningful expressions of human motions, e.g. embroiling hands, arms, face, head and body. It has a great significance in intelligent and efficient human-computer interfacing. Gesture recognition is the process through which gestures, made by different individuals, are recognized by the receiver.

It is an interesting topic in computer vision and pattern recognition (CVPR) technology which deals with the mathematical interpretation of human gestures via a computing device. It enables human being to interact with machines without any mechanical contact, e.g. signaling a finger on a computer screen such that the cursor would move accordingly, which makes such conventional input devices like mouse, keyboard obsolete.

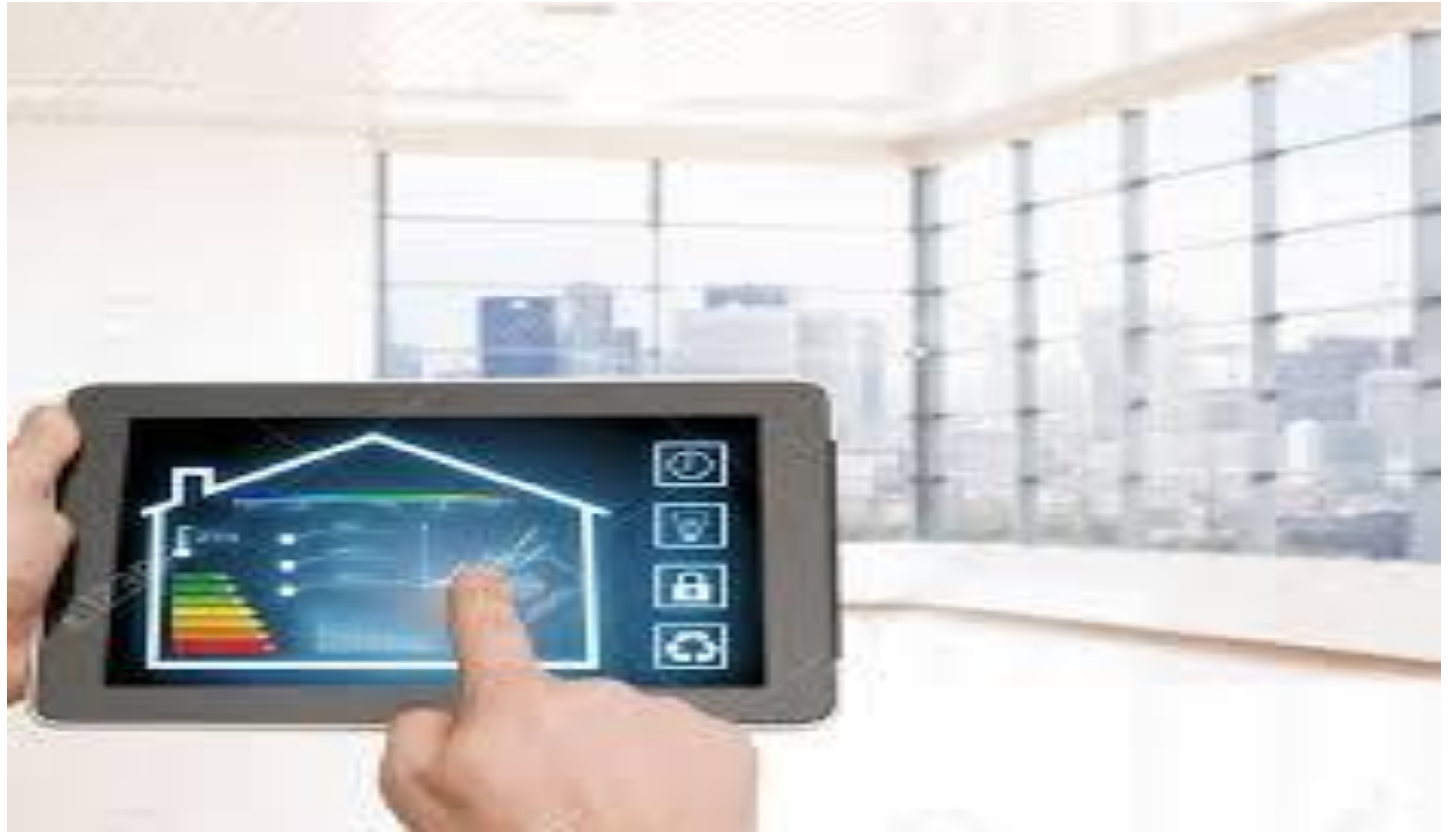

Fig 1.3 Gesture recognition using smart technology

Static gesture recognition [3] can be carried out by template matching, pattern recognition and neural networks (NN); while dynamic gesture recognition [4] involves techniques such as dynamic time warping (DTW), hidden Markov model (HMM), time delay neural network (TDNN).

### 1.2.1 Gesture Recognition System

Gesture recognition system is a combination of four different units [5]. They are data acquisition, gesture modeling, feature extraction and ultimately recognition. Different images of body gestures are captured by specific devices. Then the images are segmented from the cluttered background followed by preprocessing of the images to eliminate noises. After that several features are extracted from the preprocessed image and finally the input images are identified as meaningful body gestures. The detail procedure is discussed as followed by a schematic diagram.

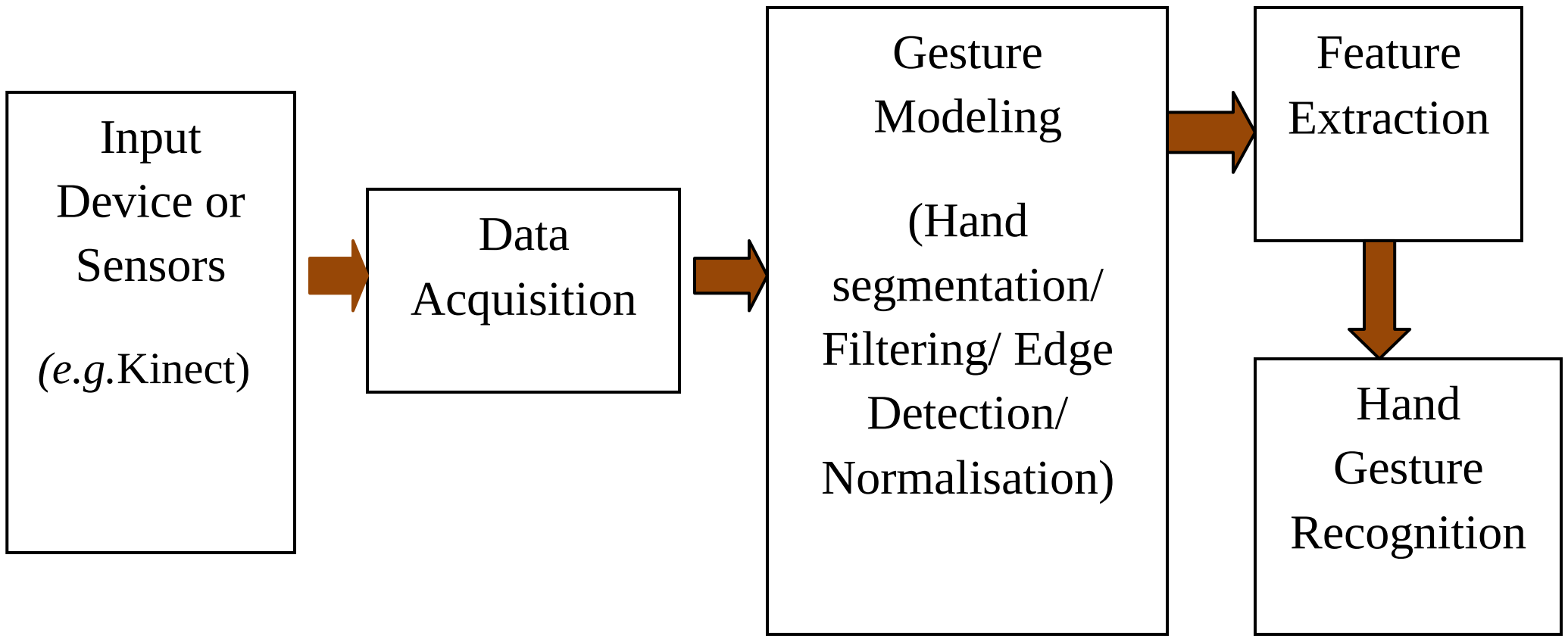

Fig 1.4 System Architecture for Hand Gesture Recognition

### *A. Data Acquisition*

For an effective gesture recognition, data acquisition system should be as much superior as possible. There are a suitable number of data collection devices, such as marker, gloves, stereo camera, webcam, Kinect sensor and so on.

Markers, with colors, are attached to human skin and hand localization is done by color. Gloves are the devices with highest precision and accuracy. Now a days, wireless data gloves are so handy to eliminate hindrance made by cables. Stereo cameras have been effective such that it can capture data automatically without any contact. Kinect sensor has been added a boost in this list, widely used in surveillance and gaming.

### *B. Gesture Modeling*

After the data acquisition is performed via different input devices, gesture modeling serves an important role in this scheme. It has four steps, visually segmentation, filtering, edge detection and normalization.

Segmentation is important to locate the object within cluttered background. Filtering is required for recognizing gestures successfully. Edge detection is necessary for capturing the significant properties of the object image. Normalization is done to enhance the processing for pointing the geometric features.

### *C. Feature Extraction*

Features are the most effective elements for gesture recognition. Features like shape, texture, orientation, contour, distance, angle, centre of gravity etc. can be employed for gesture recognition. Specially features like hand contour, finger tips, finger detection can be used for hand gesture recognition.

### *D. Gesture Recognition*

Once the features have been extracted, a suitable data set are selected to recognize the gestures. Some standard machine learning algorithms, template matching, statistical models, neural networks have been applied for recognizing the gestures.

### 1.2.2 Gesture Recognition Tools

There are different tools for gesture recognition [6] depending upon

1. ***Statistical modeling:*** hidden Markov model (HMM), Kalman filtering, hidden conditional random field (HCRF)

2. ***Computer vision and pattern recognition techniques*:** feature extraction, object detection, clustering and classification.

3. ***Image processing techniques*:** analysis and detection of shape, texture, color, optical flow, image enhancement, image segmentation and contour modeling.

4. ***Connectionist approaches:*** multilayer perceptron (MLP), time delay neural network (TDNN), radial basis function network (RBFN).

### 1.2.3 Applications of Gesture Recognition

The main application areas [7] of gesture recognition lie at automotive sectors, electronics sectors, transit sectors, healthcare, retail sector and so on. It has various applications such as robot control [8], smart surveillance [9], medical systems, sign language recognition [10,11], virtual environments [12], television control [13], gaming [14] and so on.

***A. Robot Control***

Controlling robot using various gestures is recognized one of the interesting application in this arena. Different hand poses (one, two, five) have been used to perform suitable tasks like move forward, move backward, stop respectively.

***B. Sign Language Recognition***

The sign language is an efficient mode of interacting for hearing and talking impaired. Many systems have been introduced to recognize gestures with the help of different sign languages, e.g American Sign Language (ASL) employing bounded histogram, recurrent Neural network based Japanese Sign language (JSL) etc.

### *C. Smart Surveillance*

Surveillance is the act of carefully monitoring someone's activities. It is done by means of electronic equipment such as CCTV camera or interception of electronically transmitted information like internet traffic, phone calls. It is used for intelligence gathering, crime prevention and investigation.

### *D. Television control*

Hand gestures are used for television control. Turning television on-off, volume increasing-decreasing, sound changing, channel changing can be done employing a set of gestures.

### *E. Medical Systems*

In medical technology too, various gestures have been introduced for performing several crucial tasks.

### *F. Virtual Reality*

Gestures for virtual reality have taken a new level in advance computing. Gestures are used to handle 3D and 2D display interactions.

### *G. Gaming*

Gaming is the act of playing games. It refers to manipulate a system's rules to achieve a desired output. It may be playing a role-playing game, where players assume fictional roles or playing with video game, *i.e.* an electronic game with a video interface.

## 1.3 Kinect Sensor

Kinect [15] is the official name of Xbox 360 Console. It is produced by Prime Sense company accompanying with Microsoft on June, 2010. A Windows version has been released on February, 2012. While in 2009, its announcement caused a great impact in the Computer Vision and Computer Graphics communities. This Microsoft product deciphers a versatile way for interaction in gaming, depending completely on voice and gestures. Gradually Kinect has become a popular device used in gaming, theater performances, robotics [16], natural interfacing etc.

The word ‘Kinect’ originates from the word ‘Kinematics’ because it senses object skeletal in running condition.

Kinect possesses both hardware and software-based technology that deciphers the RGB and depth signals [17]. The hardware portion of this sensor consists of a RGB camera, a three dimensional depth sensor (an infrared camera and a projector), an accelerometer, an array of microphone and a tilt motor which can produce RGB images, depth images, acceleration forces due to gravity, audio signals and rotation of head position respectively. As far as the software tools are concerned, they are able to catch human motions in three dimensional (3D) spaces.

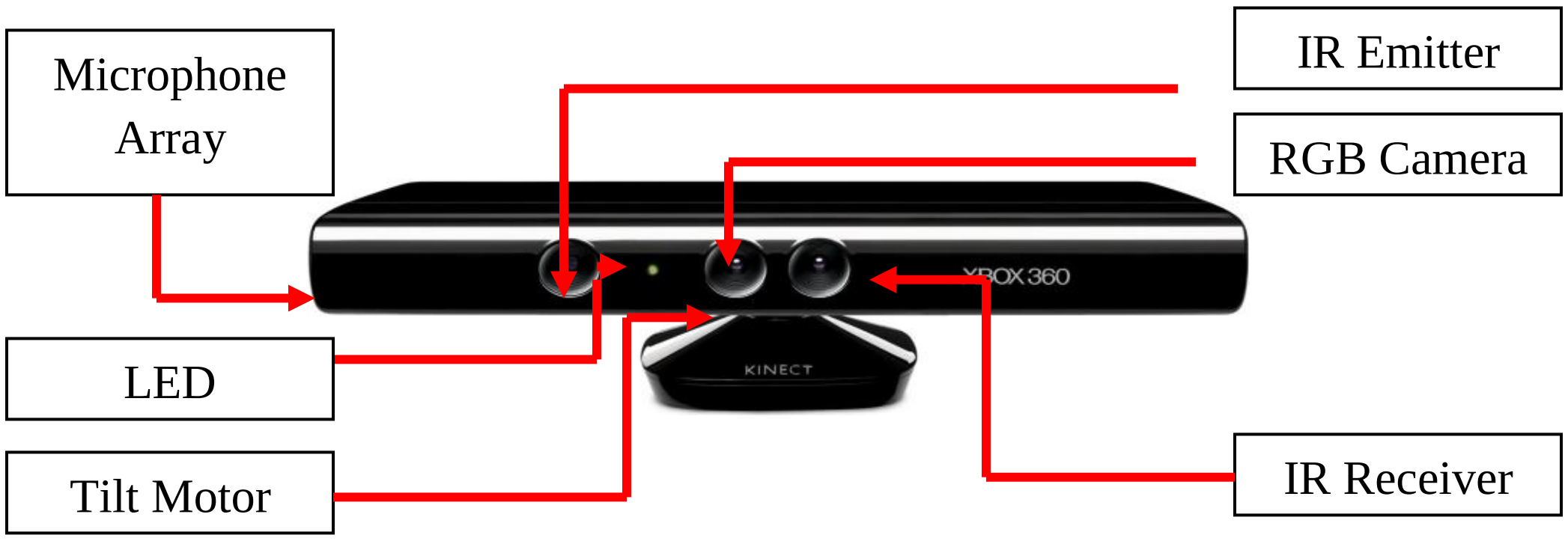

Fig 1.5 Kinect sensor depicting the hardware parts

### 1.3.1 Kinect Hardware Tools:

The Kinect is a black horizontal bar sensor and it looks like a webcam. Figure depicts the arrangement of a Kinect sensor, consisting of a color camera, an infrared (IR) laser projector, an IR camera. The IR projector projects known light speckle pattern into the 3-D images where as the IR camera captures the reflected IR speckles. The speckle is invisible to the color (RGB) camera but it is viewed by the IR camera.

Each component of the Kinect hardware is described below.

**a.** ***RGB Camera*****:** It provides three fundamental color components (Red-Green-Blue) of a video. The RGB camera can capture images at a resolution of 640×480 pixels, with a channel rate of 8 bits, operating at 30 Hz. It also has the ability to switch the camera to a higher resolution of 1280×1024 pixels, working at 10 frames per second. The features of RGB camera are automatic white balancing, flicker avoidance, black reference, defect correction and color saturation. The RGB camera has been chosen as the origin of the reference frame. The x direction is along the length of the Kinect sensor, while the y-direction points vertically up and down and the z-direction is the depth measured by the sensor.

**b.** ***Depth Sensor:*** The three dimensional (3D) depth sensor possesses an infrared (IR) emitter and an IR camera. The IR emitter produces a noisy pattern of builded IR light. The projector along with the camera jointly create a depth image, which allows the distance between the camera and an object. The sensing range of the Kinect device (depth sensor) can be adjusted to a specified limit. The IR camera can seize images at a resolution of 1200×960 pixels, operating at a frequency of 30 Hz. The images are further sampled to 640×480 pixels with a channel rate of 11 bits. Thus it has the sensitivity of 2048 levels.

**c.** ***Accelerometer:*** An electro-mechanical device that can measure acceleration forces. The forces can be of static or dynamic type. The accelerometer detects the path of acceleration forces due to gravity. This incorporates the system fixing its head at an exact level and calibrating the value such that the head can be able to move at specific angles.

**d.** ***Microphone Array:*** The array of microphone consists of four channels. Each channel can process an audio of 16 bit. The sampling rate of microphone is 16 kHz.

**e.** ***Motor Section:*** The Kinect sensor system consists of two sub-systems operated by motor with some gearing. One is to tilt the Kinect sensor up and down upto 27 degrees. The other is the accelerometer, used for determination of the head position of Kinect sensor.

The distance range within which the object should be placed in front of the Kinect sensor is nearly 1.2 to 3.5 m or 3.2 to 11 ft. The depth resolution diminishes, with an increment, at an amount of 1 cm at 2 m distance away from the Kinect.

The sensor system has an angular field of view of 45 degrees vertically, 58 degrees horizontally and 70 degrees diagonally [18].

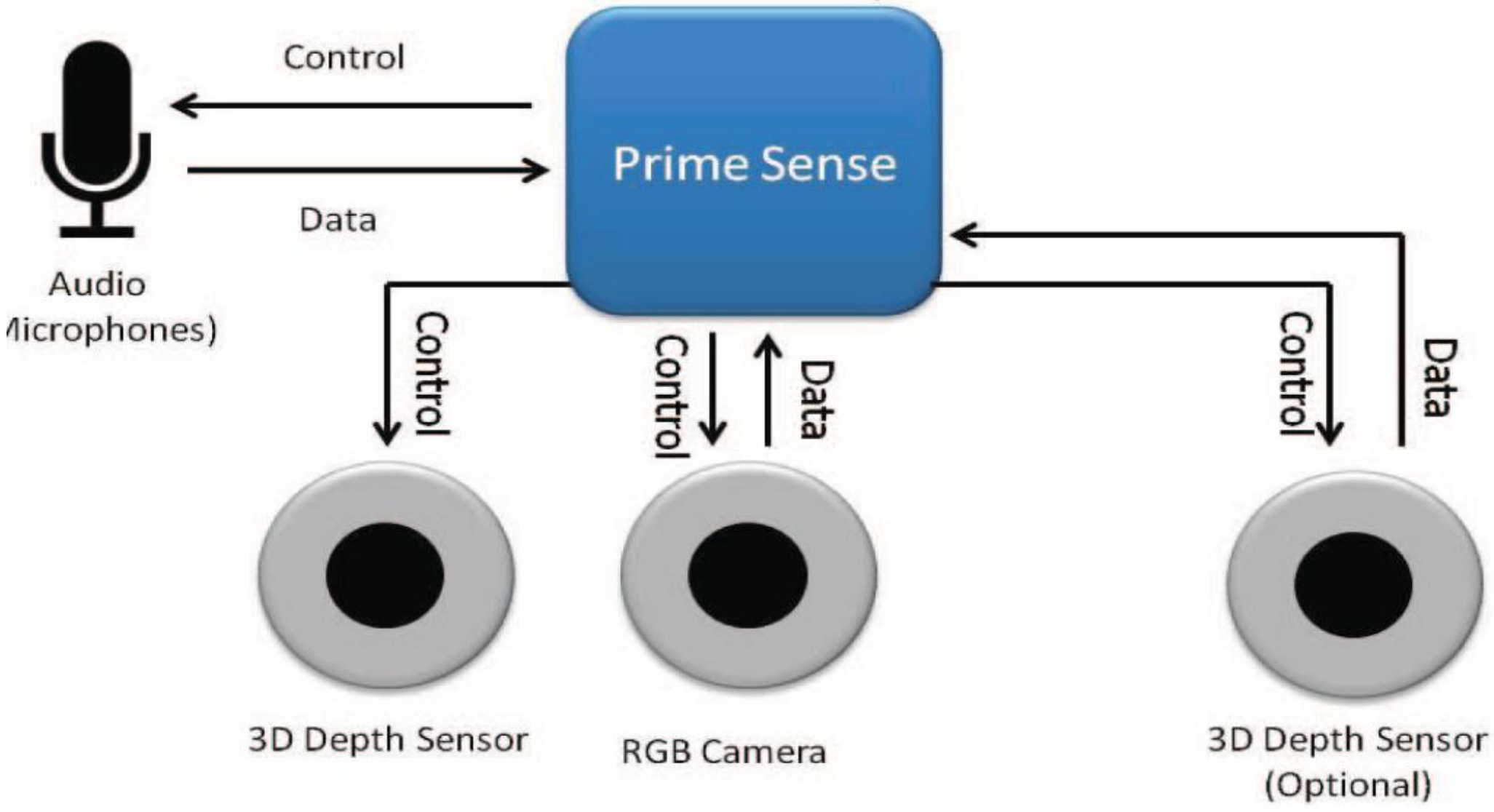

Fig 1.6 Kinect sensor structure diagram

### 1.3.2 Kinect Software Tools:

Kinect software tools include the Kinect development library along with the algorithmic components present in the library. Presently, there are various software tools such as OpenNI, Microsoft Kinect Software Development Kit (SDK) and OpenKinect(LibFreeNect). Most of the users use the first two tools.

OpenNI always works with a Compliant middleware called NITE, which has a highest version of 2.0. It is a multiplatform and open source tool.

Microsoft Kinect SDK was released by Microsoft and has a latest version of 1.7. It is only available for Windows.

OpenKinect is a free, open source library maintained by an open community of Kinect people.

The table below provides a comparison between OpenNI and Microsoft Kinect SDK;

**Table 1.1** Comparison between OpenNI and KinectSDK

| | **OpenNI** | **Microsoft Kinect SDK** |
|---|---|---|
| Automatic body calibration | No | Yes |
| Body gesture recognition | Yes | Yes |
| Camera calibration | Yes | Yes |
| Facial tracking | Yes | Yes |
| Hand gesture analysis | Yes | Yes |
| Motor control | Yes | Yes |
| Scene analysis | Yes | Yes |
| Seated skeleton | No | Yes |
| Standing skeleton | Yes(15 joints) | Yes (20 joints) |
| 3-D scanning | Yes | Yes |

It is worth noticeable that the new version of OpenNI (2.0) permits users to install Microsoft Kinect SDK [19, 20] on the same machine and run both packages using the Microsoft Kinect driver.

The Kinect sensor senses objects at a sampling rate of 30 frames per second.. Each frame data consists of twenty Cartesian coordinates $J_i$ of three dimensions (3D) associated with twenty different human joints as shown in the figure below.

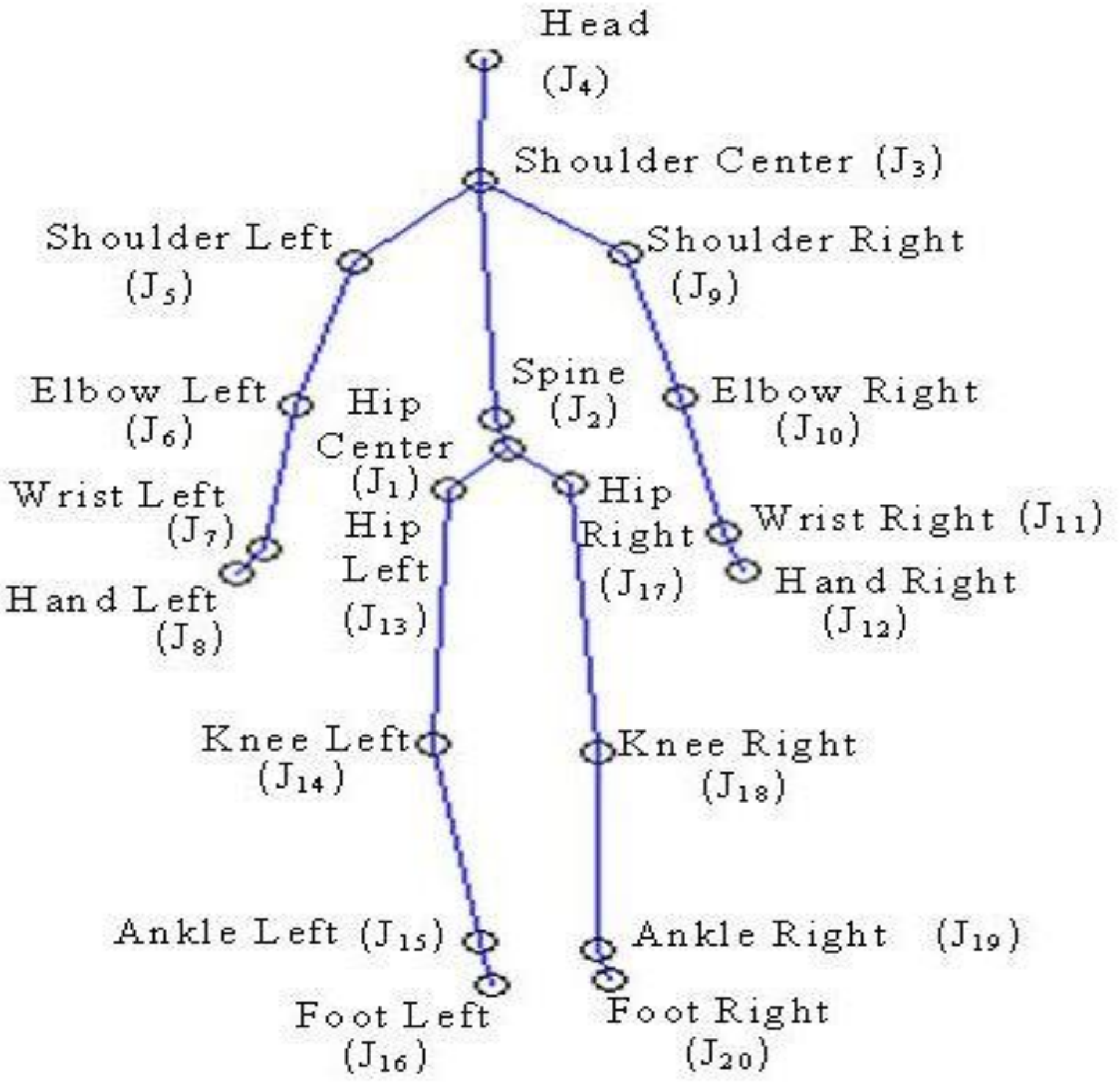

Fig 1.7 Skeleton produced by Kinect along with twenty body joints

### 1.3.3 Performance of Kinect

Kinect is more advantageous than the time of flight (TOF) camera in terms of accuracy and near about to a medium-resolution stereo camera. The performance of the Kinect sensor [21] is synonymous to that of the laser for short range distances of less than 3.5 metres.

The main advantage [22] of using Kinect is due to its versatility, size and optimum cost. It has the ability to work throughout a whole day (24 hrs). Also, Kinect reduces the effect of shadows and multiple source illumination that might affect the 3D scenes being captured by the Kinect device. Kinect sensor has a wide range of application in face recognition and voice gesturing. Subject dress is invariant in front of Kinect.

### 1.3.4 Gesture recognition through Kinect Sensor

The procedure of gesture recognition is as follows; a subject is placed in front of a camera fixing tool. An invisible Infrared ray, emitting from the tool, penetrates on the subject which is reflected back to the camera enabled tool and also on a gesture recognition chip. The chip along with a gesture recognition software (SDK) [23] produces a depth map of pictures taken by the camera and responds almost similarly to the movements taken in front of the camera.

The Kinect sensor provides three outputs; an RGB image, a skeleton and a depth image. All three outputs have shown together.

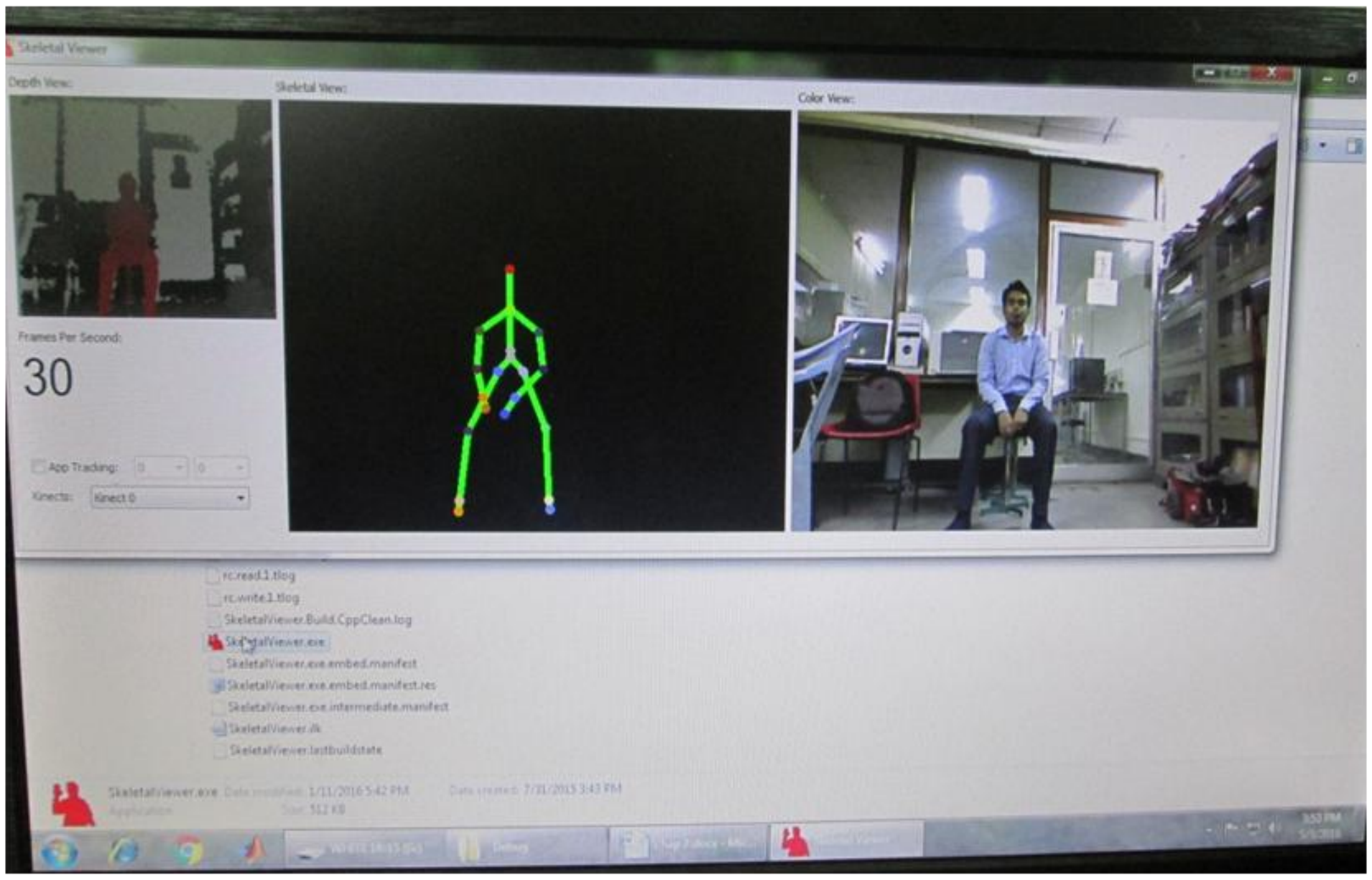

Fig 1.8 Three outputs of Kinect sensor

The pros of the gesture recognition technology are that there is no physical contact between the individual and the gesture recognition tool.

## 1.4 Literature Survey

Oszust and Wysocki [24] have proposed a method for identification of signed expressions recorded by a camera enabled tool named Kinect. Skeletal structures of the body and hand positions are the two variants of this work. In this paper Polish Sign Language (PSL) words are studied. Seven features for each hand have been analyzed with the help of three clustering algorithms such as K-means, K-medoids and minimum entropy clustering (MEC). The classifier used here is k-nearest neighbor (kNN) [25] classifier with $k$=1 as the best choice. A recognition rate of 95% is obtained for skeletal images. This work helps impaired people to hear and interact globally.

The kNN algorithm searches the k-nearest neighbors among the training dataset and puts an object into the class that is most common among its k nearest neighbors, where k is a small positive integer.

Chikkana and Guddeti [26] have proposed an algorithm for gesture spotting for Indian Sign Language (ISL) involving single hand motion. The Microsoft's Kinect sensor is employed for hand tracking. Three basic features: location, orientation and velocity are selected for gesture spotting. The hand gesture trajectory is smoothed followed by normalization of the extracted features. K-means clustering algorithm is used to classify the gesture features into k clusters on the feature space. Then Hidden Conditional Random Field (HCRF) [27] is used for training the system and gesture recognition. A recognition rate of 95.20% and 93.20% is achieved for test data gesture spotting and real-time gesture spotting respectively. Authors have decided to focus on recognizing gestures of both hands in future.

Scientists [28] have presented an algorithm named as dynamic hand gesture recognition algorithm. Here the palm is represented as a node to get the palm position coordinates. The hidden Markov model (HMM) [29] technology is used to perform the dynamic gesture model training. Seven kinds of gesture trajectories (up, down, left, right, ”S”,”O”,”E”) are recorded with the help of Kinect sensor. The feature extraction is done upon the tangent angle. It has been found that the recognition rate is about 95%. The application of this research work lies in controlling the keyboard, achieving photo views and so on.

Hidden Markov model is a dual stochastic process combining a Markov chain with definite number of states and a collection of random functions, each accompanying with a single state.

Le *et. al.* [30] have proposed a recognition algorithm regarding human posture using human skeleton. Here Kinect sensor is used to capture color, depth and motion information from the input images. Four postures e.g. standing, bending, lying and sitting are recorded in front of the Kinect. Joint angles are the extracted features in this work. Seven different experiments were performed using the joint angles with and/or without scaling via online and offline modes. Support vector machine [31] is used for human posture recognition from the extracted features. The accuracy of this work achieves an average rate of 90 %. This work finds usage in health monitoring.

Linear support vector machines (SVM) distinguishes two classes of data, known by ‘support vectors’, by creating a hyper plane within the training dataset, such that the distance between the support vectors is maximized.

Lai *et. al.* [32] proposed a paper where Kinect camera has been used for hand gesture recognition. Here Kinect sensor along with software development kit (SDK) provides the skeleton models taken from human postures. Two recognition methods have been developed; firstly recognition in feature covariance space using log-Euclidean distance and secondly recognition in feature space using Euclidean distance. Nearest neighbor with Frobenius norm as a metric is used as a classifier. Eight hand gestures (arm swing, arm push, arm back (both for right and left hand), zoom in and zoom out) are recognized in real time with an accuracy of 99.5%.

Feng and Lin [33] have proposed an alarm system for the senior using a series of human body postures. Here, background difference algorithm is used to achieve the motion area segmentation and to determine whether the motion area is a person. For human body posture recognition, various angles like front projection, side projection, back projection have been considered and seven invariant moments were extracted from the sample set to form human body eigen vector. Support vector machine was employed to classify the human body postures. The developed system is mainly used to help as well as guide old person to give an alarm when they are supposed to fall in their home. The main drawback for incorrect identification of postures may be due to shadow and light direction changing.

Parajuli *et. al.* [34] have proposed an elderly health monitoring system to monitor senior people providing Kinect device. Gestures are recognized when elders are supposed to fall by estimating their gait or changes in their postures. Data are collected depending upon the viewing angle and further the data are adjusted with respect to height and shoulder width. SVM is used to decompose the gait and posture data obtained from the Kinect sensor. This method has the ability to work with high dimensional data.

Galna *et. al.* [35] have drawn a fine approach for testing the accuracy of Microsoft Kinect for gesture recognition to measure movements in people suffering from Parkinson's Disease(a neurodegenerative disorder marked by trembling, stiffness in muscles and slowness of movements). The Kinect device along with a Vicon three dimensional motion analysis system (gold-standard) is used to assess the functional and clinically apposite crusades in people with PD. A sample of ten healthy controls followed by nine people with more or less severe PD have been participated in this experiment. Standing still, reaching forward and sideways, stepping forwards and sideways, walking on the spot are the movements recorded during the experiment. Mean bias has been assessed between the Kinect-Vicon pair using a sequence of repeated-measure two sided t-tests. Pearson's r correlation is used to estimate the relative agreement between the two systems. Absolute accuracy has been measured using intra class correlation and 95% limits of agreement. The precision of upper limb kinematics has been explored by Kinect, as PD has an adverse effect on reaching and grasping. This method is further analyzed and progressed to mend the tracking of smaller movements and evolve user-friendly software to oversee PD indication in home.

Zhou *et. al.* [36] have proposed a paper to explore a real time system for in-home activity monitoring and working imposement of elderly persons. By embedding a video sensor along with intelligent computer vision and learning algorithm in the living environment , continuous home activities can be assessed and efficiency can be enhanced of elders. Data is collected and it is classified with the help of silhouette extraction. Silhouette extraction serves two main purposes; one is privacy protection in the living room and the other is human motion tracking and spatio-temporal feature extraction for intelligent automation. Here high level knowledge is blended with low level feature related clustering results to deal time varying changes in ambient light conditions. Moving speed is the extracted feature for recognizing human actions; such as standing, sitting, walking etc. This algorithm accomplishes smaller error rate of 7.3% than the algorithm in [37]. Furthermore advanced data mining algorithm has been developed to examine several patterns and interlink them for automated functional assessment and early recognition of health problems.

Oniga *et. al.* [38] have proposed a hand gesture recognition module where intelligent man-machine interfacing (MMI) has been introduced. The proposed module consists of two subsystems where the first subsystem is a bracelet that recognizes the hand movements using accelerometers, and the second one is a control box used to process the captured data. Five simple hand gestures like up, down, idle, left and right have been recognized with the help of feed-forward Neural Network with two layer architecture.

Ren *et. al.* [39] have developed a system regarding hand gesture recognition using Kinect sensor. Here a robust part based technique have been introduced to recognize several hand gestures. The Kinect sensor captures the color images at a resolution of 640×480. Then hand detection is performed by hand segmentation and shape representation followed by drawing the time series curves between distance and angle. Gesture recognition is carried out using dissimilarity measure and template matching. The pros of this method is that it can handle noisy hand shapes with an accuracy of 93.2 %. It has been demonstrated in real life applications such as arithmetic computation and rock-paper-scissors game.

Sadhu *et. al.* [40] have designed a simple system for person identification using their gait posture. They have used Kinect sensor to produce the twenty body joint coordinates, out of which only nine are considered in this algorithm. Eight features (movements of eight body joints with respect to hip center) have been extracted followed by coordinate normalization based on the hip center. A high accuracy rate of 92.48% is achieved using this algorithm which is better than SVM and kNN. The algorithm is reliable and robust and finds application in surveillance.

Saha *et. al.* [41] have presented a paper where emotion recognition has been carried out from different body gestures using a cost effective sensor known as Kinect. This work is concerned with the recognition of gestures corresponding to five basic human emotional states viz. anger, sadness, fear, happiness and relaxation. Nine unique features, based on the angles-displacements-acceleration between the different joints, have been extracted. Five types of classifiers (binary decision tree, ensemble decision tree, k-nearest neighbor, SVM with radial basis function kernel and neural network with back propagation learning) have been used for data classification. The best result is obtained by ensemble decision tree [42] classifier with an accuracy of 90.83% and a computation time of 15.59 sec. The time requirement is best for binary decision tree classifier with 0.9 sec. It finds application in controlling devices automatically according to a person's emotion.

Yu *et. al.* [43] have proposed an archetype system, where children tantrum behavior in video has been analyzed depending upon Microsoft Kinect sensor. This paper utilizes medical knowledge, questionnaire for attitudes investigation and assessment, Kinect for data collection and stochastic grammar based behavior analysis algorithm. Some special behavior; such as push, shout, attack have been detected and classified by means of machine learning based object tracking and behavior analysis algorithm. Principal Component Analysis (PCA) [44] is applied for dimensionality reduction and Euclidean method is used for distance measurement between behaviors. Finally K-means clustering algorithm [45] is implemented to extract features. Three groups of experiments have been conducted to illustrate our system with other two conventional methods such as camera recording and memory based handwriting to register the frequency of childrens' attitudes of intense emotion changes, intense degree, lasting period, possible triggering events and the reaction of parents. Our proposed system finds high correlation (0.731) in some measuring factors which shows its effectiveness and works better in workload and subject feeling. In future behavior analysis algorithm is supposed to be optimized for fully use of Kinect.

## 1.5 Scope of the Thesis

The proposed work aims at recognition of unknown gestures by some standard machine learning algorithms. The objectives of the thesis are,

1. Human Machine Intelligent Interaction (HMII).
2. Advance Driver Assistance Systems (ADAS).
3. Intelligent Robotics.
4. Monitoring and Surveillance.
5. Gaming.
6. Research on pain and depression.
7. Health care appliances.
8. Advance Computing Technologies.

## 1.6 Organization of the Thesis

In chapter 1, gestures and gesture recognition tools are explained. An improved Human-Computer Interaction (HCI) using hand gesture recognition using Kinect sensor is proposed in chapter 2. Two person interaction has been detected via Ensemble decision tree in chapter 3. Finally, thesis is summarized and concluded in chapter 4. Computer programs along with outputs in the accompanying CD-ROM are given in Appendix section. Bibliography is given at the end of each chapter.

## References

1. Mitra, Sushmita, and Tinku Acharya. "Gesture recognition: A survey."*Systems, Man, and Cybernetics, Part C: Applications and Reviews, IEEE Transactions on* 37, no. 3 (2007): 311-324.
2. Hasan, Mokhtar M., and Pramod K. Mishra. "Hand gesture modeling and recognition using geometric features: a review." *Canadian Journal on Image Processing and Computer Vision* 3, no. 1 (2012): 12-26.
3. Ahuja, Mandeep Kaur, and Amardeep Singh. "A Survey of Hand Gesture Recognition." *International Journal* 3, no. 5 (2015).
4. Ionescu, Bogdan, Didier Coquin, Patrick Lambert, and Vasile Buzuloiu. "Dynamic hand gesture recognition using the skeleton of the hand."*EURASIP Journal on Advances in Signal Processing* 2005, no. 13 (2005): 1-9.
5. Sarkar, Arpita Ray, G. Sanyal, and S. Majumder. "Hand gesture recognition systems: a survey." *International Journal of Computer Applications* 71, no. 15 (2013).
6. Khan, Rafiqul Zaman, and Noor Adnan Ibraheem. "HAND GESTURE RECOGNITION: A Literature." (2012).
7. Hasan, Haitham Sabah, and S. Abdul Kareem. "Human computer interaction for vision based hand gesture recognition: a survey." In *Advanced Computer Science Applications and Technologies (ACSAT), 2012 International Conference on*, pp. 55-60. IEEE, 2012.
8. Malima, Asanterabi, Erol Özgür, and Müjdat Çetin. "A fast algorithm for vision-based hand gesture recognition for robot control." In *Signal Processing and Communications Applications, 2006 IEEE 14th*, pp. 1-4. IEEE, 2006.
9. Hu, Weiming, Tieniu Tan, Liang Wang, and Steve Maybank. "A survey on visual surveillance of object motion and behaviors." *Systems, Man, and Cybernetics, Part C: Applications and Reviews, IEEE Transactions on* 34, no. 3 (2004): 334-352.
10. Vijay, Paranjape Ketki, Naphade Nilakshi Suhas, Chafekar Suparna Chandrashekhar, and Deshpande Ketaki Dhananjay. "Recent developments in sign language recognition: a review." *Int J Adv Comput Eng Commun Technol* 1 (2012): 21-26.
11. Starner, Thad, and Alex Pentland. "Real-time american sign language recognition from video using hidden markov models." In *Motion-Based Recognition*, pp. 227-243. Springer Netherlands, 1997.
12. Murthy, G. R. S., and R. S. Jadon. "A review of vision based hand gestures recognition." *International Journal of Information Technology and Knowledge Management* 2, no. 2 (2009): 405-410.

13. Freeman, William T., and Craig Weissman. "Television control by hand gestures." In *Proc. of Intl. Workshop on Automatic Face and Gesture Recognition*, pp. 179-183. 1995.
14. Konrad, T., Demirdjian, D. & Darrell, T. “Gesture + Play: Full-Body Interaction for Virtual Environments”. In: CHI ’03 Extended Abstracts on *Human Factors in Computing Systems.* ACM Press, (2003) 620–621.
15. Han, Jungong, Ling Shao, Dong Xu, and Jamie Shotton. "Enhanced computer vision with microsoft kinect sensor: A review." *Cybernetics, IEEE Transactions on* 43, no. 5 (2013): 1318-1334.
16. Gu, Ye, Ha Do, Yongsheng Ou, and Weihua Sheng. "Human gesture recognition through a Kinect sensor." In *Robotics and Biomimetics (ROBIO), 2012 IEEE International Conference on*, pp. 1379-1384. IEEE, 2012.
17. Smisek, Jan, Michal Jancosek, and Tomas Pajdla. "3D with Kinect." In *Consumer Depth Cameras for Computer Vision*, pp. 3-25. Springer London, 2013.
18. Chen, Lingchen, Feng Wang, Hui Deng, and Kaifan Ji. "A survey on hand gesture recognition." In *Computer Sciences and Applications (CSA), 2013 International Conference on*, pp. 313-316. IEEE, 2013.
19. Xia, Lu, Chia-Chih Chen, and Jake K. Aggarwal. "Human detection using depth information by kinect." In *Computer Vision and Pattern Recognition Workshops (CVPRW), 2011 IEEE Computer Society Conference on*, pp. 15-22. IEEE, 2011.
20. Rafibakhsh, Nima, Jie Gong, Mohsin K. Siddiqui, Chris Gordon, and H. Felix Lee. "Analysis of xbox kinect sensor data for use on construction sites: depth accuracy and sensor interference assessment." In *Constitution research congress*, pp. 848-857. 2012.
21. Li, Yi. "Hand gesture recognition using Kinect." In *Software Engineering and Service Science (ICSESS), 2012 IEEE 3rd International Conference on*, pp. 196-199. IEEE, 2012.
22. Zhang, Zhengyou. "Microsoft kinect sensor and its effect." *MultiMedia, IEEE*19, no. 2 (2012): 4-10.
23. Ballester, Jorge, and Chuck Pheatt. "Using the Xbox Kinect sensor for positional data acquisition." *American journal of Physics* 81, no. 1 (2013): 71-77.
24. Oszust, Mariusz, and Marian Wysocki. "Recognition of signed expressions observed by Kinect Sensor." In *Advanced Video and Signal Based Surveillance (AVSS), 2013 10th IEEE International Conference on*, pp. 220-225. IEEE, 2013.
25. P. Cunningham and S. J. Delany, "k-Nearest neighbour classifiers," *Multiple Classifier Systems*, pp. 1-17,2007.
26. Chikkanna, Mahesh, and Ram Mohana Reddy Guddeti. "Kinect based real-time gesture spotting using hcrf." In *Advances in Computing, Communications and Informatics (ICACCI), 2013 International Conference on*, pp. 925-928. IEEE, 2013.

27. Wang, Sy Bor, Ariadna Quattoni, Louis-Philippe Morency, David Demirdjian, and Trevor Darrell. "Hidden conditional random fields for gesture recognition." In *Computer Vision and Pattern Recognition, 2006 IEEE Computer Society Conference on*, vol. 2, pp. 1521-1527. IEEE, 2006.
28. *Wang, Youwen, Cheng Yang, Xiaoyu Wu, Shengmiao Xu, and Hui Li. "Kinect based dynamic hand gesture recognition algorithm research."* In Intelligent Human-Machine Systems and Cybernetics (IHMSC), 2012 4th International Conference on, vol. 1, pp. 274-279. IEEE, 2012.
29. Prabhu, Srikanth, R. N. Banerjee, A. K. Ray, N. S. Giridhar, N. V. Subbareddy, and K. V. Prema. "Gesture recognition using hmm." (2005).
30. Le, Thi-Lan, Minh-Quoc Nguyen, and Thi-Thanh-Mai Nguyen. "Human posture recognition using human skeleton provided by Kinect." In*Computing, Management and Telecommunications (ComManTel), 2013 International Conference on*, pp. 340-345. IEEE, 2013.
31. Mavroforakis, Michael E., and Sergios Theodoridis. "A geometric approach to support vector machine (SVM) classification." *IEEE transactions on neural networks* 17, no. 3 (2006): 671-682.
32. Lai, Kam, Janusz Konrad, and Prakash Ishwar. "A gesture-driven computer interface using Kinect." In *Image Analysis and Interpretation (SSIAI), 2012 IEEE Southwest Symposium on*, pp. 185-188. IEEE, 2012.
33. Feng, Gui, and Qiwei Lin. "Design of Elder alarm system based on body posture reorganization." In *Anti-Counterfeiting Security and Identification in Communication (ASID), 2010 International Conference on*, pp. 249-252. IEEE, 2010
34. Parajuli, Monish, Dat Tran, Wanli Ma, and Divya Sharma. "Senior health monitoring using Kinect." In *Communications and Electronics (ICCE), 2012 Fourth International Conference on*, pp. 309-312. IEEE, 2012.
35. Galna, Brook, Gillian Barry, Dan Jackson, Dadirayi Mhiripiri, Patrick Olivier, and Lynn Rochester. "Accuracy of the Microsoft Kinect sensor for measuring movement in people with Parkinson's disease." *Gait & posture* 39, no. 4 (2014): 1062-1068.
36. Zhou, Zhongna, Wenqing Dai, Jay Eggert, Jarod T. Giger, James Keller, Marilyn Rantz, and Zhihai He. "A real-time system for in-home activity monitoring of elders." In *Engineering in Medicine and Biology Society, 2009. EMBC 2009. Annual International Conference of the IEEE*, pp. 6115-6118. IEEE, 2009.
37. C. Wren, A. Azarbayejani, T. Darrell, and A.P. Pentland, “Pfinder: Real-Time Tracking of the Human Body,” IEEE Trans. Pattern Analysis and Machine Intelligence, vol. 19, no. 7, pp. 780-785, July 1997
38. Oniga, Stefan, János Vegh, and Ioan Orha. "Intelligent human-machine interface using hand gestures recognition." In *Automation Quality and Testing Robotics (AQTR), 2012 IEEE International Conference on*, pp. 559-563. IEEE, 2012.

39. Ren, Zhou, Junsong Yuan, Jingjing Meng, and Zhengyou Zhang. "Robust part-based hand gesture recognition using kinect sensor." *Multimedia, IEEE Transactions on* 15, no. 5 (2013): 1110-1120.
40. Sadhu, A.K., Saha, S., Konar, A. and Janarthanan, R., 2014, January. Person identification using Kinect sensor. In *Control, Instrumentation, Energy and Communication (CIEC), 2014 International Conference on* (pp. 214-218). IEEE.
41. Saha, S., Datta, S., Konar, A. and Janarthanan, R., 2014, April. A study on emotion recognition from body gestures using Kinect sensor. In *Communications and Signal Processing (ICCSP), 2014 International Conference on* (pp. 056-060). IEEE.
42. R. Polikar, "Ensemble based systems in decision making," *Circuits and Systems Magazine, IEEE,* vol. 6, no. 3, pp. 21-45, 2006.
43. Yu, Xiaoyi, Lingyi Wu, Qingfeng Liu, and Han Zhou. "Children tantrum behaviour analysis based on Kinect sensor." In *Intelligent Visual Surveillance (IVS), 2011 Third Chinese Conference on*, pp. 49-52. IEEE, 2011.
44. Wold, Svante, Kim Esbensen, and Paul Geladi. "Principal component analysis." *Chemometrics and intelligent laboratory systems* 2, no. 1-3 (1987): 37-52.
45. Kanungo, Tapas, David M. Mount, Nathan S. Netanyahu, Christine D. Piatko, Ruth Silverman, and Angela Y. Wu. "An efficient k-means clustering algorithm: Analysis and implementation." *Pattern Analysis and Machine Intelligence, IEEE Transactions on* 24, no. 7 (2002): 881-892.

# CHAPTER 2

# GESTURE BASED IMPROVED HUMAN-COMPUTER INTERACTION

# USING

# MICROSOFT'S KINECT SENSOR

*This chapter provides a simple and robust gesture recognition system for better human-computer interaction using Microsoft's Kinect sensor. The Kinect is employed to construct skeletons for a subject in the 3D space using twenty body joint coordinates. From this skeletal information, ten joints are required and six triangles have been constructed along with six respective centroids. The feature space corresponds to the Euclidean distances between spine joint and the centroids for each frame. For classification purpose, support vector machine is used using a kernel function. The proposed work is widely applicable for several gesture driven computer applications and produces an average accuracy rate of 88.7%.*

## 2.1 Introduction

Hand gesture recognition [1] has been granted as one of the emerging fields of research today providing a natural way of communication between man and a machine. Gestures are some forms of body motions which a person expresses when doing a work or giving a reply. Gestures can be of static type (certain pose or configuration, or still body posture) or dynamic type (movements of different body parts). Presently some gestures are used for communication purpose.

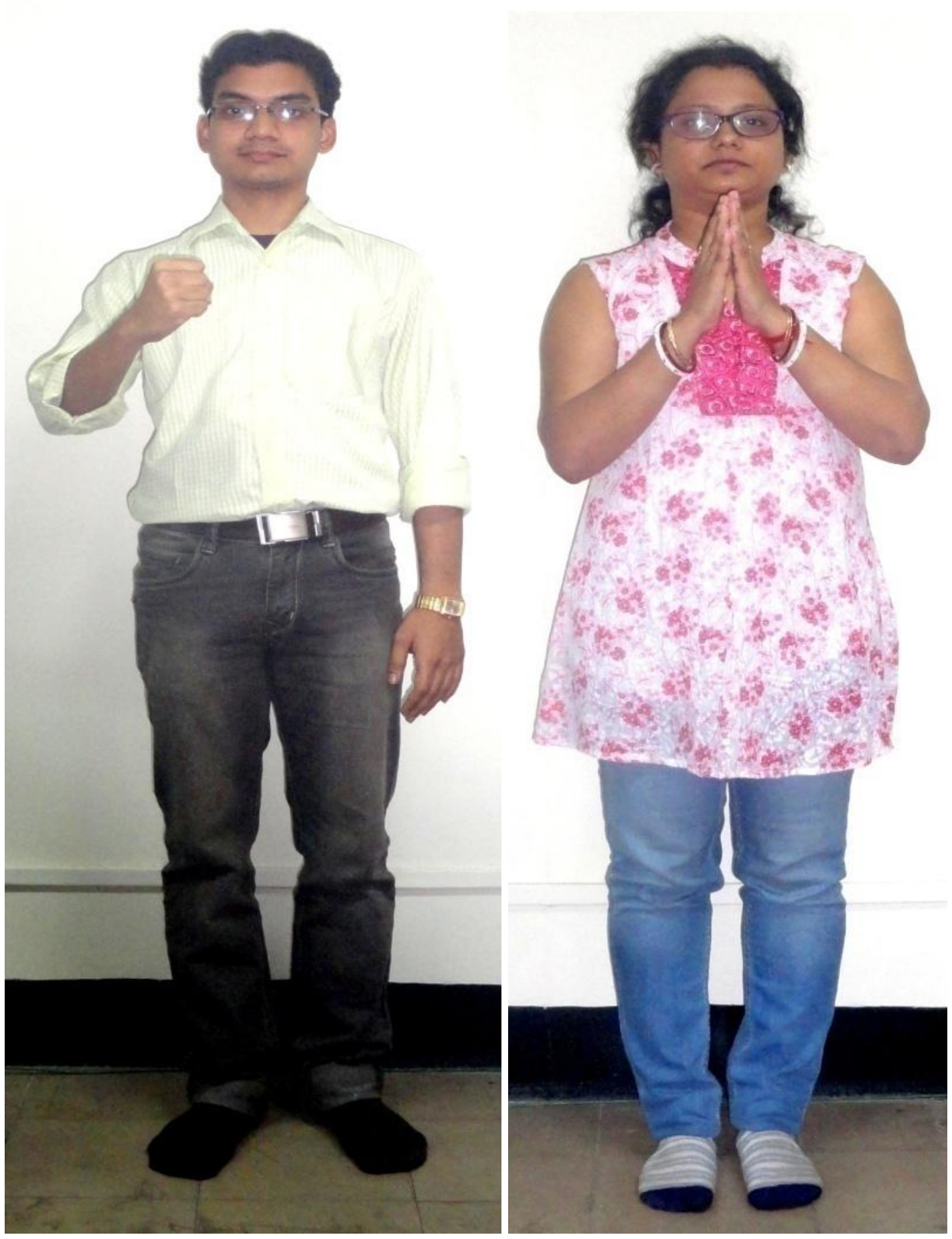

(a) (b)

Fig 2.1 Gestures (a) 'puching' (b) 'greeting'

Also gestures, various types of actions, have been used to convey information without speaking anything, *e.g.* traffic/ umpiring in the field *etc.*

For recognition of gestures, we have used Microsoft's Kinect sensor [2,3]. Kinect is the official name of XBOX360 issued by Microsoft on June, 2010. The hardware portion of this sensor comprises of three cameras such as RGB camera, an infrared (IR) camera and an IR projector; and an array of microphone, which can produce RGB images, depth images and audio signals respectively. As far as the software tools are concerned, they are able to catch human motions in three dimensional (3D) spaces.

Gesture recognition technology has been employed for several domains, like hand gesture, dynamic hand gesture tracking, posture recognition, sign language recognition, robot control, person identification, healthcare etc.

Saha *et al.*[4] have designed a simple hand gesture recognition module where feelings and expressions of single person have been analyzed. Microsoft's Kinect sensor is used to track the human body joint coordinates in three dimensional spaces. A total of ten hand gestures (six single hand and four double hand) have been processed. Left hand or right hand coordinate is processed for one hand gesture while both hand coordinates are processed for double hand gestures. Variation for right and left hand per frame is taken as the features for this work. After that, normalization is performed using hip center coordinate. Support vector machine [5] is applied for classification purpose achieving a recognition rate of 94.3%.

Malima *et al.*[6] have proposed a simple algorithm mainly for hand gesture recognition problem through vision. This algorithm is mainly used in navigation problem such as robot navigation where robot movements have been manipulated by some sort of hand signs provided by human being captured through a camera. This approach of hand gesture recognition consists steps like hand region segmentation, which is performed by skin color detection; finger identification by locating centroid, constructing circle, counting number of black to white transitions; and finally gesture classification by determining the number of intersecting fingers. The derived algorithm does not vary with rotation, scaling and translation of the hand. An accuracy of 91% is achieved throughout the experiment.

Alsheakhali *et al*. [7] have introduced a new hand gesture recognition technique for recognizing dynamic gestures in a complex background. The system consists of hand location detection, trajectory tracking of moving hand and then hand location variations are analyzed. Angles and distances of center region of hand in each frame and the movement counter in each direction are the extracted features. The algorithms of hand movement tracking, single hand detection and both hand detection are executed using optimized C# code. Twelve different hand gestures are recognized with a recognition rate of 94.21%. Its performance is good under various scene backgrounds and illumination conditions.

Singha and Das [8] have designed a gesture recognition system based on various hand gestures. Karhunen-Loeve transform [9] technique has been applied to reduce dimensions and to eliminate correlated data. The designed gesture recognition technique possesses mainly five steps; skin filtering, hand cropping, edge detection, feature extraction and finally classification. Detection of hand has been performed by skin filtering followed by cropping the palm of the hand. One of the finest edge detection technique, namely Canny edge detection, has been employed to draw the outline of the hand. Plotting of eigen vector by K-L transform is selected as the feature of this work. Classification has been done depending upon the angle created between eigen vector and a reference axis, and considering the Euclidean distance between test image and database images. A recognition rate of 96% has been obtained for ten unique hand gestures.

Rantaray and Agrawal [10] have designed a real time gesture recognition system with hands performing dynamic works. Images are taken from camera followed by background elimination procedure. Further hands are detected and Camshift technique have been applied for hand tracking. A simple detection technique named Haar has been employed as a recognizer for hand position locating and gesture classification. Seven hand gestures have been defined in this experiment for object manipulation. The pros of this method are its robustness, scalability and computational efficiency. It can work well even under noisy environment. This proposed technology is used for mouse controlling in the virtual environments.

Chen *et al*. [11] have introduced a recognition system regarding hand gestures for recognizing continuous gestures. First of all, tracking and shape extraction of the hands have been executed by skin color detection followed by edge detection followed by region of interest processing followed by background subtraction. Then suitable features have been extracted by Fourier descriptor and motion analysis. After that hidden Markov model training [12] has been executed for each input gestures evaluating the suitable parameters by respective algorithms. Finally gestures have been recognized separately using different HMMs. Twenty different gestures have been tested in this experiment with a recognizing rate of above 90%.

Correa *et al*. [13] have proposed an article where hand gesture recognition technique has been employed to communicate with a human robot. The proposed method is a collection of five modules. Firstly, face detection and tracking has been done using statistical classifiers and mean shift algorithm. The skin segmentation and motion analysis has been implemented using skin-doff algorithm and background elimination respectively. Hand detection and tracking section is done by same as face detection. RGB color histograms and rotation invariant LBP histograms are the extracted features for mean shift algorithm. Then static gesture recognition module is instrumented by some boosted and multiclass classifiers. Four static hand gestures like fist, palm, pointing and five were recognized, out of which 'five' has been recognized the best with 84.3% accuracy. Finally dynamic gesture recognition module evaluates temporal features to recognize gestures using Bayes classifier [14]. Ten dynamic gestures such as ten digits are recognized with a recognition rate of 84% in real-time, dynamic environments. The novelty is using context information for adapting skin model and restricting image's region.

We have used Microsoft's Kinect sensor to recognize the body gestures with the help of some specified features. Among twenty body joint co-ordinates collected by Kinect, only ten (spine, shoulder center, shoulder left, elbow left, wrist left, hand left, shoulder right, elbow right, wrist right hand right) of them are employed for this work. A total of six triangles per frame are formed by taking the 3D joint information by Kinect sensor. In the left side, three triangles are formed by shoulder center and four left arm joints. Similarly, three triangles are formed at the right section of the skeleton. For each six triangles, we get six centroids whose coordinate information is estimated by averaging the three suitable vertices of a particular triangle. The Euclidean distances from spine joint to the six centroids are the estimated features for this work. For a particular gesture, 3s video information is carried out to produce a skeleton. Support vector machine (SVM) is employed using a Gaussian kernel function for classification purpose in a one-against-all approach with an average accuracy rate of 88.7%.

This chapter is presented as follows. We have discussed some preliminary theories, i.e. Kinect sensor and SVM in section 2.2.1 and 2.2.2 respectively. Section 2.3 presents the overall proposed system, whereas detailed experimental results are given in section 2.4. Conclusion is drawn in section 2.5.

## 2.2 Preliminaries

This portion provides the two preliminary knowledge required regarding the Microsoft's Kinect sensor and support vector machine.

### 2.2.1 Kinect Sensor

The Kinect sensor [2,3] has been developed by Microsoft and the name is coined from the word 'kinematics', i.e., whenever a human is in front of the Kinect, the device is able to capture the human along with his twenty cartesian body joint coordinates in three dimensional (3D) space. Kinect is a black horizontal bar sensor and it looks like a webcam.

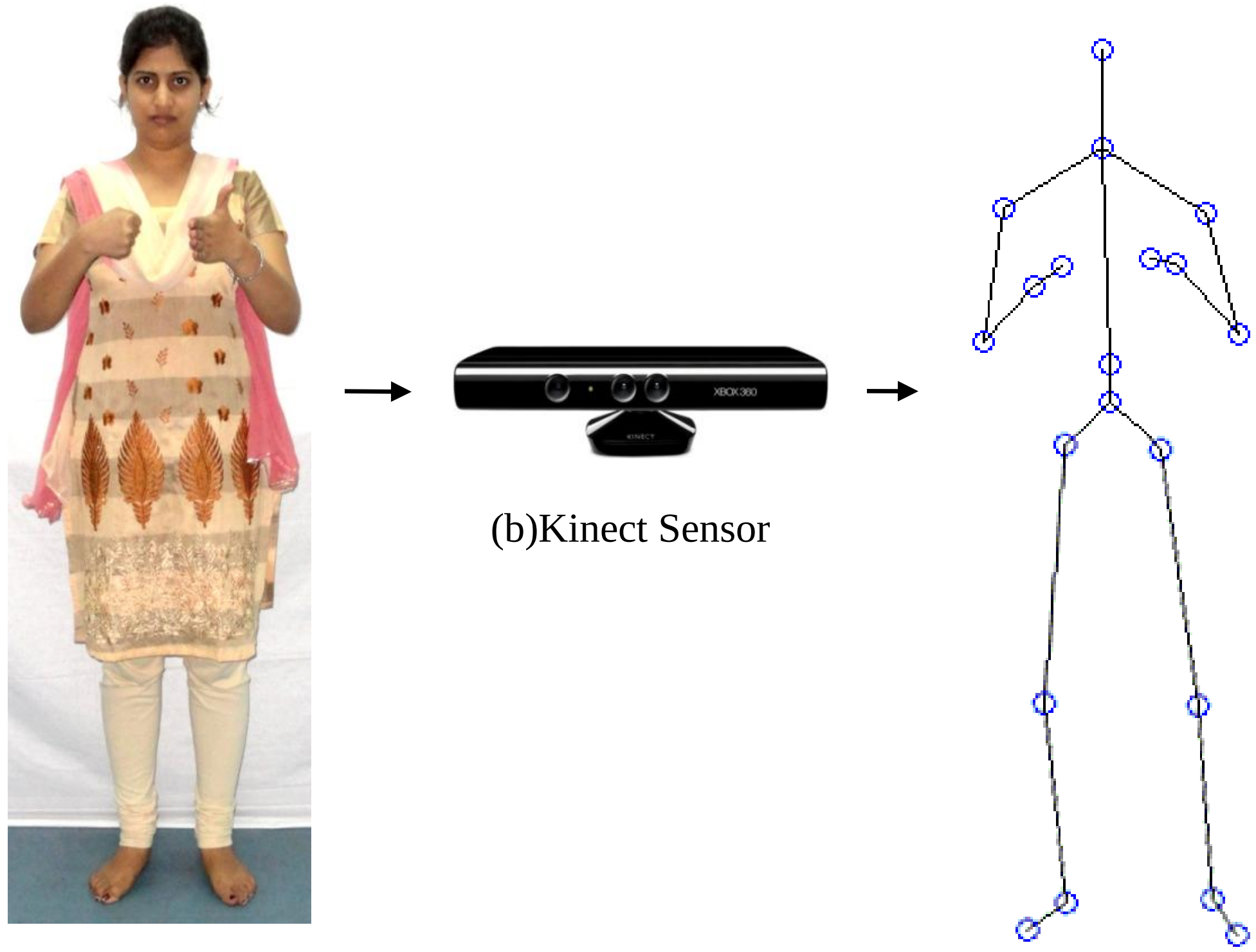

(a) RGB Image

(b)Kinect Sensor

(c)Skeletal Image

Fig 2.2 Kinect sensor with (a) RGB image, (b) sensor and (c) skeleton

### 2.2.2 Support Vector Machine

Support Vector Machine (SVM) [15] is a popular machine learning algorithm. We can design models for permitting computer to learn via several machine learning algorithms. Over the years, there have been so many machine learning algorithms proposed for classification purposes.

SVM was proposed by Vapnik [16] and has achieved a lot of attention because of its better performance. Till now, SVM is chosen to be the best classifier. Let us consider, the following two classes; class I represents the red stars while blue triangles are under class II. There are numerous lines for boundary, but SVM selects the superior decision boundary taking into account the maximum possible margin. SVM is a maximum margin type classifier as it creates a maximum margin hyper plane.

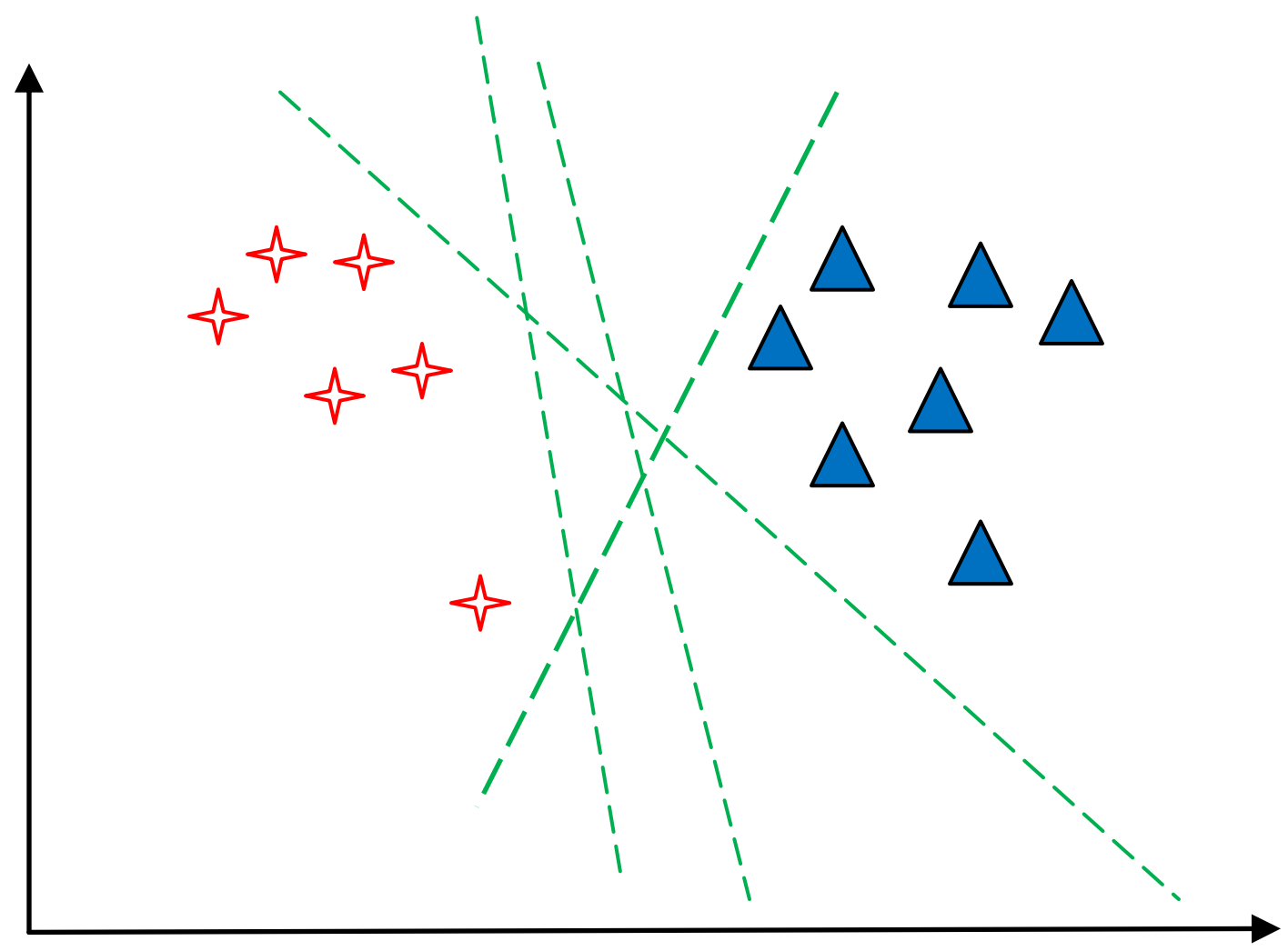

Fig 2.3 Introducing Support Vector Machine

The basic support vector machine (SVM) algorithm functions as a non-probabilistic binary linear classifier by separating a set of input vectors into two classes. Linear SVM divides the classes of data by building a hyper plane depending on 'support vectors', within the input training data set, such that the distance between the support vectors is maximized. Fig. 2.3 is provided for classification using linear SVM. Assuming, these feature vectors belong to either of the two classes $w_1$ and $w_2$ which are linearly separable. The hyper plane is represented by $g(x)$ in (2.1) where $w_0$ is called the threshold.

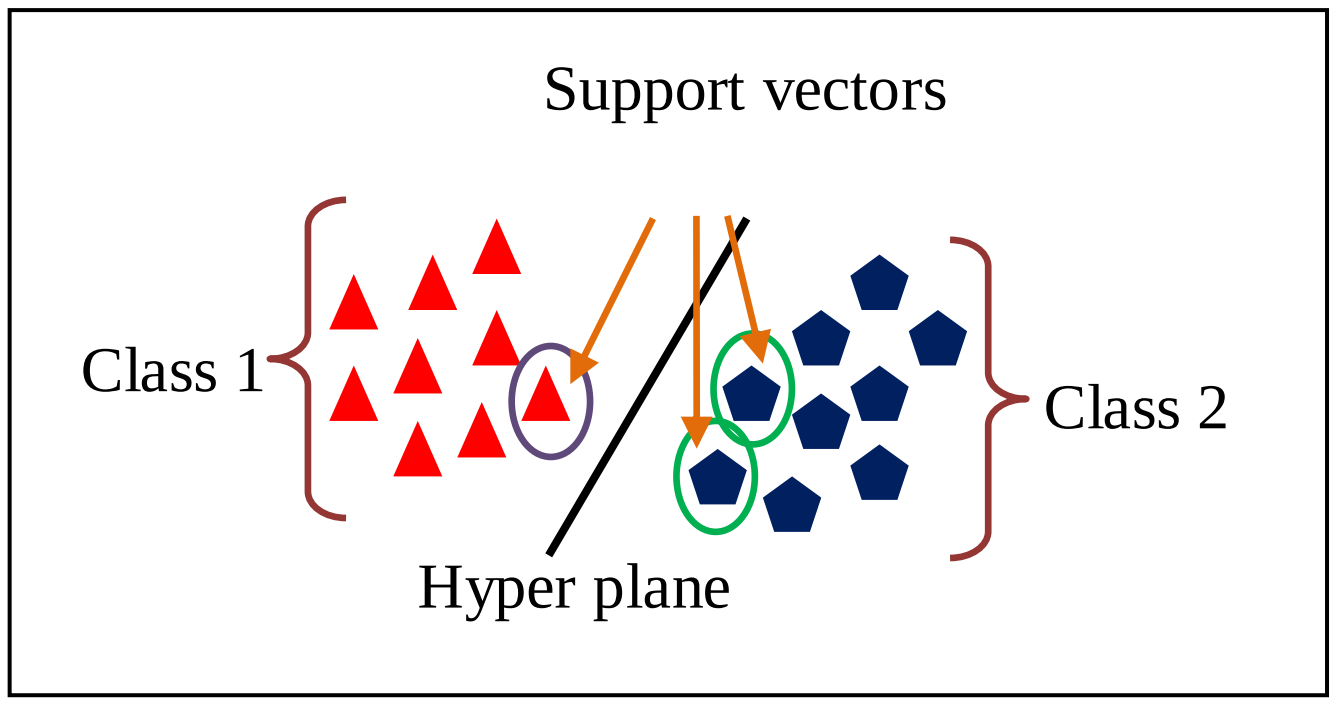

Fig 2.4 Separation of two classes using hyper plane.

$$g(x) = w^T x + w_o = 0 \qquad \textbf{(2.1)}$$

$$w^T x + w_o \geq 1 \forall x \in w_1 \qquad \textbf{(2.2)}$$

$$w^T x + w_o \leq 1 \forall x \in w_2 \qquad \textbf{(2.3)}$$

But linear SVM is effective if the data are linearly distinguishable. This condition can only be accomplished while mapping the data into a larger dimensional space accounting a kernel function [17,18]. The kernel function used in this work is Gaussian kernel having the width of the Gaussian as unity. Binary SVM classification is accomplished for each and every class of emotion in a one-to-all approach.

## 2.3 Details of the Proposed Work

The structural view of the entire work is given in Fig 2.5

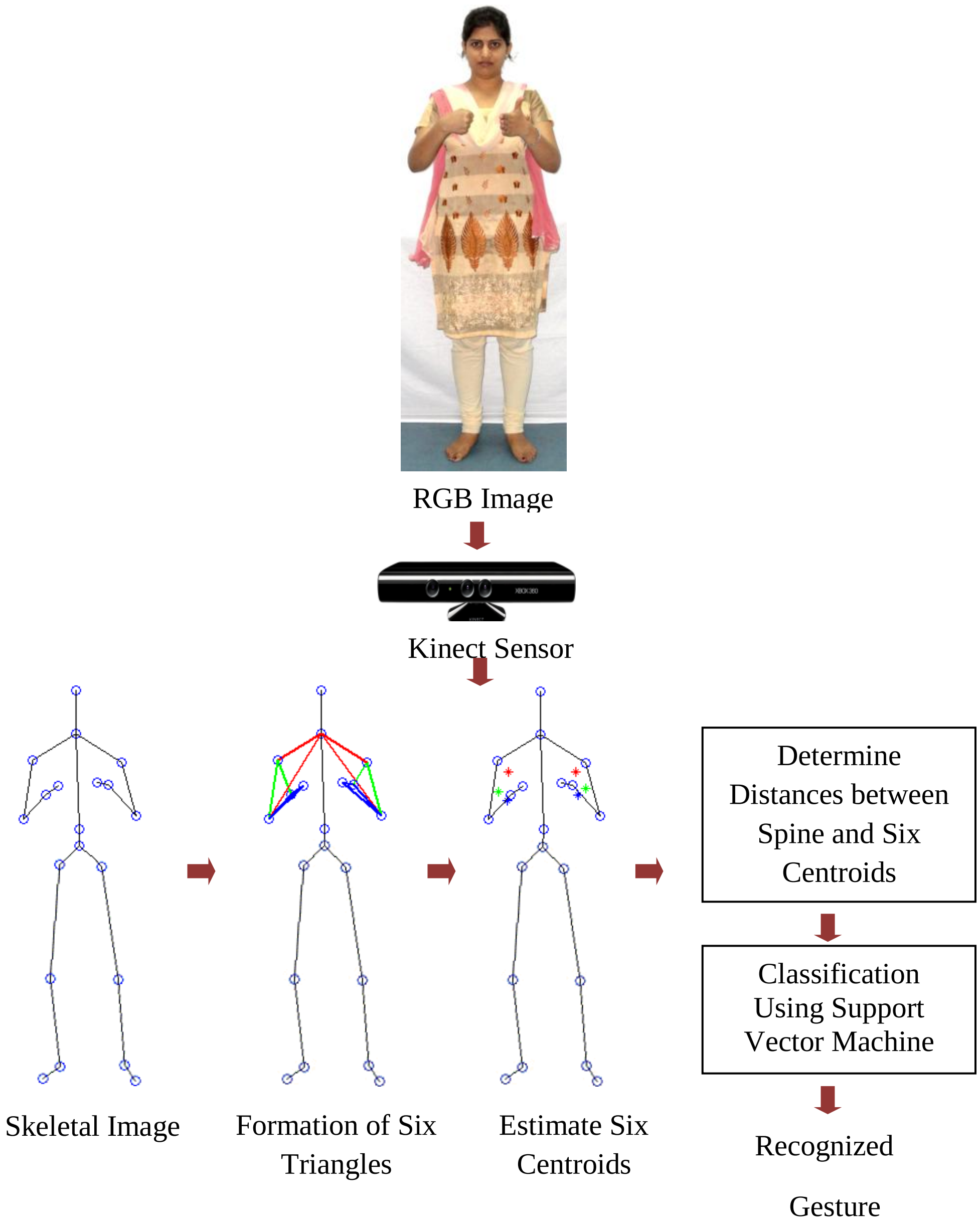

Fig 2.5 Block diagram of the proposed work

Suppose for our proposed work, we have *N* number of subjects and *G* number of gestures. Now each subject is asked to give data for each gesture for a total of *T* number of frames. Now for each *t*-th ($1 \leq t \leq T$) frame, we have twenty joints information in three dimensional space. From these twenty joints, only ten joints are needed for this proposed work. These required joints are spine (*S*), shoulder center (*SC*), shoulder left (*SL*), shoulder right (*SR*), elbow left (*EL*), elbow right (*ER*), wrist left (*WL*), wrist right (*WR*), hand left (*HL*) and hand right (*HR*) as shown in the figure below marked with red circles.

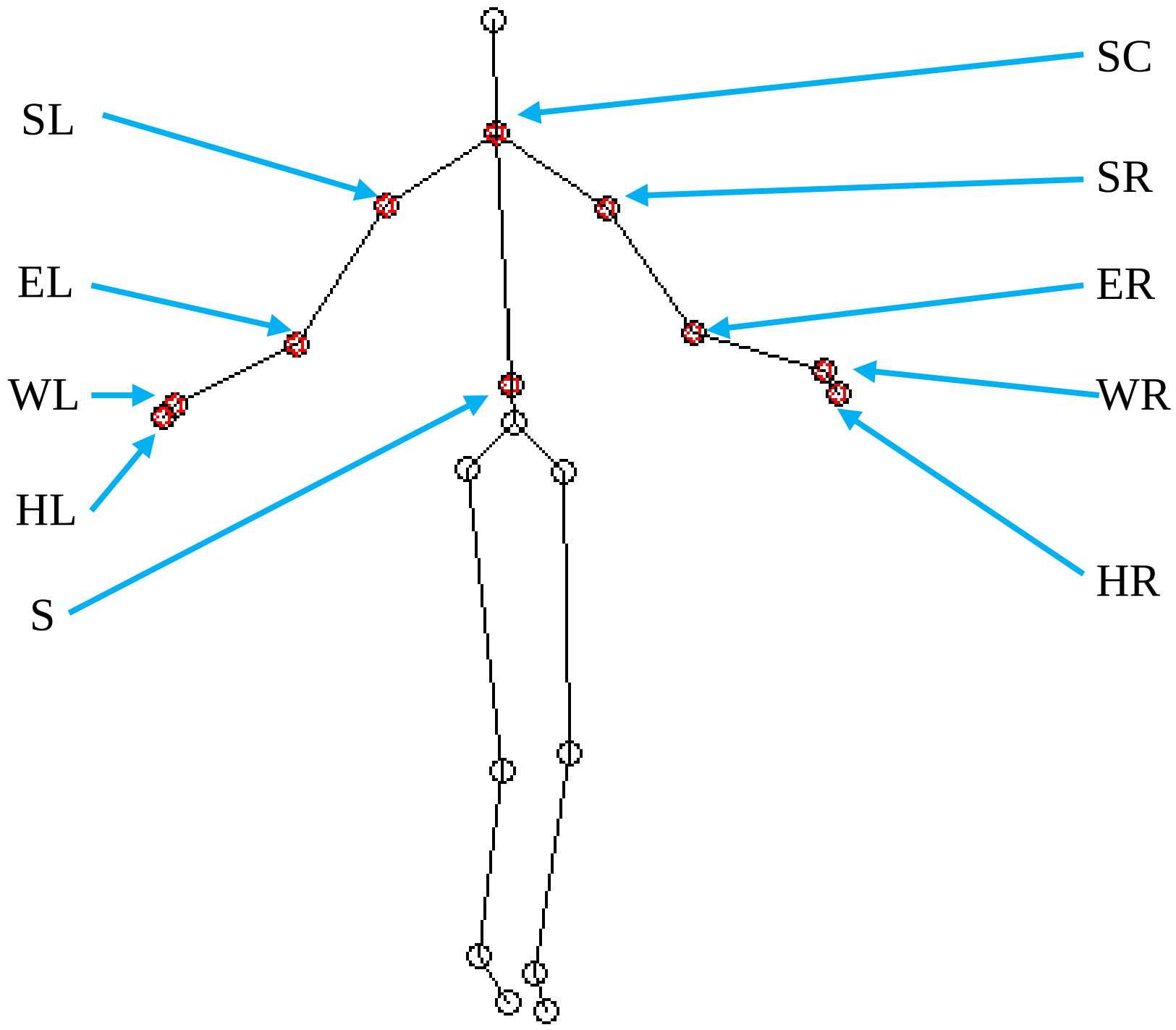

Fig 2.6 Required ten body joints for the proposed work

Using these ten joints, for each skeleton produced by Kinect sensor, six triangles have been constructed taking three joints at a time. These triangles are;

Triangle 1: $\Delta SCSEL^{t}$ formed using SC, SL and EL joints

Triangle 2: ΔSCSER$^{t}$ formed using SC, SR and ER joints

Triangle 3: ΔSEWL$^{t}$ formed using SL, EL and WL joints

Triangle 4: ΔSEWR$^{t}$ formed using SR, ER and WR joints

Triangle 5: ΔEWHL$^{t}$ formed using EL, WL and HL joints

Triangle 6: ΔEWHR$^{t}$ formed using ER, WR and HR joints

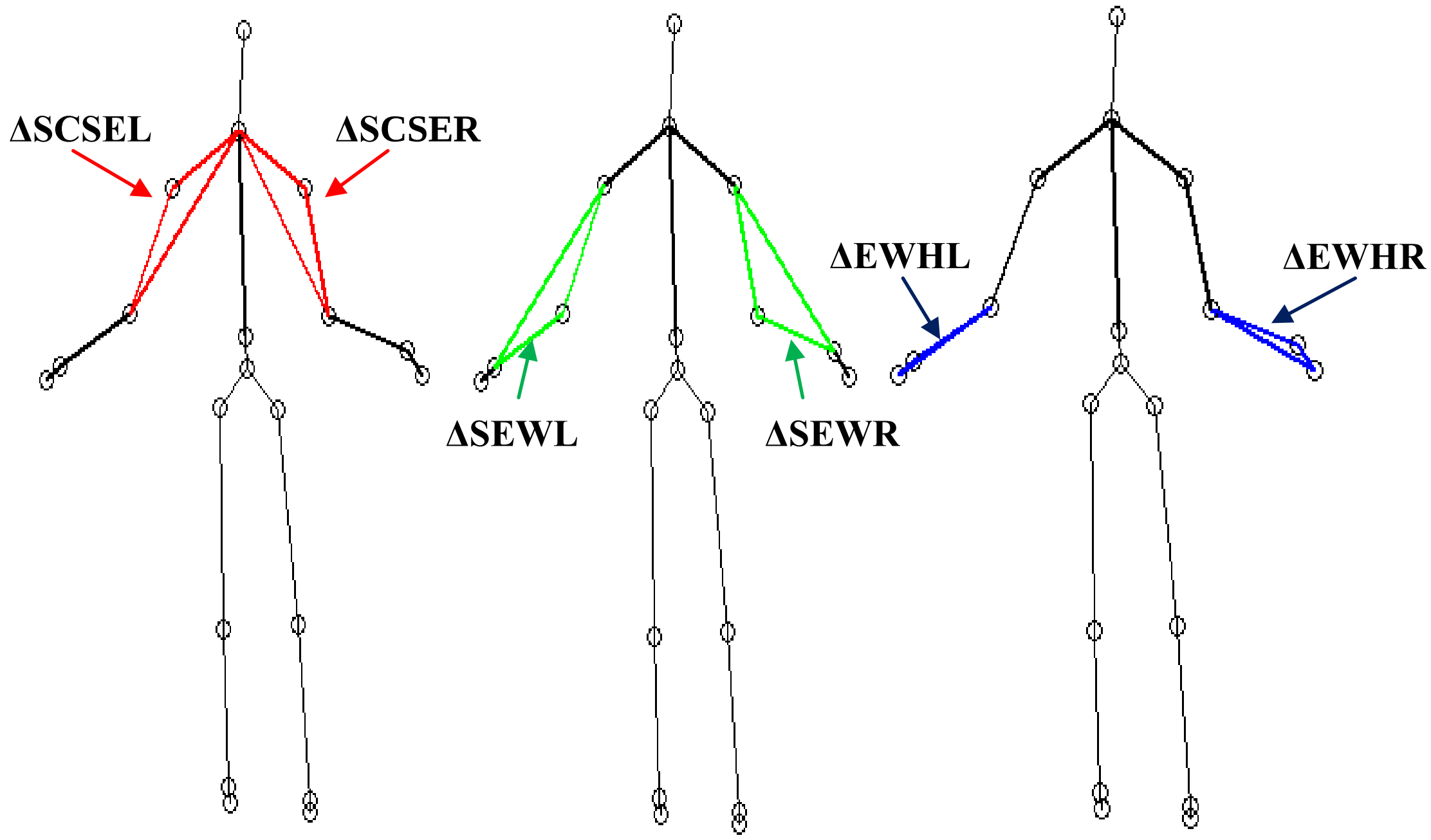

Fig 2.7 Formation of six triangles

From each of six triangles, centroid ($C_i^t$, 1≤i≤6) is calculated using equations (**2.4-2.6**). For the sake of simplicity, the centroid ($C_{SCSEL}{}^t$) [19] calculation for *ΔSCSEL* for *t*-th frame is provided here.

$$C_{SCSEL}^{t,x} = \frac{SC^{t,x} + SL^{t,x} + EL^{t,x}}{3} \tag{2.4}$$

$$C_{SCSEL}^{t,y} = \frac{SC^{t,y} + SL^{t,y} + EL^{t,y}}{3} \tag{2.5}$$

$$C_{SCSEL}^{t,z} = \frac{SC^{t,z} + SL^{t,z} + EL^{t,z}}{3} \tag{2.6}$$

where *x*, *y*, *z* are 3D usual axes.

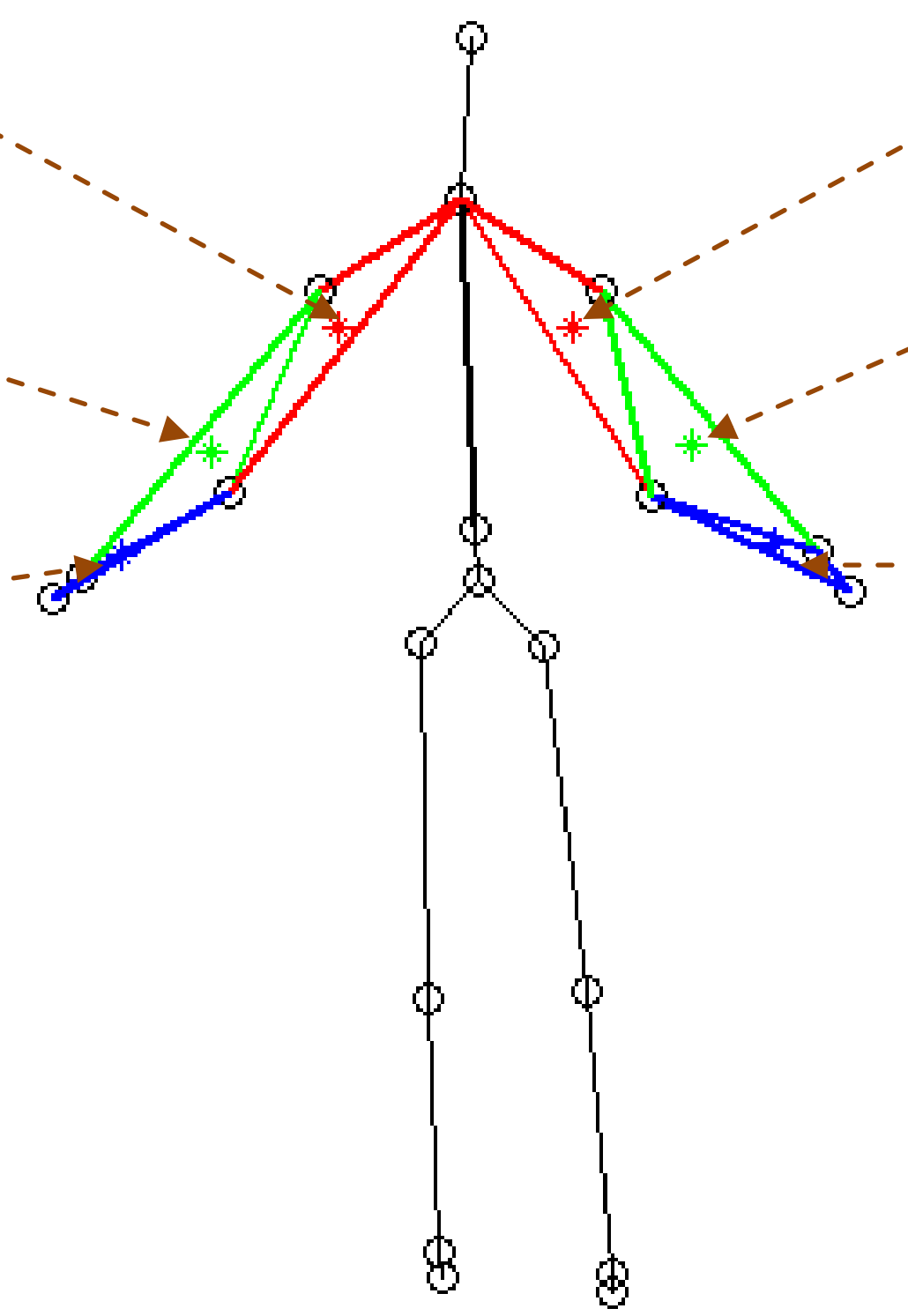

Fig 2.8 Six centroids along with their respective triangles

The features are calculated using Euclidean distances ($D^{i,t}$, 1≤i≤6) [20] between six centroids and spine ($S^t$) joint for that particular frame. As the distance between the subject and the Kinect plays a pivotal role in measuring distance, so normalization is applied based on the average value of the *z* information (i.e. depth) from $S^t$ and respective centroid ($C^t$). Thus for *t*-th frame, the six features are:

$$D^{1,t} = \frac{\left\| C^{t}_{SCSEL} - S^{t} \right\|}{\left( C^{t,z}_{SCSEL} + S^{t,z} \right) / 2} = 2 \times \frac{\left\| C^{t}_{SCSEL} - S^{t} \right\|}{C^{t,z}_{SCSEL} + S^{t,z}} \qquad \textbf{(2.7)}$$

$$D^{2,t} = 2 \times \frac{\left\| C_{SCSER}^{t} - S^{t} \right\|}{C_{SCSER}^{t,z} + S^{t,z}} \quad \textbf{(2.8)}$$

$$D^{3,t} = 2 \times \frac{\left\| C_{SEWL}^{t} - S^{t} \right\|}{C_{SEWL}^{t,z} + S^{t,z}} \quad \textbf{(2.9)}$$

$$D^{4,t} = 2 \times \frac{\left\| C_{SEWR}^{t} - S^{t} \right\|}{C_{SEWR}^{t,z} + S^{t,z}} \quad \textbf{(2.10)}$$

$$D^{5,t} = 2 \times \frac{\left\| C_{EWHL}^{t} - S^{t} \right\|}{C_{EWHL}^{t,z} + S^{t,z}} \quad \textbf{(2.11)}$$

$$D^{6,t} = 2 \times \frac{\left\| C_{EWHR}^{t} - S^{t} \right\|}{C_{EWHR}^{t,z} + S^{t,z}} \quad \textbf{(2.12)}$$

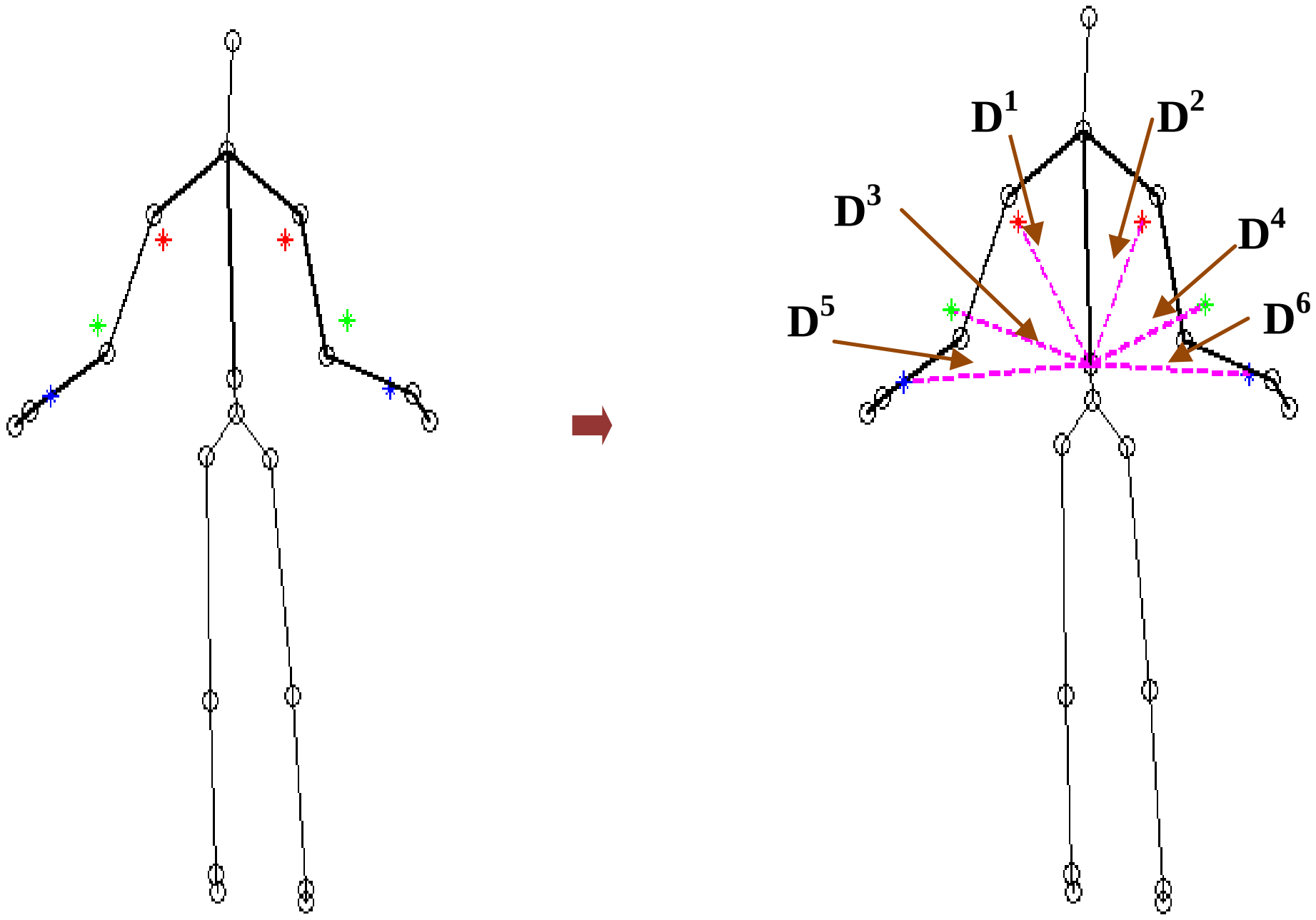

Six centroids

Six Euclidean distances between *S* and six centroids

Fig 2.9 (a) Six centroids (b) Six Euclidean distances between *S* and centroids

For each gesture given by each subject, we have $T$ frames and from each frame 6 features are extracted. Thus for each gesture given by each subject, the feature space dimension becomes $T\times 6$. As the total training database is constituted from all the $N$ subjects and $G$ gestures, thus the total dimension is $N\times G\times T\times 6$. Whenever an unknown gesture is captured by the Kinect sensor, then the testing data dimension is $T\times 6$. For the classification purpose, SVM is executed as already stated in section 2.2.2

## 2.4 Experimental Results

For the proposed system, we have taken into account both single and double hand gestures. A list of twenty gestures is given in Table 2.1

**Table 2.1** List of gestures

| ***Single hand gestures*** | ***Double hand gestures*** |
|---|---|
| 1. Waving | 1. Disgust |
| 2. Answering a call | 2. Clap |
| 3. Stop | 3. Greeting |
| 4. Slide | 4. Please |
| 5. Punching | 5. Push |
| 6. Picking up an object | 6. Grab |
| 7. Move up | 7. Zoom in |
| 8. Move down | 8. Zoom out |
| 9. Move left | 9. Move front |
| 10. Move right | 10. Move back |

### 2.4.1 Preparation of Training Dataset

Each gesture is taken for 3s duration thus T is 90 (=3s×30 frames/s) frames. Three distinct datasets are prepared as follows:

- Jadavpur university research scholar dataset with 15 female subjects (age 30±5yrs).
- Jadavpur university research scholar dataset with 15 male subjects (age 34±8yrs).
- Jadavpur university research scholar dataset with 8 female and 7 male subjects (age 36±9yrs).

Thus N is 15. The dimension for each gesture is 540 (=90×6). The total training dataset becomes 162000 (=15×20×90×6). For the testing purpose, we have asked separate twenty subjects (age 32±8yrs) to enact the gestures for 3s duration. Here the male and female subject ratio is 1:1.

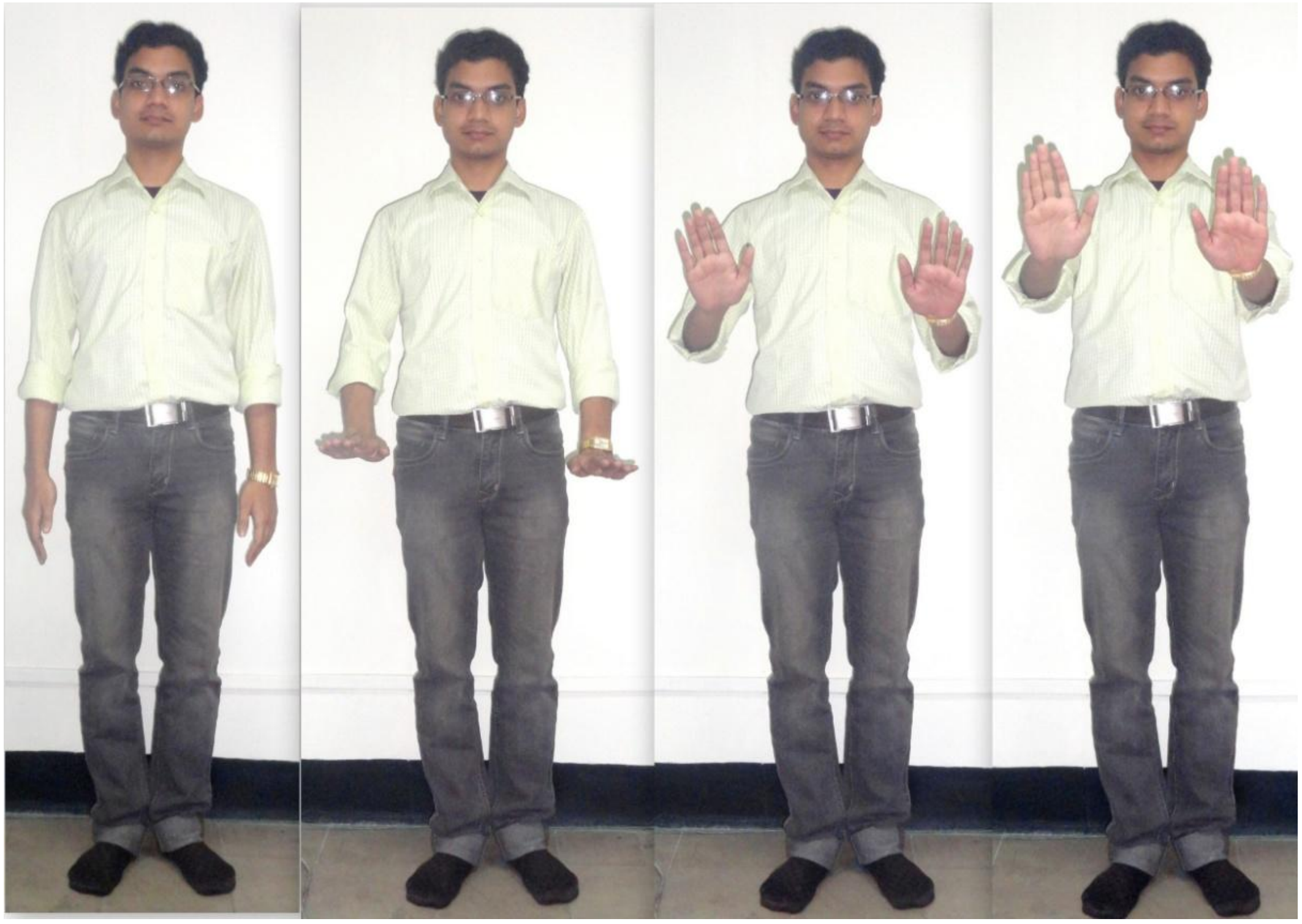

Fig 2.10 RGB images for gesture ‘push’ for four different frame numbers

### 2.4.2 Recognition of an Unknown Subject

The description of the feature extraction process for an unknown gesture is given in the previous section. From each frame, 6 features are extracted. For the lack of space, the features obtained for four different frames are provided in Table II. After feeding the SVM algorithm with the features obtained from unknown gesture for 3s duration, the output is obtained as ‘zoom out’. Thus the proposed work is able to correctly recognize the unknown gesture correctly.

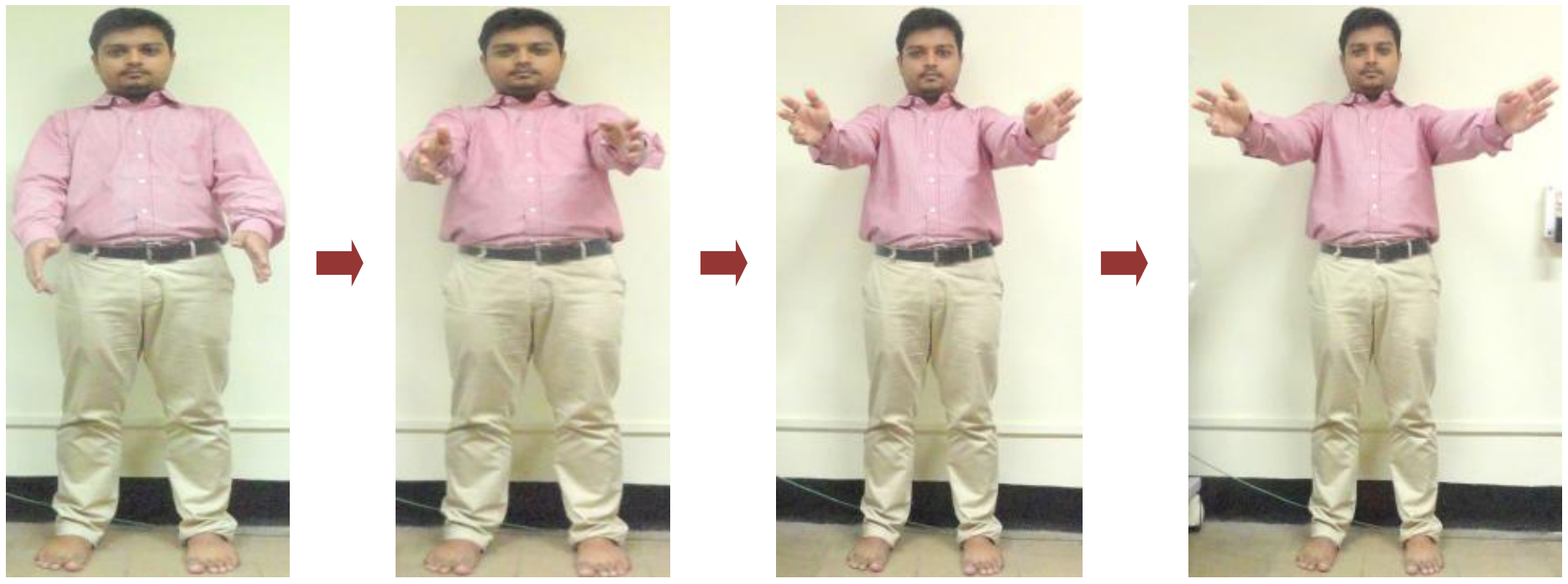

(a) RGB images for frame numbers 10, 30, 60 and 85

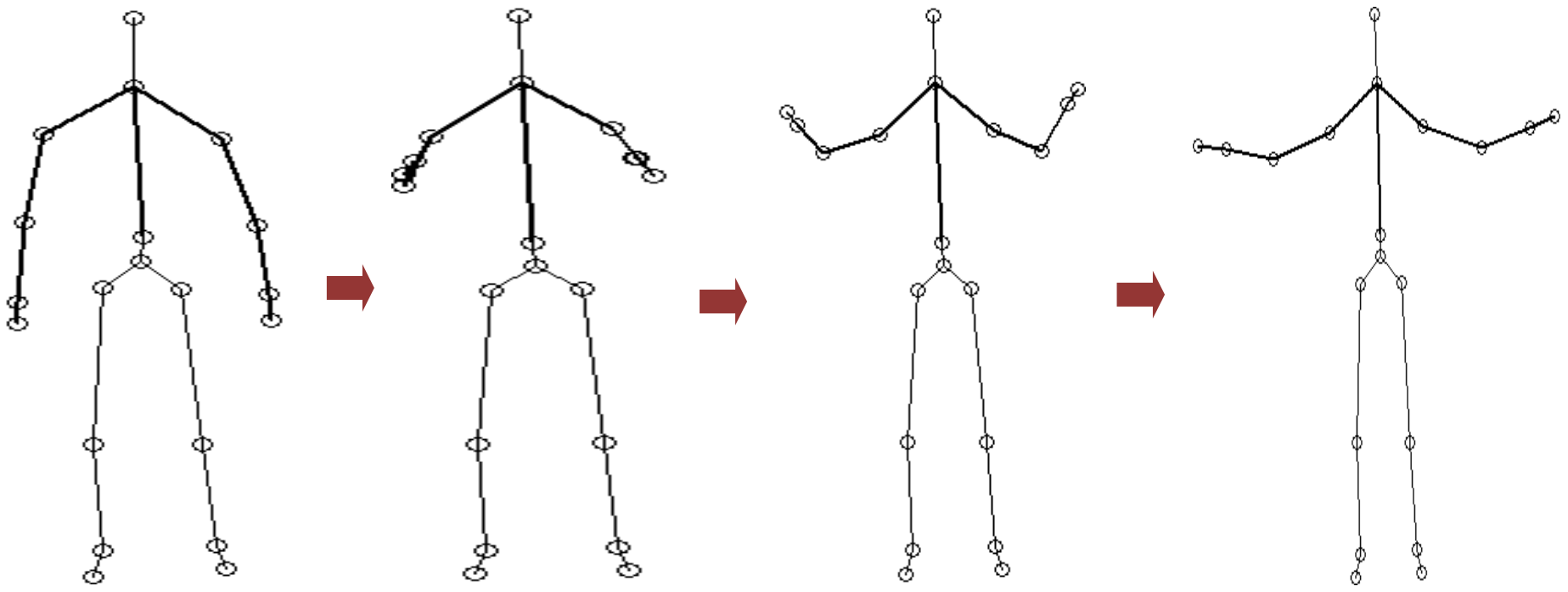

(b) Skeletons obtained using Kinect sensor for corresponding frame numbers

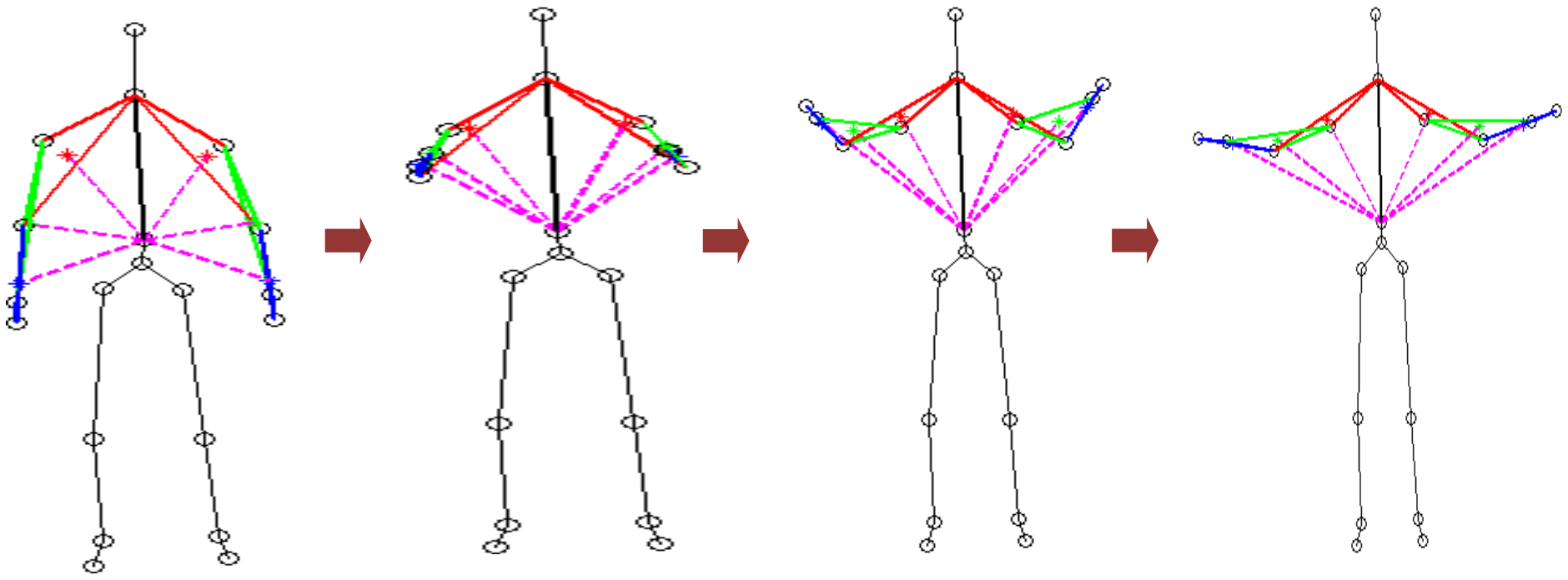

(c) Six Euclidean distances with six triangles and centroids

Fig 2.11 (a), (b), (c) Feature extraction process

**Table 2.2** Features obtained for unknown gesture

| *Frame No.* | *Feature No.* | *Euclidean distance (m)* | *Depth (m)* | *Normalized Euclidean Distance (m)* |
|---|---|---|---|---|
| 10 | 1 | 0.1850 | 2.2069 | 0.0838 |
| | 2 | 0.1574 | 2.1936 | 0.0718 |
| | 3 | 0.2029 | 2.1808 | 0.0930 |
| | 4 | 0.1931 | 2.1606 | 0.0894 |
| | 5 | 0.2375 | 2.1512 | 0.1104 |
| | 6 | 0.2438 | 2.1303 | 0.1144 |
| 30 | 1 | 0.2008 | 2.1948 | 0.0915 |
| | 2 | 0.1802 | 2.1919 | 0.0822 |
| | 3 | 0.2689 | 2.1080 | 0.1276 |
| | 4 | 0.2585 | 2.1012 | 0.1230 |
| | 5 | 0.3525 | 2.0374 | 0.1730 |
| | 6 | 0.3544 | 2.0240 | 0.1751 |
| 60 | 1 | 0.2462 | 2.1906 | 0.1124 |
| | 2 | 0.2128 | 2.2022 | 0.0966 |
| | 3 | 0.3828 | 2.1064 | 0.1818 |
| | 4 | 0.3439 | 2.1225 | 0.1620 |
| | 5 | 0.5577 | 2.0310 | 0.2746 |
| | 6 | 0.5100 | 2.0491 | 0.2489 |
| 85 | 1 | 0.2650 | 2.2237 | 0.1192 |
| | 2 | 0.2497 | 2.2166 | 0.1126 |
| | 3 | 0.4218 | 2.1792 | 0.1936 |
| | 4 | 0.4205 | 2.1704 | 0.1937 |
| | 5 | 0.5920 | 2.1405 | 0.2766 |
| | 6 | 0.6035 | 2.1282 | 0.2836 |

### 2.4.3 Comparative Algorithms

The performance of the proposed system is examined here with respect to the three datasets considered. The comparative framework includes *k*-Nearest Neighbour (*k*-NN) classification [21] ensemble decision tree (EDT) [22] and Neural Network using Levenberg–Marquardt Algorithm (LMA-NN) [23,24].

### 2.4.4 Performance Metrics

For machine learning or for statistical classification, a binary confusion matrix [25] or an error matrix is a table layout which depicts the visualization of the performance of the said algorithm. The average of performance metrics over all classes is considered. For a binary confusion matrix if the true positive, false negative, false positive and true negative samples are denoted by *TP*, *FN*, *FP* and *TN* respectively then performance metrics [26,27] are as follows;

$$\text{Precision} = \frac{TP}{TP + FP} \tag{2.13}$$

$$\text{Recall} = \frac{TP}{TP + FN} \tag{2.14}$$

$$\text{Accuracy} = \frac{TP + TN}{TP + FN + FP + TN} \tag{2.15}$$

$$\text{Error Rate} = \frac{FP + FN}{TP + FN + FP + TN} \tag{2.16}$$

$$\text{F1 Score} = 2 \times \left( \frac{Precision \times Recall}{Precision + Recall} \right) \tag{2.17}$$

### 2.4.5 Performance Analysis

The comparison of performance metrics among all the classification techniques is illustrated in fig 2.12 for three datasets. The best result obtained is for the proposed algorithm.

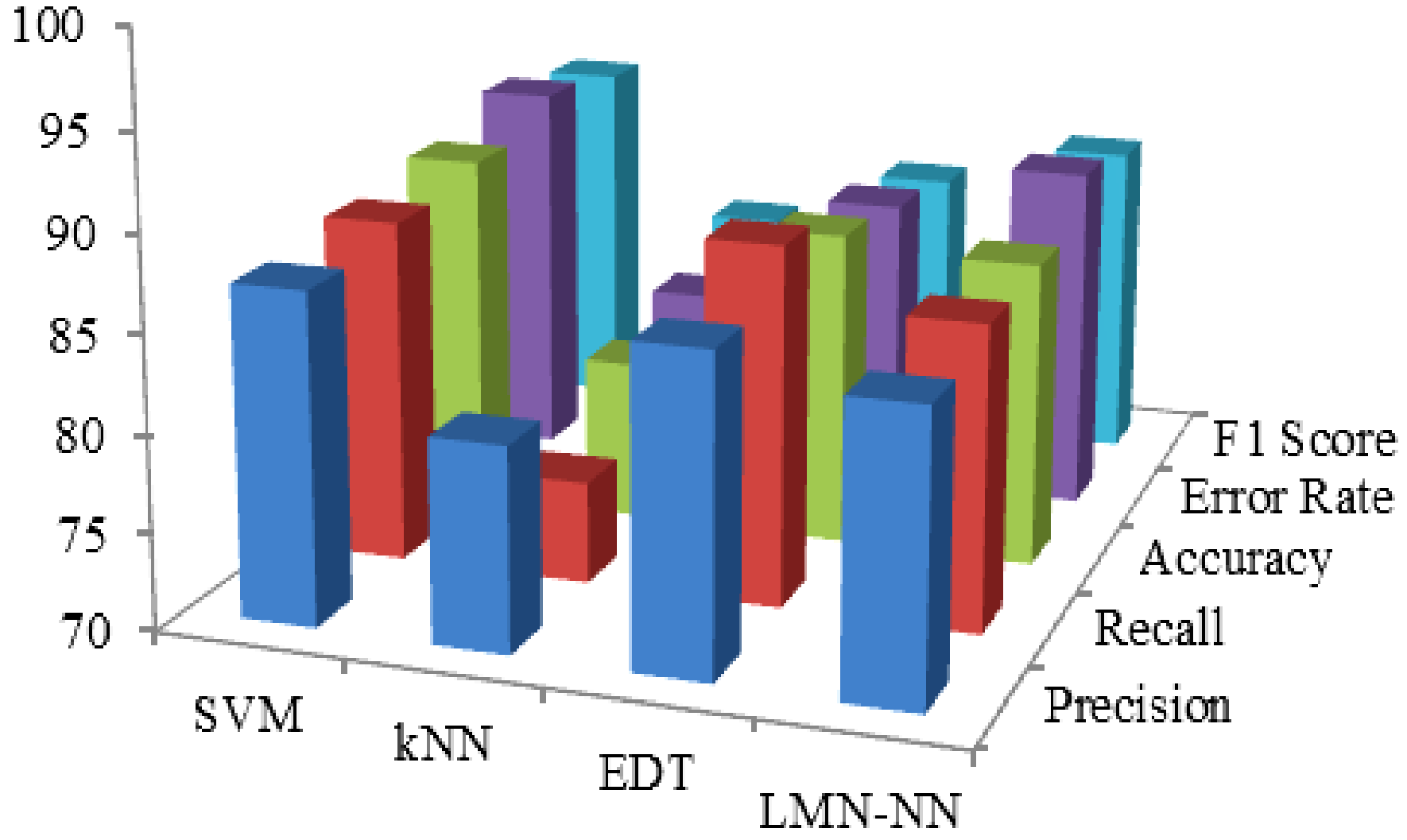

Fig 2.12 Comparison of performance metrices for four classifiers

### 2.4.6 Statistical Analysis Using Friedman Test

The best of all the *c* algorithms, i.e., $1 \leq c \leq C$, is given rank 1 and the worst is given rank *D*. The average ranking acquired by the *c*-th algorithm over all *d* ($1 \leq d \leq D$) datasets is $R_c$ [28].

$$\chi^2 = \frac{12D}{C(C+1)}\left[\sum_{c=1}^{C} R_c^2 - \frac{C(C+1)^2}{4}\right] \tag{2.18}$$

In this paper *D*=3 and *C*=4. In Table 2.3, it is shown that the null hypothesis [29] has been rejected, as $\chi^2_F$ = 8.2000 is greater than 7.815, the critical value of $\chi^2$ distribution for $C-1=3$ degrees of freedom at probability of 0.05.

**Table 2.3** Performance analysis using Friedman test

| Algorithm | Dataset 1 | Dataset 2 | Dataset 3 | $R_c$ | $\chi^2$ |
|---|---|---|---|---|---|
| SVM | 1 | 1 | 1 | 1 | 8.2000 |
| kNN | 4 | 4 | 4 | 4 | |
| EDT | 2 | 2 | 3 | 2.3333 | |
| LMA-NN | 3 | 3 | 2 | 2.6667 | |

## 2.5 Conclusion

The new technique elucidated in this chapter deals with gestures by estimating six different features in a simple and effective way with an accuracy of 88.7%. Gestures are detected for 3s interval with a nearly constant speed. The pros of the algorithm are its simplicity, robustness, and effectiveness. A wireless, easy to use system has been implemented which is also effective in cost.

Although gesture recognition is not a new topic, but Kinect based gesture recognition has achieved a major breakthrough in the era of Human-Computer Interaction. HCI deals with the methods; by which human being make use of computational systems and infrastructures by improving the usability of computer interfaces. The proposed work finds application in public place surveillance and also in behavior modeling. The way the hearing and talking impaired people interact to the environment is only via gestures through vision. The smart technologies found in mobile phones and tablets for communication can also be performed by specific gestures in this generation.

## References

1. Smith, Anthony Vernon Walker, Alistair Ian Sutherland, Arnaud Lemoine, and Sean Mcgrath. "Hand gesture recognition system and method." U.S. Patent 6,128,003, issued October 3, 2000.
2. Zhang, Zhengyou. "Microsoft kinect sensor and its effect." *MultiMedia, IEEE*19, no. 2 (2012): 4-10.
3. Ballester, Jorge, and Chuck Pheatt. "Using the Xbox Kinect sensor for positional data acquisition." *American journal of Physics* 81, no. 1 (2013): 71-77.
4. Saha, Sriparna, Amit Konar, and Jayashree Roy. "Single Person Hand Gesture Recognition Using Support Vector Machine." In *Computational Advancement in Communication Circuits and Systems*, pp. 161-167. Springer India, 2015.
5. Theodoridis, Sergios, Aggelos Pikrakis, Konstantinos Koutroumbas, and Dionisis Cavouras. *Introduction to Pattern Recognition: A Matlab Approach: A Matlab Approach*. Academic Press, 2010.
6. Malima, Asanterabi, Erol Özgür, and Müjdat Çetin. "A fast algorithm for vision-based hand gesture recognition for robot control." In *Signal Processing and Communications Applications, 2006 IEEE 14th*, pp. 1-4. IEEE, 2006.
7. Alsheakhali, Mohamed, Ahmed Skaik, Mohammed Aldahdouh, and Mahmoud Alhelou. "Hand Gesture Recognition System." *Information & Communication Systems* 132 (2011).
8. Singha, Joyeeta, and Karen Das. "Hand gesture recognition based on Karhunen-Loeve transform." *arXiv preprint arXiv:1306.2599* (2013).
9. Jain, Anil K. "A fast Karhunen-Loeve transform for a class of random processes." *NASA STI/Recon Technical Report A* 76 (1976): 42860.
10. Rautaray, Siddharth S., and Anupam Agrawal. "Real time hand gesture recognition system for dynamic applications." *International Journal of UbiComp* 3, no. 1 (2012): 21.
11. Chen, Feng-Sheng, Chih-Ming Fu, and Chung-Lin Huang. "Hand gesture recognition using a real-time tracking method and hidden Markov models."*Image and vision computing* 21, no. 8 (2003): 745-758.
12. Blunsom, Phil. "Hidden markov models." *Lecture notes, August* 15 (2004): 18-19.
13. Correa, Mauricio, Javier Ruiz-del-Solar, Rodrigo Verschae, Jong Lee-Ferng, and Nelson Castillo. "Real-time hand gesture recognition for human robot interaction." In *RoboCup 2009: Robot Soccer World Cup XIII*, pp. 46-57. Springer Berlin Heidelberg, 2009.

14. Rish, Irina. "An empirical study of the naive Bayes classifier." In *IJCAI 2001 workshop on empirical methods in artificial intelligence*, vol. 3, no. 22, pp. 41-46. IBM New York, 2001.
15. Suykens, Johan AK, and Joos Vandewalle. "Least squares support vector machine classifiers." *Neural processing letters* 9, no. 3 (1999): 293-300.
16. Cortes, Corinna, and Vladimir Vapnik. "Support vector machine." *Machine learning* 20, no. 3 (1995): 273-297.
17. H.-C. Kim, S. Pang, H.-M. Je, D. Kim, and S. Yang Bang, "Constructing support vector machine ensemble," *Pattern Recognit.*, vol. 36, no. 12, pp. 2757–2767, 2003.
18. J. A. K. Suykens and J. Vandewalle, "Least squares support vector machine classifiers," *Neural Process. Lett.*, vol. 9, no. 3, pp. 293–300, 1999.
19. Yiu Paul. "A tour of triangle geometry." *Reflections* 6 (2004): 29.
20. Breu, Heinz, Joseph Gil, David Kirkpatrick, and Michael Werman. "Linear time Euclidean distance transform algorithms." *Pattern Analysis and Machine Intelligence, IEEE Transactions on* 17, no. 5 (1995): 529-533.
21. Zhang, Hao, Alexander C. Berg, Michael Maire, and Jitendra Malik. "SVM-KNN: Discriminative nearest neighbor classification for visual category recognition." In *Computer Vision and Pattern Recognition, 2006 IEEE Computer Society Conference on*, vol. 2, pp. 2126-2136. IEEE, 2006.
22. T. G. Dietterich, "An experimental comparison of three methods for constructing ensembles of decision trees: Bagging, boosting, and randomization," *Mach. Learn.*, vol. 40, no. 2, pp. 139–157, 2000.
23. S. Saha, M. Pal, A. Konar, and R. Janarthanan, "Neural Network Based Gesture Recognition for Elderly Health Care Using Kinect Sensor," in *Swarm, Evolutionary, and Memetic Computing*, Springer, 2013, pp. 376–386.
24. M. Buscema, "Back propagation neural networks," *Subst. Use Misuse*, vol. 33, no. 2, pp. 233–270, 1998.
25. https://en.wikipedia.org/wiki/Confusion_matrix
26. Sokolova, Marina, and Guy Lapalme. "A systematic analysis of performance measures for classification tasks." *Information Processing & Management*45, no. 4 (2009): 427-437.
27. Demšar, Janez. "Statistical comparisons of classifiers over multiple data sets." *The Journal of Machine Learning Research* 7 (2006): 1-30.
28. Dietterich, Thomas G. "Approximate statistical tests for comparing supervised classification learning algorithms." *Neural computation* 10, no. 7 (1998): 1895-1923.
29. García, Salvador, Daniel Molina, Manuel Lozano, and Francisco Herrera. "A study on the use of non-parametric tests for analyzing the evolutionary algorithms' behaviour: a case study on the CEC'2005 special session on real parameter optimization." *Journal of Heuristics* 15, no. 6 (2009): 617-644.

# CHAPTER 3

# GESTURE RECOGNITION FROM TWO-PERSON INTERACTIONS USING ENSEMBLE DECISION TREE

*The evolution of depth sensors has furnished a new horizon for human-computer interaction (HCI). An efficient two person interaction detection system is proposed for an improved HCI using Kinect sensor. This device is able to identify 20 body joint coordinates in 3D space among which 16 joints are selected and those have been adapted with certain weights to form 4 average points. The direction cosines of these four average points are evaluated followed by the angles made by x, y and z axes respectively, i.e., 12 angles have been constructed for each frame. For recognition purpose, ensemble of tree classifiers with bagging mechanism is used. This novel work is widely acceptable for various gesture based computer appliances and yields a recognition rate of 87.15%.*

## 3.1 Introduction

Human body tracking is a well-studied topic in today's era of Human Computer Interaction (HCI) [1] and it can be formed by the virtue of human skeleton structures. These skeleton structures have been detected successfully due to the smart progress of some devices, used to measure depth (e.g. Sony PlayStation, Kinect sensor etc.). Human body movements have been viewed using these depth sensors [2] which can provide sufficient accuracy while tracking full body in real-time mode with low cost.

In reality action and reaction activities are hardly periodic in a multi-person perspective situation. Also, recognizing their complex, a-periodic gestures are highly challenging for detection in surveillance system [3]. Interaction detections like pushing, kicking, punching, exchanging objects are the essence of this work. Here, two person interactions have been recognized by an RGB-D sensor, named as Kinect [4,5]

Wachter and Nagel [6] have presented an approach of person tracking in monocular image strings. They have employed certain 3D models of persons, where each model is comprised of a collection of right elliptical cones, which are formed by some body parts (head, neck, trunk, thighs, shanks, feet). A Cartesian coordinate system has to camera coordinate system followed by perspective projection where the been formed at each body part and the human kinematic model is depicted by a transformation tree, where the root denotes the person coordinate system and leaves represent the coordinate system of different body parts. The person coordinate system has been transformed camera coordinates are forecasted into image coordinates. This method determines the number of degrees of freedom (DOF) of the body joints and also their changing rate quite easily, which is solved by Kalman filtering [7]. One remarkable feature of the technique is its regular integration in edge and region information.

Kalman filtering or linear quadratic estimation (LQE), is an algorithm that uses a series of measurements observed over time, consisting statistical noise and other inaccuracies, and generates estimates of unknown variables that tend to be more precise than those based on a single measurement alone.

Saha *et al*. [8] have proposed a superior detection technique of two person interaction with the help of Microsoft's Kinect sensor. The sensor is able to capture the twenty body joint coordinates in 3D space followed by constructing skeletons per frame. Five average points have been configured per person to form a pentagon per frame. The Euclidean distances between the vertices of two pentagons of two persons are taken as the feature space for this work. Eight interactions have been modeled and highest recognition rate of 90% is achieved through multiclass support vector machine [9] with rotation invariance case.

Park and Aggarwal [10] have presented a method regarding two person interactions via natural language descriptors at a semantic level. They have adapted 'verb argument structure' for representing human actions by forming triplets. Static poses and dynamic gestures are recognized through hierarchical Bayesian network [11,12]. This method affords a cordial natural language elaboration for different gestures and distinguishes into positive, negative and neutral interactions.

Yun *et al*. [13] have recognized several interactions acted by two persons using an RGB-D sensor. The color and depth image information are extracted from the skeleton model captured from the sensor. Six different body pose features, viz. plane features, joint features and velocity features are considered as the feature set whereas the joint features work better than others and the velocity features are affected to noise. A comparative study between SVM and Multiple Instance Learning Boost classifier is drawn and it is observed that MIL Boost classifier [14,15] is superior than SVM. Yao *et al*. [16] have indicated velocity features to posses the best accuracy while recognizing single person activity.

Freeman and Weissman [17] have shown how hand gestures can be used to control television set. Only one hand gesture, open hand, is used by the user facing the sensor or camera. Operations such as, television turning on-off and channel control-change are performed by moving single hand. The open hand image is detected and tracked followed by template normalized correlation. The television set is controlled by a computer via serial port communication.

Here, we have implemented Kinect sensor to identify 8 interactions using skeleton structures with some predefined features. As Kinect sensor captures 20 body joint coordinates in 3D, out of which 16 of those have been adapted with some specified weights to form 4 mean points. The direction cosines of these 4 average points, i.e., 12 angles per frame are the feature set of our proposed work. For a specific interaction, 3s video stream is captured to detect the skeleton. Ensemble decision tree (EDT) [18,19] with bagging technology [20] is employed for recognition purpose with a recognition rate of 87.15%.

In this chapter, fundamental ideas regarding Kinect sensor and Ensemble Decision Tree classifier have been discussed in Section 3.2.1 and Section 3.2.2 respectively. Section 3.3 elucidates the proposed algorithm. Section 3.4 describes the experimental results. Finally, Section 3.5 draws the conclusion and future work.

## 3.2 Fundamental Ideas

This section explains the Kinect sensor and the ensemble decision tree classifier respectively.

### 3.2.1 Kinect Sensor

Kinect [21,22] is a combination of a camera, an infrared (IR) emitter-receiver, a microphone block and a tilt motor. The RGB camera captures three dimensional data at 30 frames per second in a 640×480 resolution. The IR camera estimates the reflected beam and measures the depth of the subject from the Kinect sensor in 1.2 to 3.5m range. The figure below shows the three dimensional axes constructed by the sensor itself to estimate the direction cosines of the mean points. The green arrow for positive *x* axis, blue for positive *y* axis and red for measuring the depth *i.e. z*-axis.

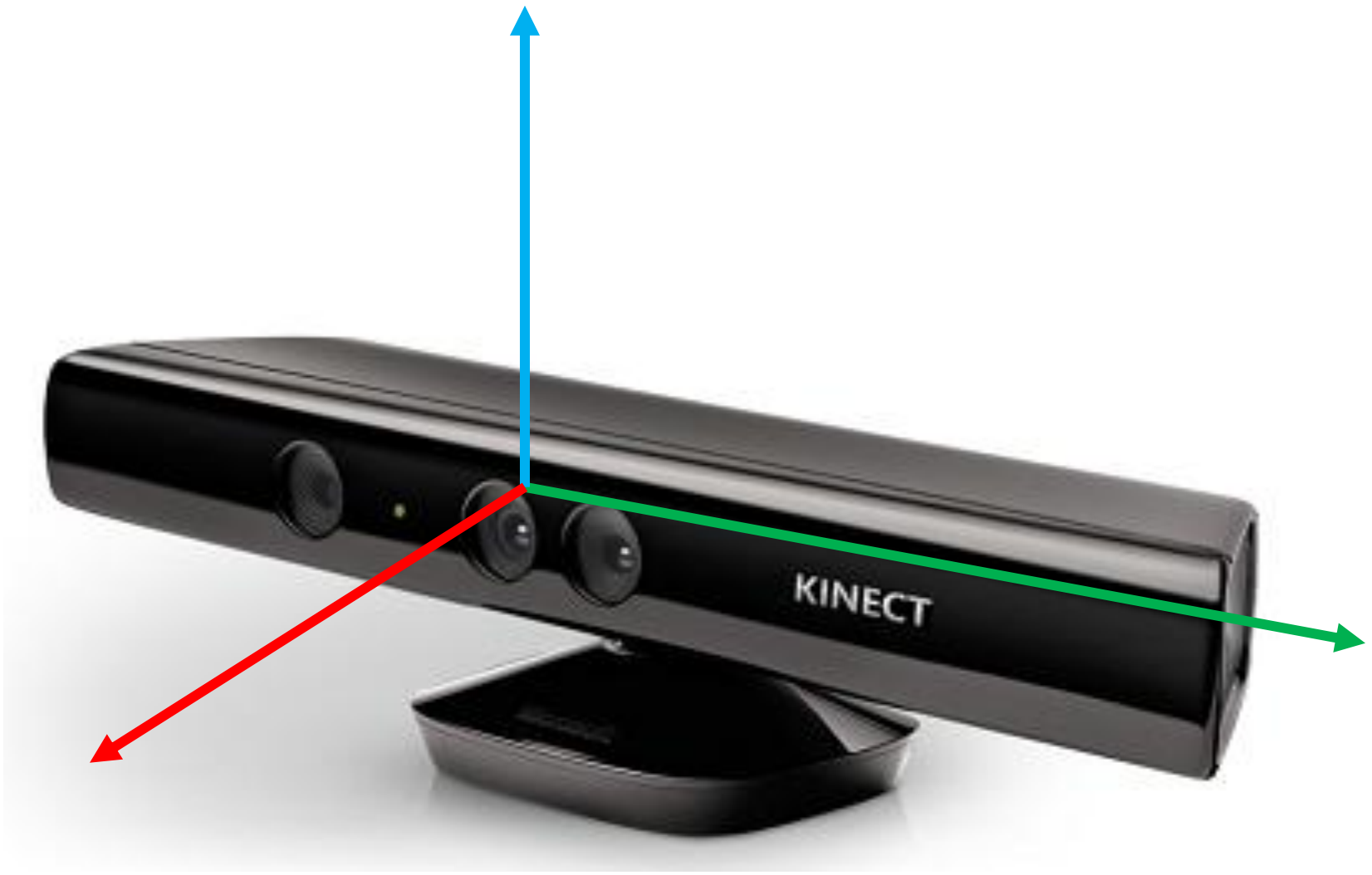

Fig 3.1 Kinect Sensor and its three dimensional axes

### 3.2.2 Ensemble Decision Tree

An ensemble tree classifier [18] or frequently known as multiple classifier system is a combination of a number of ‘base’ classifiers, where each classifier distinguishes the training data set separately. The output of the ensemble tree classifier is a collection of the decisions of the base classifiers. It is seen that the performance the classifier is superior than the base classifiers, mentioning the errors of the base classifiers to be uncorrelated. Among the base classifiers, a ‘tree’ classifier has been selected, where each node of it decides on each of its features in the dataset to estimate the class of the data sample. Two popular algorithms, *i.e.* bagging and boosting [20], have been used to demonstrate the classifier. In case of ‘boosting’, initially the weights of all the data sample are assumed equal. In each iteration, a new base classifier is produced employing the full dataset and simultaneously the weights of all the data sample are updated. This weight adaptation is carried out by increasing the sample weights misclassified and decreasing the sample weights correctly classified. Thus it is known as ‘adaptive’ boosting. A weighted voting procedure estimates the class of a new sample. In bagging, classifiers are trained by various datasets, obtained from the original dataset via bootstrapping. The divergence among the weak learners is examined by this re-sampling process, repeated *T* times. Then majority voting is taken to predict the class of the unknown. Here, *T* has been chosen to be 100 and bootstrap size (*n*) is taken 30% of the total dataset.

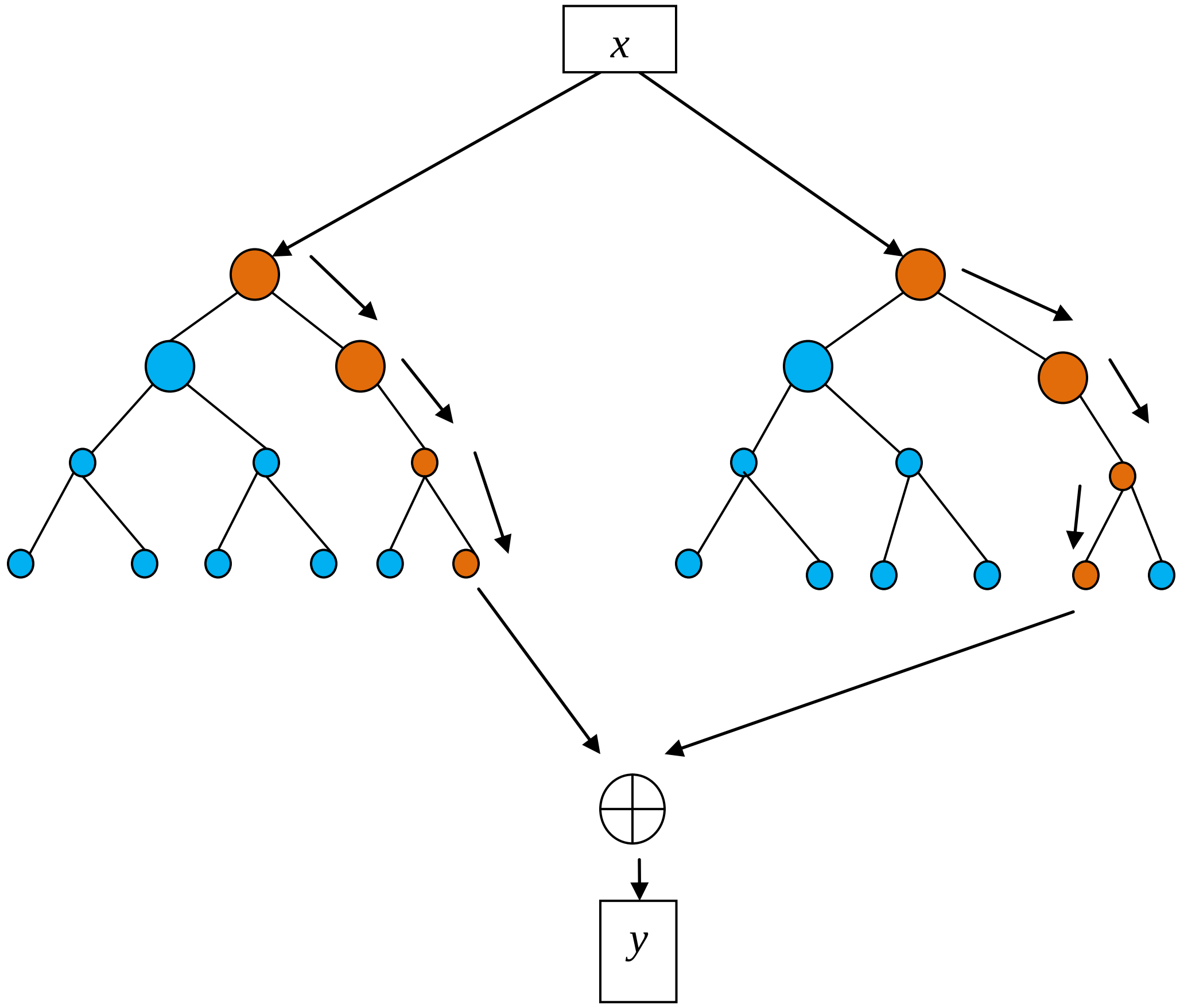

Fig 3.2 Ensemble Decision Tree classifier

## 3.3 Proposed Algorithm

The block diagram of the proposed algorithm is given below,

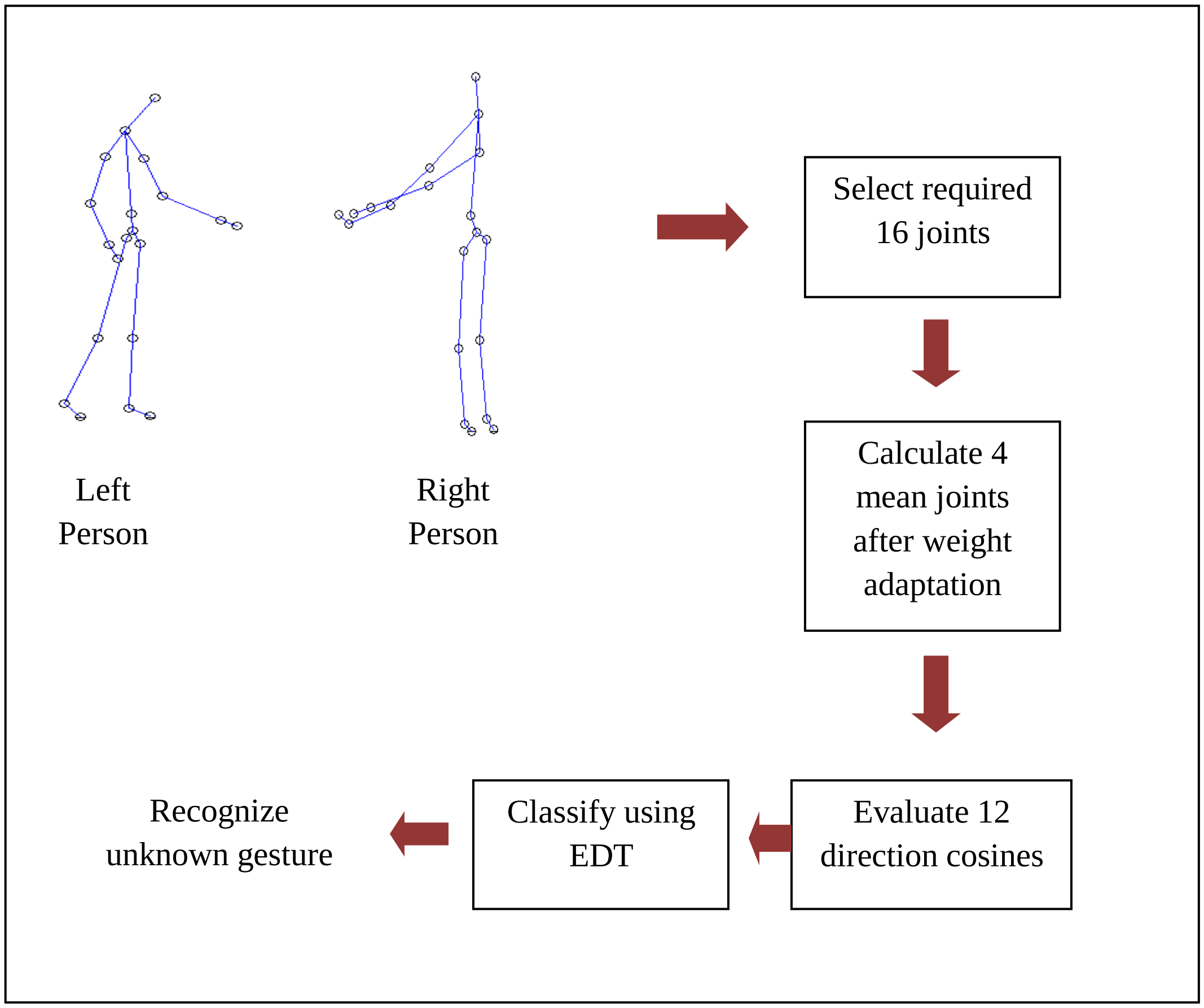

Fig 3.3 Block diagram for gesture recognition for two-person interactions (as same steps need to be followed for both the persons, thus steps required to be followed for right person is given only).

Suppose for our proposed algorithm, number of subjects to be chosen are $N$ and number of actions to be executed for a single person are $G$. Thus the total interactions possible between the two persons are $GG$. Now when one specific subject $n$ ($1 \leq n \leq N$) is asked to perform a particular action $g$ ($1 \le g \le G$), we have captured a total number of $T$ frames. Now for each $t^{th}$ ($1 \le t \le T$) frame, we have twenty 3D body joints information out of which 16 joints are selected for this suggested work. These selected joints are shoulder left (*SL*), shoulder right (*SR*), elbow left (*EL*), elbow right (*ER*), wrist left (*WL*), wrist right (*WR*), hand left (*HaL*), hand right (*HaR*), hip left (*HL*), hip right (*HR*), knee left (*KL*), knee right (*KR*), ankle left (*AL*), ankle right (*AR*), foot left (*FL*) and foot right (*FR*) as shown in the following figure with colored squares.

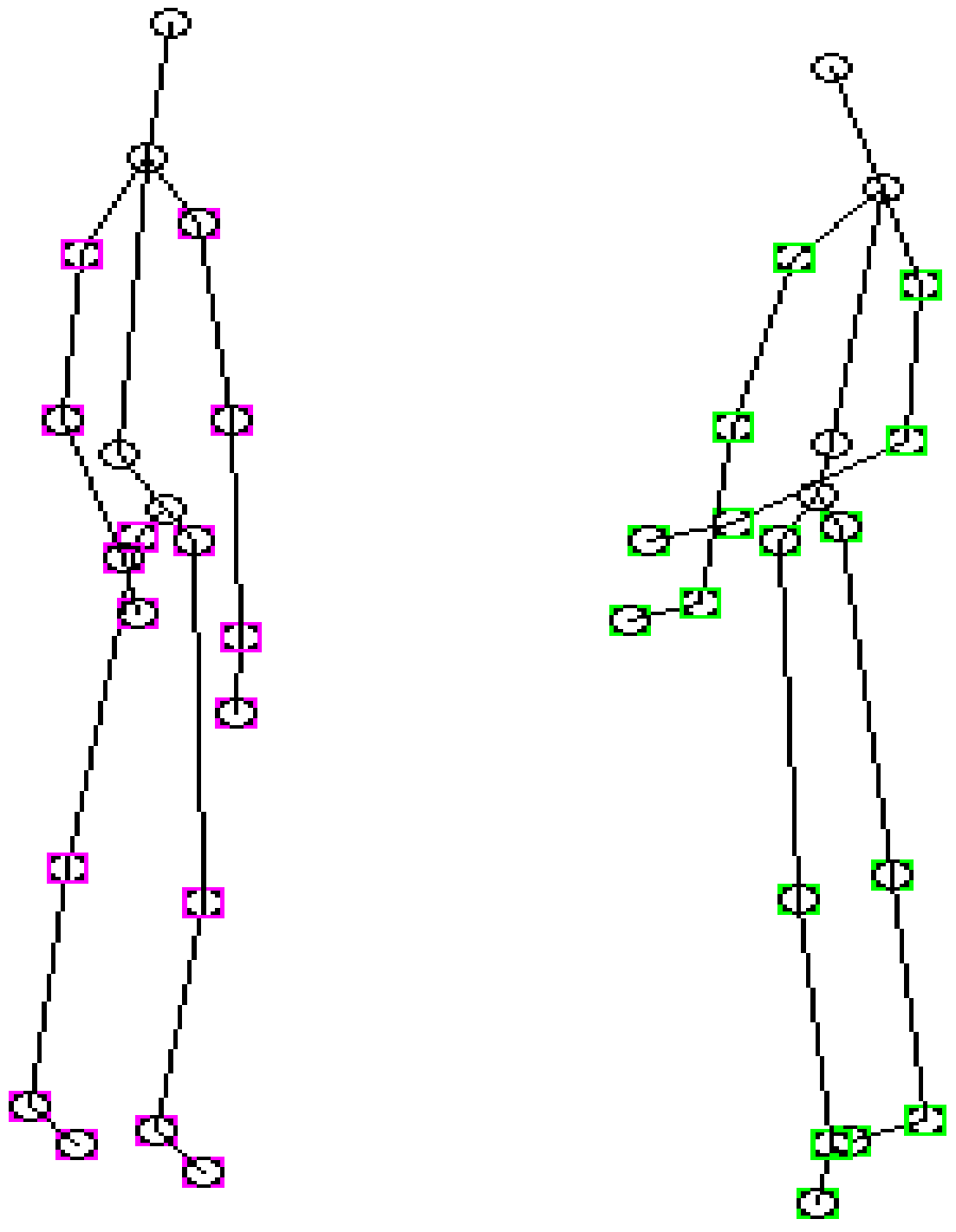

Fig 3.4 Sixteen required joints for the proposed work (for left and right person)

Now 4 mean joints are formulated taking 4 body joints from arm or leg at a time. But while calculating these mean joints we have given weights to the corresponding body joints based on the distances between them and our proposed work outperforms [23].

Considering left arm, the distance between *SL* and *EL* are the highest, while the distance between *WL* and *HaL* is the lowest. The weightage given to the 16 body joints should be according to these ratios. The weights, by which the joints are adjusted, are $w^{SL}=w^{SR}=0.271$, $w^{EL}=w^{ER}=0.449$, $w^{WL}=w^{WR}=0.149$, $w^{HaL}=w^{HaR}=0.131$, $w^{HL}=w^{HR}=0.348$, $w^{KL}=w^{KR}=0.437$, $w^{AL}=w^{AR}=0.119$, $w^{FL}=w^{FR}=0.096$. Finally, the 4 mean joints become,

$$J_1^t = \frac{w^{SL} \times SL^t + w^{EL} \times EL^t + w^{WL} \times WL^t + w^{HaL} \times HaL^t}{4} \qquad \textbf{(3.1)}$$

$$J_2^t = \frac{w^{SR} \times SR^t + w^{ER} \times ER^t + w^{WR} \times WR^t + w^{HaR} \times HaR^t}{4} \qquad \textbf{(3.2)}$$

$$J_3^t = \frac{w^{HL} \times HL^t + w^{KL} \times KL^t + w^{AL} \times AL^t + w^{FL} \times FL^t}{4} \qquad \textbf{(3.3)}$$

$$J_4^t = \frac{w^{HR} \times HR^t + w^{KR} \times KR^t + w^{AR} \times AR^t + w^{FR} \times FR^t}{4} \qquad \textbf{(3.4)}$$

where, *w* are the respective weights given to the body joints according to the superscript values. These weight values are same irrespective of time or frame number. The weights have been adopted in such a fashion that $w^{SL}+w^{EL}+w^{WL}+w^{HaL}\approx1$, $w^{SR}+w^{ER}+w^{WR}+w^{HaR}\approx1$, $w^{HL}+w^{KL}+w^{AL}+w^{FL}\approx1$ and $w^{HR}+w^{KR}+w^{AR}+w^{FR}\approx1$ [10]. From each $J_i$ ($1\leq i\leq4$) bearing 3D coordinate information, direction cosines [24] ($\cos\alpha_{Ji}^t$, $\cos\beta_{Ji}^t$, $\cos\gamma_{Ji}^t$) are evaluated followed by the angles $\alpha$, $\beta$, $\gamma$; that the mean joints make with positive *x*, *y* and *z* axes respectively using equations (**3.5-3.7**).

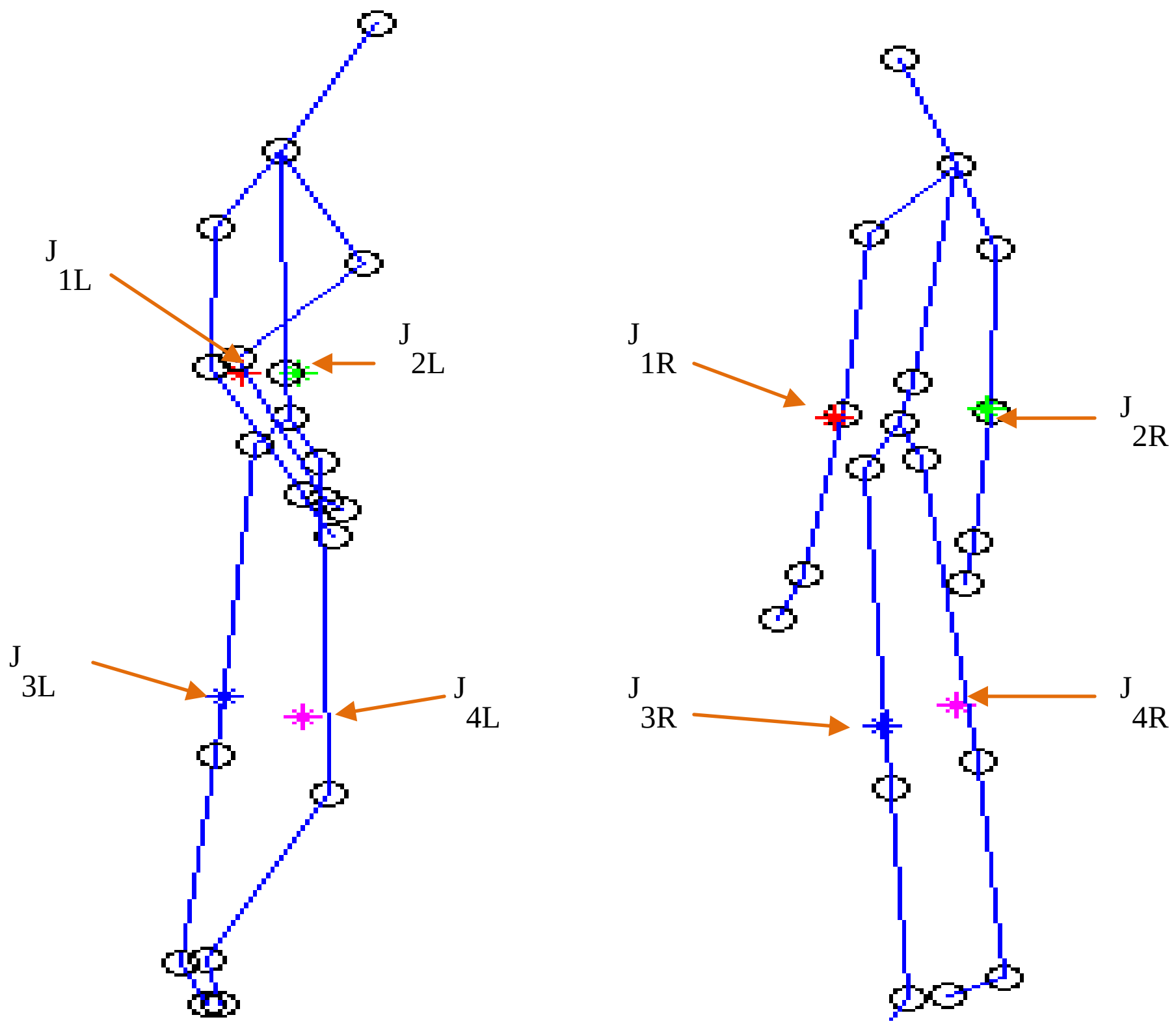

Fig 3.5 Four mean joints for left-right person after weight adaptation

$$\alpha_{Ji}^{t} = \cos^{-1} \frac{x_{Ji}^{t}}{\sqrt{(x_{Ji}^{t})^2 + (y_{Ji}^{t})^2 + (z_{Ji}^{t})^2}} \tag{3.5}$$

$$\beta_{Ji}^{t} = \cos^{-1} \frac{y_{Ji}^{t}}{\sqrt{(x_{Ji}^{t})^2 + (y_{Ji}^{t})^2 + (z_{Ji}^{t})^2}} \tag{3.6}$$

$$\gamma_{Ji}^{t} = \cos^{-1} \frac{z_{Ji}^{t}}{\sqrt{(x_{Ji}^{t})^2 + (y_{Ji}^{t})^2 + (z_{Ji}^{t})^2}} \tag{3.7}$$

where $x$, $y$, $z$ represents the 3D axes. Thus for a specific subject $n$ to interact with a particular action $g$, we have a total of 12 angles (3 angles from each $J_i$) for a particular frame, which forms the feature space of our modeled work. For each action performed by each person, we have $T$ number of frames and 12 features have been extracted per frame. Therefore the dimension of the feature space becomes $T \times 12$. Since the total training dataset is composed of $N$ subjects and 8 actions, the total dimension is $N \times 8 \times T \times 12$. Whenever an unknown interaction is delivered by two persons, we segregate the two actions performed by two subjects and each subject's body gestures are recognized using EDT classifier already specified in Section 3.2.2

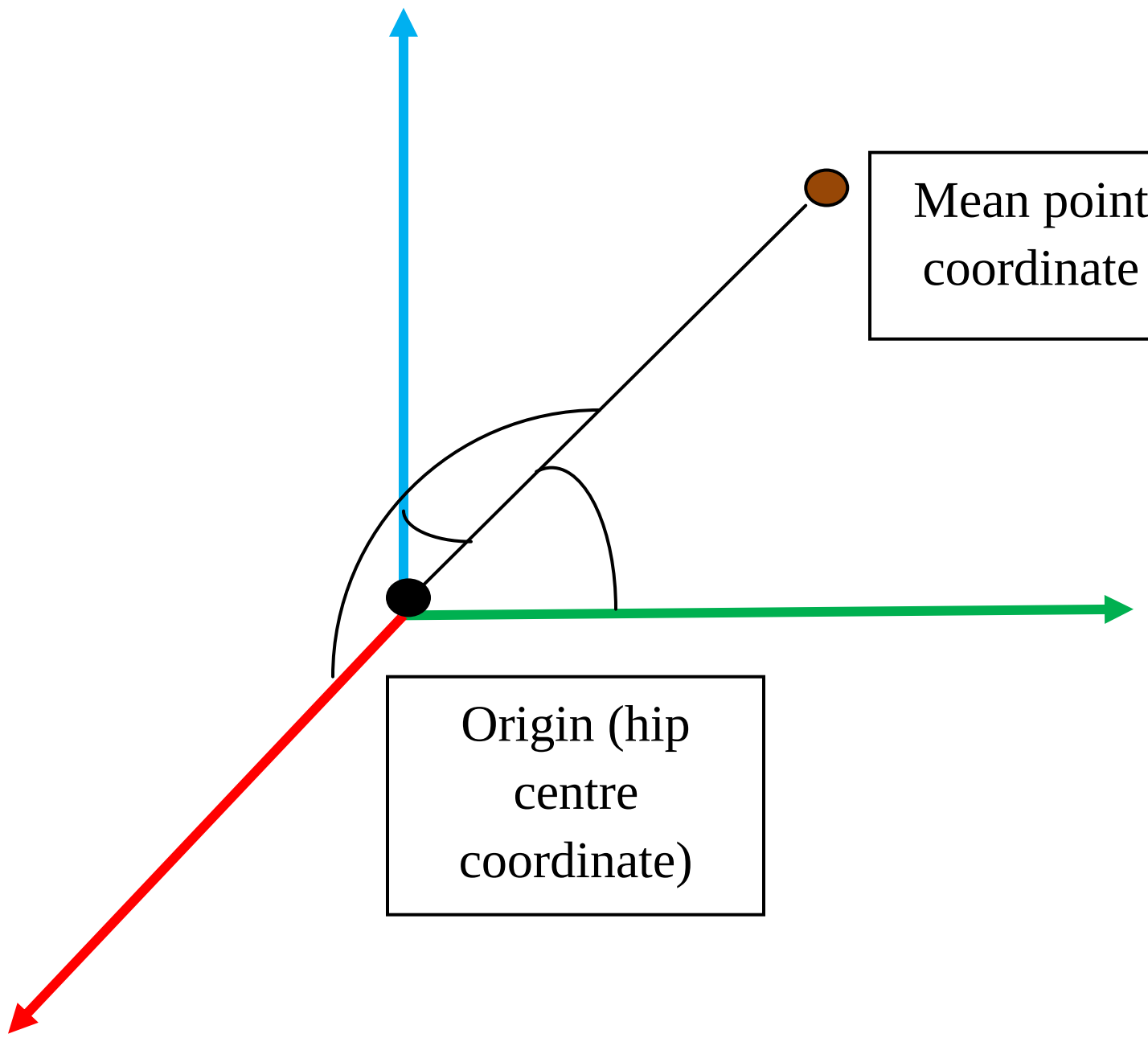

Fig 3.6 Estimation of direction cosines of the points

**Table 3.1** Procedure for calculation of feature space for the subject in the right side from Fig 3.3

| 3D Co-ordinates obtained | | | | | | | | Direction cosines | |
|---|---|---|---|---|---|---|---|---|---|
| $SL^{48}$ | -0.423 | 0.440 | 3.048 | | | | | | |
| $EL^{48}$ | -0.468 | 0.199 | 2.933 | $J_1^{48}$ | -0.109 | 0.049 | 0.736 | $\cos\alpha_{J1}^{48}$ | -0.146 |
| $WL^{48}$ | -0.411 | -0.009 | 2.878 | | | | | $\cos\beta_{J1}^{48}$ | 0.066 |
| $HaL^{48}$ | -0.388 | -0.086 | 2.858 | | | | | $\cos\gamma_{J1}^{48}$ | 0.986 |
| $SR^{48}$ | -0.258 | 0.407 | 3.289 | | | | | | |
| $ER^{48}$ | -0.233 | 0.110 | 3.161 | $J_2^{48}$ | -0.059 | 0.029 | 0.812 | $\cos\alpha_{J2}^{48}$ | -0.072 |
| $WR^{48}$ | -0.212 | -0.136 | 3.321 | | | | | $\cos\beta_{J2}^{48}$ | 0.036 |
| $HaR^{48}$ | -0.218 | -0.179 | 3.386 | | | | | $\cos\gamma_{J2}^{48}$ | 0.996 |
| $HL^{48}$ | -0.336 | 0.050 | 3.039 | | | | | | |
| $KL^{48}$ | -0.379 | -0.451 | 2.980 | $J_3^{48}$ | -0.091 | -0.090 | 0.743 | $\cos\alpha_{J3}^{48}$ | -0.120 |
| $AL^{48}$ | -0.407 | -0.808 | 2.865 | | | | | $\cos\beta_{J3}^{48}$ | -0.120 |
| $FL^{48}$ | -0.355 | -0.876 | 2.840 | | | | | $\cos\gamma_{J3}^{48}$ | 0.985 |
| $HR^{48}$ | -0.260 | 0.037 | 3.179 | | | | | | |
| $KR^{48}$ | -0.272 | -0.486 | 3.202 | $J_4^{48}$ | -0.072 | -0.094 | 0.783 | $\cos\alpha_{J4}^{48}$ | -0.091 |
| $AR^{48}$ | -0.374 | -0.807 | 2.865 | | | | | $\cos\beta_{J4}^{48}$ | -0.119 |
| $FR^{48}$ | -0.340 | -0.844 | 2.975 | | | | | $\cos\gamma_{J4}^{48}$ | 0.989 |

## 3.4 Experimental Results

For the proposed work, we have granted $GG = {}^{8}C_{2}\text{-}8 = 20$ two person interactions, out of which a few of them are listed in Table 3.2.

**Table 3.2** List of two-person interactions

| ***Left person*** | ***Right person*** |
|---|---|
| 1. Approaching | 1. Departing |
| 2. Exchanging | 2. Shaking hands |
| 3. Approaching | 3. Hugging |
| 4. Punching | 4. Departing |
| 5. Shaking hands | 5. Pushing |
| 6. Pushing | 6. Kicking |
| 7. Approaching | 7. Shaking hands |
| 8. Kicking | 8. Departing |
| 9. Punching | 9. Kicking |
| 10. Exchanging | 10. Departing |

while each person is showing $G$=8; namely approaching, departing, exchanging, hugging, shaking hands, punching, pushing and kicking.

### 3.4.1 Preparation of Training Dataset

Each gesture is taken for 3s duration thus T is 90 (=3s×30 frames/s) frames. Three distinct datasets are prepared as follows:

- Jadavpur university research scholar dataset with 30 female subjects (age 30±5yrs).
- Jadavpur university research scholar dataset with 30 male subjects (age 34±8yrs).
- Jadavpur university research scholar dataset with 15 female and 15 male subjects (age 36±9yrs).

Thus N is 30. The dimension for each gesture is 1080 (=90×12). The total training dataset becomes 648000 (=30×20×90×12). For the testing purpose, we have asked separate twenty subjects (age 32±8yrs) to enact the gestures for 3s duration. Here the male and female subject ratio is 1:1.

### 3.4.2 Recognition of an Unknown Interaction

The description of the feature extraction process for an unknown interaction is given in the previous section. From each frame, 12 features are extracted for each person. For the lack of space, the features obtained for a particular frame *i.e.* frame number=48 are provided in Table 3.1. After feeding the EDT algorithm with the features obtained from unknown interaction for 3s duration, the output interaction identified to be approaching-shaking hands, *i.e.* the left person gesture is approaching while the same for right person is shaking hands. Thus the proposed work is able to recognize the performed interaction correctly. The calculation procedure for direction cosines is explained in Table 3.1

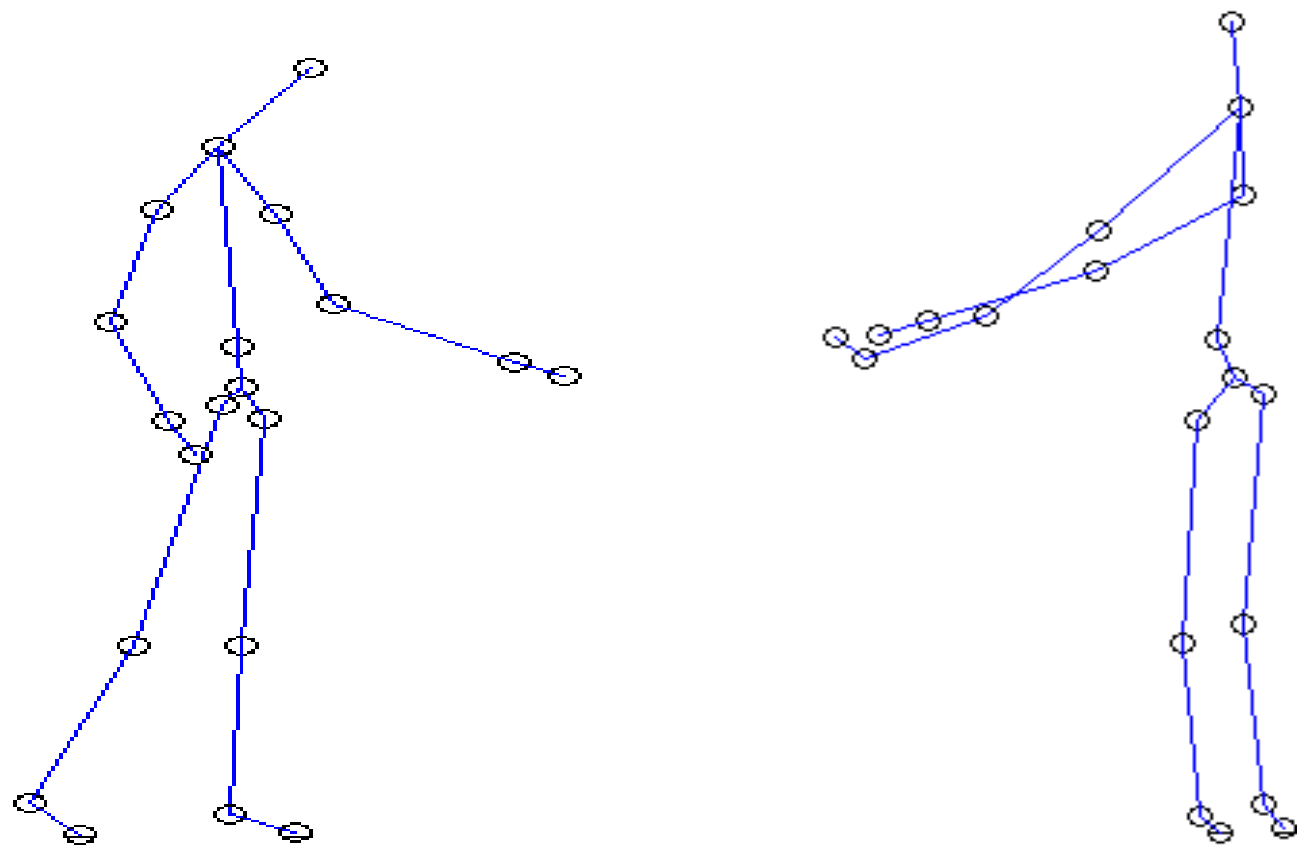

Fig 3.7 Two person interaction, approaching-shaking hands

### 3.4.3 Comparative Algorithms

The comparison of the proposed method is done with respect to the three datasets considered. The comparative framework includes support vector machine (SVM) [25]*k*-nearest neighbor (k-NN) [26] and back propagation neural network (BPNN) [27].

### 3.4.4 Performance Metrics

The average of performance metrics [28] over all classes is considered. For a binary confusion matrix if the true positive, false negative, false positive and true negative samples are denoted by *TP*, *FN*, *FP* and *TN* respectively then performance metrics are the following.

$$\text{Accuracy} = \frac{TP + TN}{TP + FN + FP + TN} \quad \textbf{(2.8)}$$

$$\text{F1 Score} = 2 \times \left( \frac{Precision \times Recall}{Precision + Recall} \right) \quad \textbf{(2.9)}$$

$$\text{Sensitivity} = \frac{TP}{TP + FN} \quad \textbf{(2.10)}$$

$$\text{Specificity} = \frac{TN}{FP + TN} \quad \textbf{(2.11)}$$

$$\text{PPV} = \frac{TP}{TP + FP} \quad \textbf{(2.12)}$$

$$\text{NPV} = \frac{TN}{TN + FN} \quad \textbf{(2.13)}$$

$$\text{Error Rate} = \frac{FP + FN}{TP + FN + FP + TN} \quad \textbf{(2.14)}$$

### 3.4.5 Performance Analysis

The comparison of performance metrics among all the classification techniques is illustrated in fig 3.8 for three datasets. The best result obtained is for the proposed algorithm.

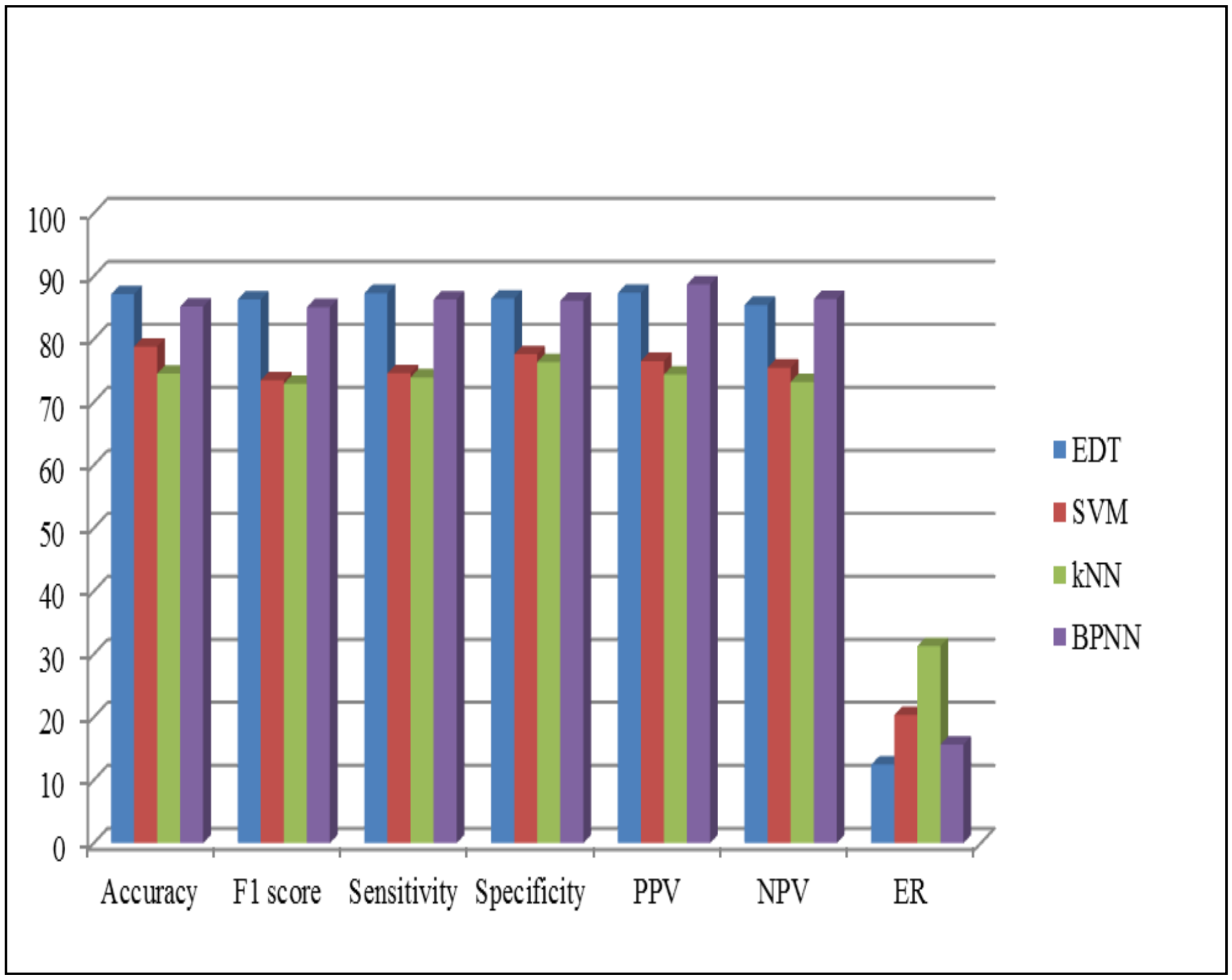

Fig 3.8 Performance analysis with standard classifiers.

## 3.5 Conclusion

Our proposed scheme recognizes 20 two-person interactions to explore an improved human computer interaction in an efficient technique using Kinect sensor. Previous research [] have been gone through calculating only average points, but in our proposed work, the body joints have been adapted with some weights based upon the distance between different body joints of human anatomy. We have obtained a better recognition rate of 87.15%. Therefore this proposed method discovers its application from vision based gesture recognition to public place surveillance. In the upcoming days, we will figure out to enhance our dataset comprising of some complicated interactions between two-persons and recognized them with some statistical models like hidden Markov model, Hidden Conditional Random Field etc.

Here, we have used MATLAB version (R2012b) as a software tool and hardware used is Intel® Core(TM) i3-2120 CPU @ 3.30 GHz processor machine, Windows 7 Home basic (64 bit), 4GB RAM. One drawback of this method is gesture recognition is carried out at constant speed.

## Acknowledgments

The research work is supported by the University Grants Commission, India, University with Potential for Excellence Program (Phase II) in Cognitive Science, Jadavpur University.

## References

1. Dix, Alan. *Human-computer interaction*. Springer US, 2009.
2. Keskin, Cem, Furkan Kıraç, Yunus Emre Kara, and Lale Akarun. "Real time hand pose estimation using depth sensors." In *Consumer Depth Cameras for Computer Vision*, pp. 119-137. Springer London, 2013.
3. Hu, Weiming, Tieniu Tan, Liang Wang, and Steve Maybank. "A survey on visual surveillance of object motion and behaviors." *Systems, Man, and Cybernetics, Part C: Applications and Reviews, IEEE Transactions on* 34, no. 3 (2004): 334-352.
4. M. R. Andersen, T. Jensen, P. Lisouski, A. K. Mortensen, M. K. Hansen, T. Gregersen, and P. Ahrendt, “Kinect depth sensor evaluation for computer vision applications,” *Tech. Rep. Electron. Comput. Eng.*, vol. 1, no. 6, 2015.
5. T. T. Dao, H. Tannous, P. Pouletaut, D. Gamet, D. Istrate, and M. C. H. B. Tho, “Interactive and Connected Rehabilitation Systems for E-Health,” *IRBM*, 2016.
6. Wachter, Stefan, and Hans-Hellmut Nagel. "Tracking of persons in monocular image sequences." In *Nonrigid and Articulated Motion Workshop, 1997. Proceedings., IEEE*, pp. 2-9. IEEE, 1997.
7. Frühwirth, Rudolf. "Application of Kalman filtering to track and vertex fitting."*Nuclear Instruments and Methods in Physics Research Section A: Accelerators, Spectrometers, Detectors and Associated Equipment* 262, no. 2 (1987): 444-450.
8. S.Saha, A. Konar, and R. Janarthanan, “Two Person Interaction Detection Using Kinect Sensor,” in *Facets of Uncertainties and Applications*, Springer, 2015, pp. 167–176.
9. Mathur, A., and G. M. Foody. "Multiclass and binary SVM classification: Implications for training and classification users." *Geoscience and Remote Sensing Letters, IEEE* 5, no. 2 (2008): 241-245.
10. Park, Sangho, and Jake K. Aggarwal. "A hierarchical Bayesian network for event recognition of human actions and interactions." *Multimedia systems*10, no. 2 (2004): 164-179.
11. Friedman, Nir, Dan Geiger, and Moises Goldszmidt. "Bayesian network classifiers." *Machine learning* 29, no. 2-3 (1997): 131-163.
12. Cheng, Jie, and Russell Greiner. "Comparing Bayesian network classifiers." In *Proceedings of the Fifteenth conference on Uncertainty in artificial intelligence*, pp. 101-108. Morgan Kaufmann Publishers Inc., 1999.
13. Yun, Kiwon, Jean Honorio, Debaleena Chattopadhyay, Tamara L. Berg, and Dimitris Samaras. "Two-person interaction detection using body-pose features and multiple instance learning." In *Computer Vision and Pattern Recognition Workshops (CVPRW), 2012 IEEE Computer Society Conference on*, pp. 28-35. IEEE, 2012.

14. Zhang, Cha, John C. Platt, and Paul A. Viola. "Multiple instance boosting for object detection." In *Advances in neural information processing systems*, pp. 1417-1424. 2005.
15. Babenko, Boris, Ming-Hsuan Yang, and Serge Belongie. "Robust object tracking with online multiple instance learning." *Pattern Analysis and Machine Intelligence, IEEE Transactions on* 33, no. 8 (2011): 1619-1632.
16. A. Yao, J. Gall, G. Fanelli, and L. J. Van Gool, "Does Human Action Recognition Benefit from Pose Estimation?.," in *BMVC*, 2011, vol. 3, p. 6.
17. Freeman, William T., and Craig Weissman. "Television control by hand gestures." In *Proc. of Intl. Workshop on Automatic Face and Gesture Recognition*, pp. 179-183. 1995.
18. R. Polikar, "Ensemble based systems in decision making," *Circuits and Systems Magazine, IEEE*, vol. 6, no. 3, pp. 21-45, 2006.
19. T. G. Dietterich, "An experimental comparison of three methods for constructing ensembles of decision trees: Bagging, boosting, and randomization," *Machine learning,* vol. 40, no. 2, pp. 139-157,2000.
20. G. Ratsch, T. Onoda, and K.-R. MUlier, "Soft margins for AdaBoost," *Machine learning,* vol. 42, no. 3, pp. 287-320, 2001.
21. Han, Jungong, Ling Shao, Dong Xu, and Jamie Shotton. "Enhanced computer vision with microsoft kinect sensor: A review." *Cybernetics, IEEE Transactions on* 43, no. 5 (2013): 1318-1334.
22. Andersen, Michael Riis, Thomas Jensen, Pavel Lisouski, Anders Krogh Mortensen, Mikkel Kragh Hansen, Torben Gregersen, and Peter Ahrendt. "Kinect depth sensor evaluation for computer vision applications." *Technical Report Electronics and Computer Engineering* 1, no. 6 (2015).
23. R. Drillis, R. Contini, and M. Maurice Bluestein, "Body segment parameters1," *Artif. Limbs*, p. 44, 1966.
24. Spurrier, Richard A. "Comment on" Singularity-Free Extraction of a Quaternion from a Direction-Cosine Matrix"." *Journal of spacecraft and rockets* 15, no. 4 (1978): 255-255.
25. Cortes, Corinna, and Vladimir Vapnik. "Support vector machine." *Machine learning* 20, no. 3 (1995): 273-297.
26. Cunningham, Padraig, and Sarah Jane Delany. "k-Nearest neighbour classifiers." *Multiple Classifier Systems* (2007): 1-17.
27. Chu, Hui, and Hui-cheng LAI. "An Improved Back-propagation NN Algorithm And Its Application [J]." *Computer Simulation* 4 (2007): 022.
28. Sokolova, Marina, and Guy Lapalme. "A systematic analysis of performance measures for classification tasks." *Information Processing & Management*45, no. 4 (2009): 427-437.

29. Dietterich, Thomas G. "Approximate statistical tests for comparing supervised classification learning algorithms." *Neural computation* 10, no. 7 (1998): 1895-1923.
30. García, Salvador, Daniel Molina, Manuel Lozano, and Francisco Herrera. "A study on the use of non-parametric tests for analyzing the evolutionary algorithms' behaviour: a case study on the CEC'2005 special session on real parameter optimization." *Journal of Heuristics* 15, no. 6 (2009): 617-644.

# CHAPTER 4

# CONCLUSION

# AND

# FUTURE DIRECTIONS

*This chapter provides a thorough self-review of the thesis with an aim to recognize several gestures introduced in the previous chapters. It also summarizes the results obtained experimentally and proposes the possible directions of future research, particularly gesture recognition through probabilistic models.*

## 4.1 Conclusion

The significance of human gesture recognition lies in proposing intelligent man-machine interaction (MMI). It has several applications, such as robot control, smart surveillance, sign language recognition, virtual environments, television control, gaming and so on.

In chapter 1, a survey regarding gestures and its recognizing procedures is illustrated. Here, we have used Microsoft's Kinect sensor for recognizing hand gestures both for single as well as double person. The architecture, performance and utility of Kinect have also been discussed.

In chapter 2, gesture based improved human-computer interaction has been defined. Here, single person hand gesture recognition has been carried out using support vector machines algorithm. The features include six Euclidean distances from spine joint to the centroids of six triangles formed from twenty body joints provided by the Kinect sensor followed by classification technique. Friedman test is used for statistical analysis.

In chapter 3, two person interaction scheme has been developed via another classifier known as Ensemble decision tree with bagging mechanism. The body joints, constructed by the sensor, have been adapted with some specified weights according to the distance between human arm and leg joints. Four mean points have selected followed by evaluation of the direction cosine angles of these mean points. The feature space consists of twelve angles per frame.

Finally a total of twenty single person single and double hand gestures, and twenty two person interactions have been recognized through our novel work.

## 4.2 Future Directions

Gesture recognition, specially dynamic gesture recognition, can also be carried out by some probabilistic classifier like hidden Markov model [1,2], hidden conditional random field model [3] *etc.*

Though various features like location, velocity, acceleration [4] can be used, but in this case orientation vector suits the best. Selecting the orientation vector as the feature, the recognizing procedure is as follows [5,6] ;

Step I : Defining the states.

Step II : Calculation of initial state probabilities.

Step III : Construction of state transition matrices.

Step IV : Declaring observation intervals.

Step V : Estimating observation symbols.

## References

1. Brand, Matthew, Nuria Oliver, and Alex Pentland. "Coupled hidden Markov models for complex action recognition." In *Computer Vision and Pattern Recognition, 1997. Proceedings., 1997 IEEE Computer Society Conference on*, pp. 994-999. IEEE, 1997.
2. Chen, Feng-Sheng, Chih-Ming Fu, and Chung-Lin Huang. "Hand gesture recognition using a real-time tracking method and hidden Markov models."*Image and vision computing* 21, no. 8 (2003): 745-758.
3. Elmezain, Mahmoud, Ayoub Al-Hamadi, Samy Sadek, and Bernd Michaelis. "Robust methods for hand gesture spotting and recognition using hidden markov models and conditional random fields." In *Signal Processing and Information Technology (ISSPIT), 2010 IEEE International Symposium on*, pp. 131-136. IEEE, 2010.
4. Elmezain, Mahmoud, and Ayoub Al-Hamadi. "Gesture Recognition for Alphabets from Hand Motion Trajectory Using Hidden Markov Models." In*Signal Processing and Information Technology, 2007 IEEE International Symposium on*, pp. 1192-1197. IEEE, 2007.
5. Wilson, Andrew D., and Aaron F. Bobick. "Parametric hidden markov models for gesture recognition." *Pattern Analysis and Machine Intelligence, IEEE Transactions on* 21, no. 9 (1999): 884-900.
6. Elmezain, Mahmoud, Ayoub Al-Hamadi, Jörg Appenrodt, and Bernd Michaelis. "A hidden markov model-based isolated and meaningful hand gesture recognition." *International Journal of Electrical, Computer, and Systems Engineering* 3, no. 3 (2009): 156-163.

# APPENDIX A

# USER GUIDE TO RUN PROGRAM CODES

*This Appendix provides a guide to execute the program codes used in this work provided in the accompanying CD. The codes are meant to be executed in the MATLAB R2012b environment.*

## A.1 User Guide for Single person hand gesture recognition

1. Install **MATLAB R2012b**
2. Select **hand_gesture** folder given in the CD.
3. Run **pushing.txt** file in the folder using **MATLAB R2012b**.

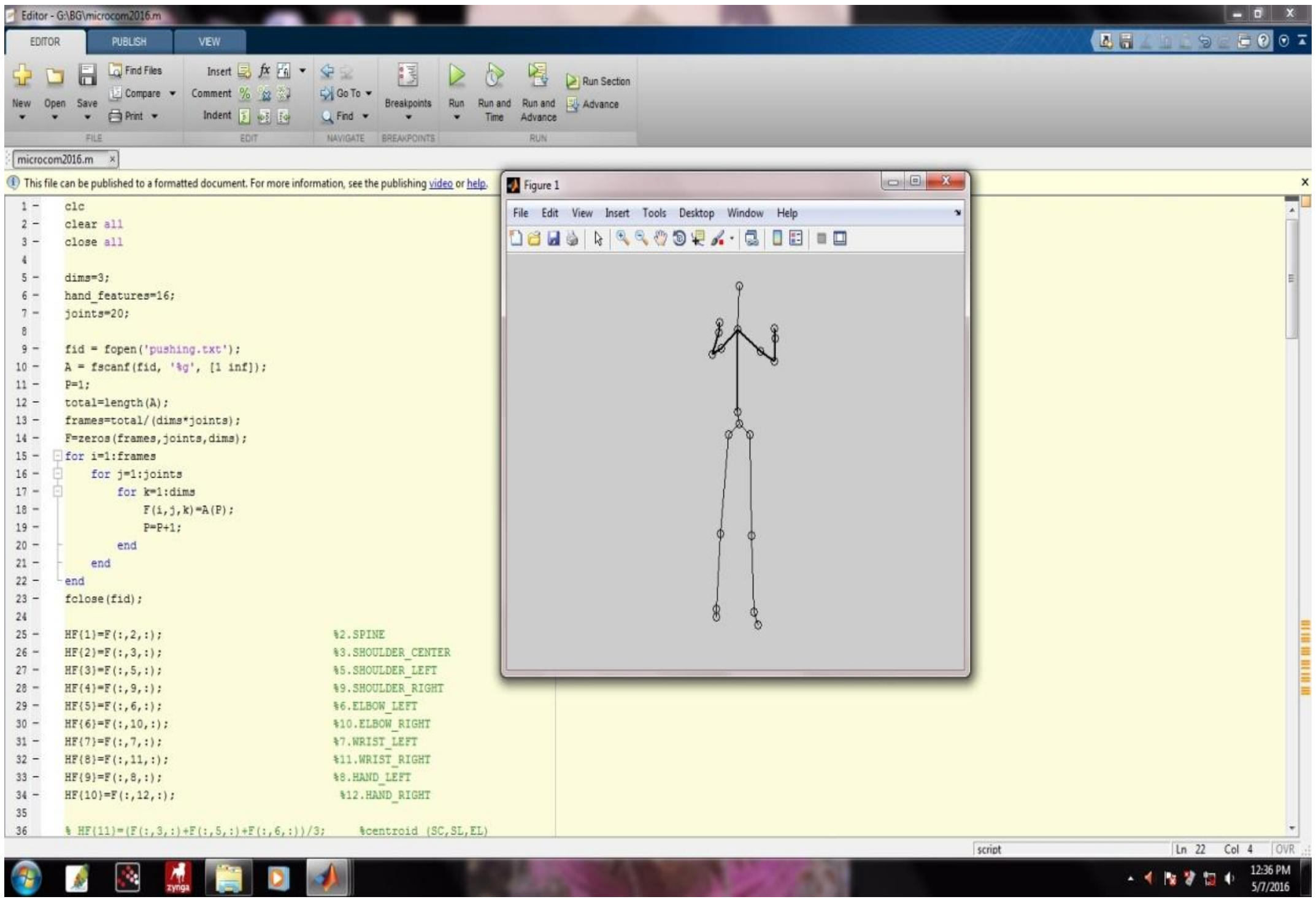

4. Line no 9 of the .m file takes the input gesture. Change the name of input file to run the code for different gestures.
5. Total 4 images are obtained as output.
6. The output images obtained are saved in **.bmp** format.

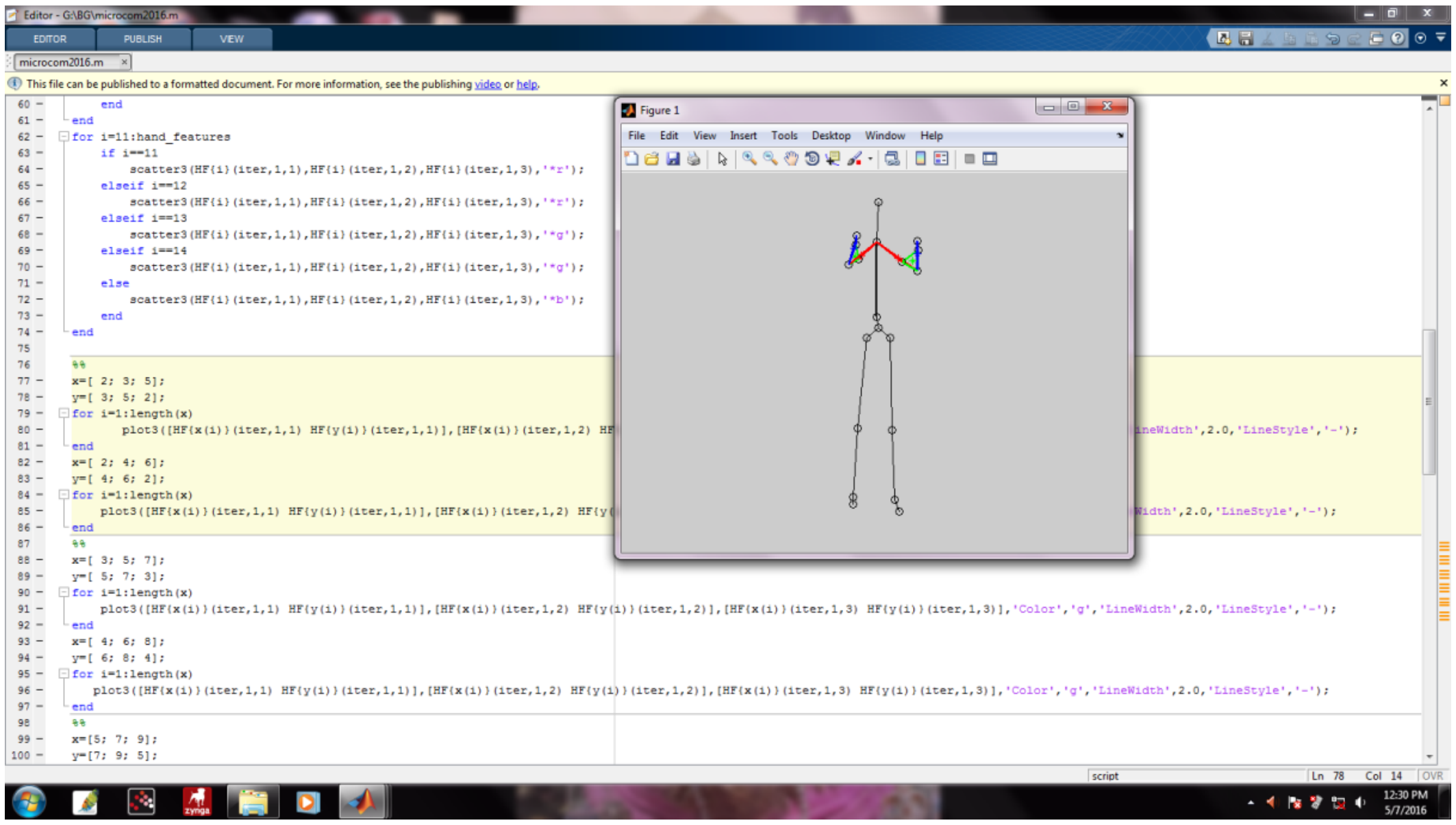

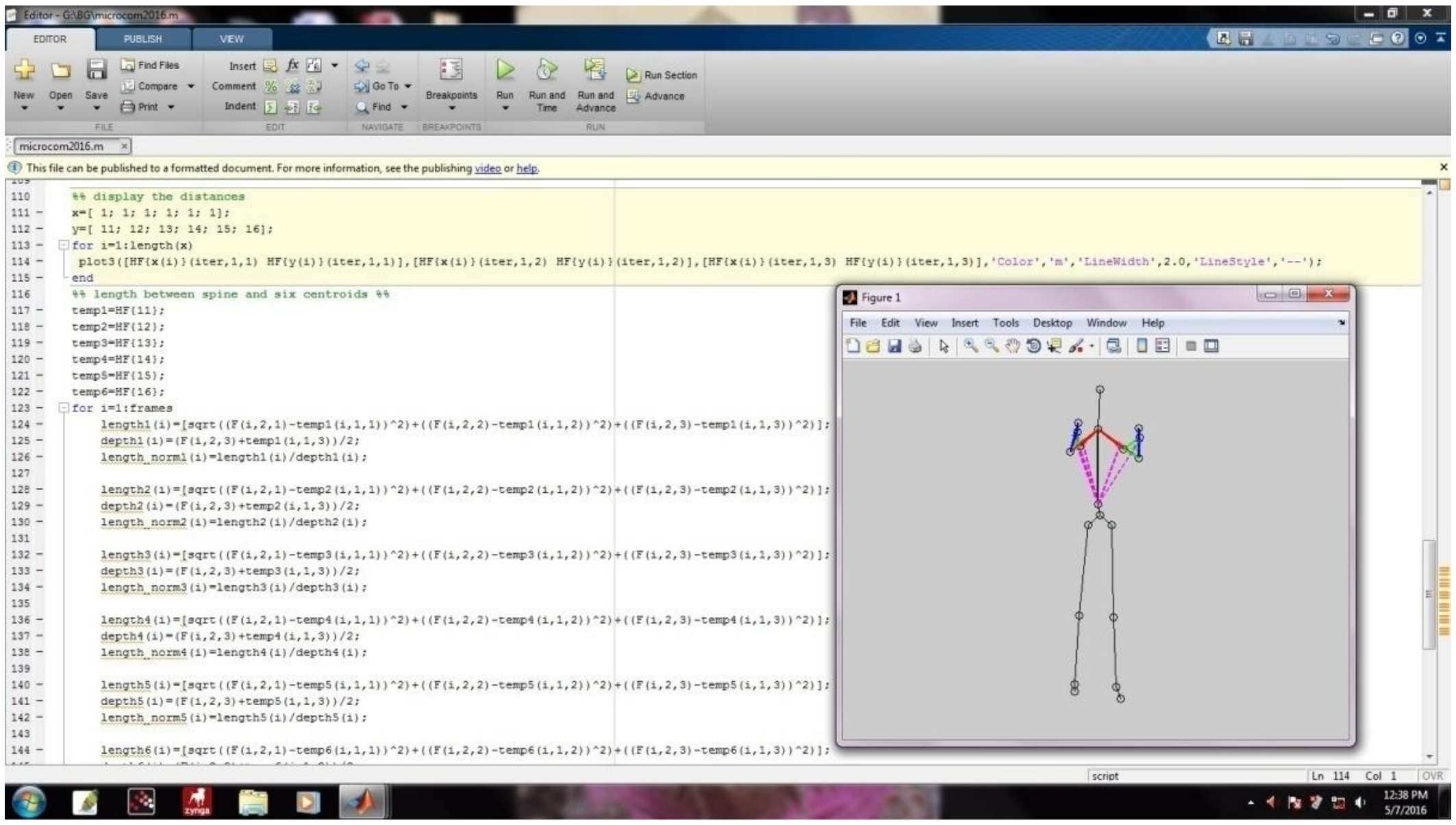

7. The command window shows the Euclidean distances obtained during the feature extraction throughout the columns.

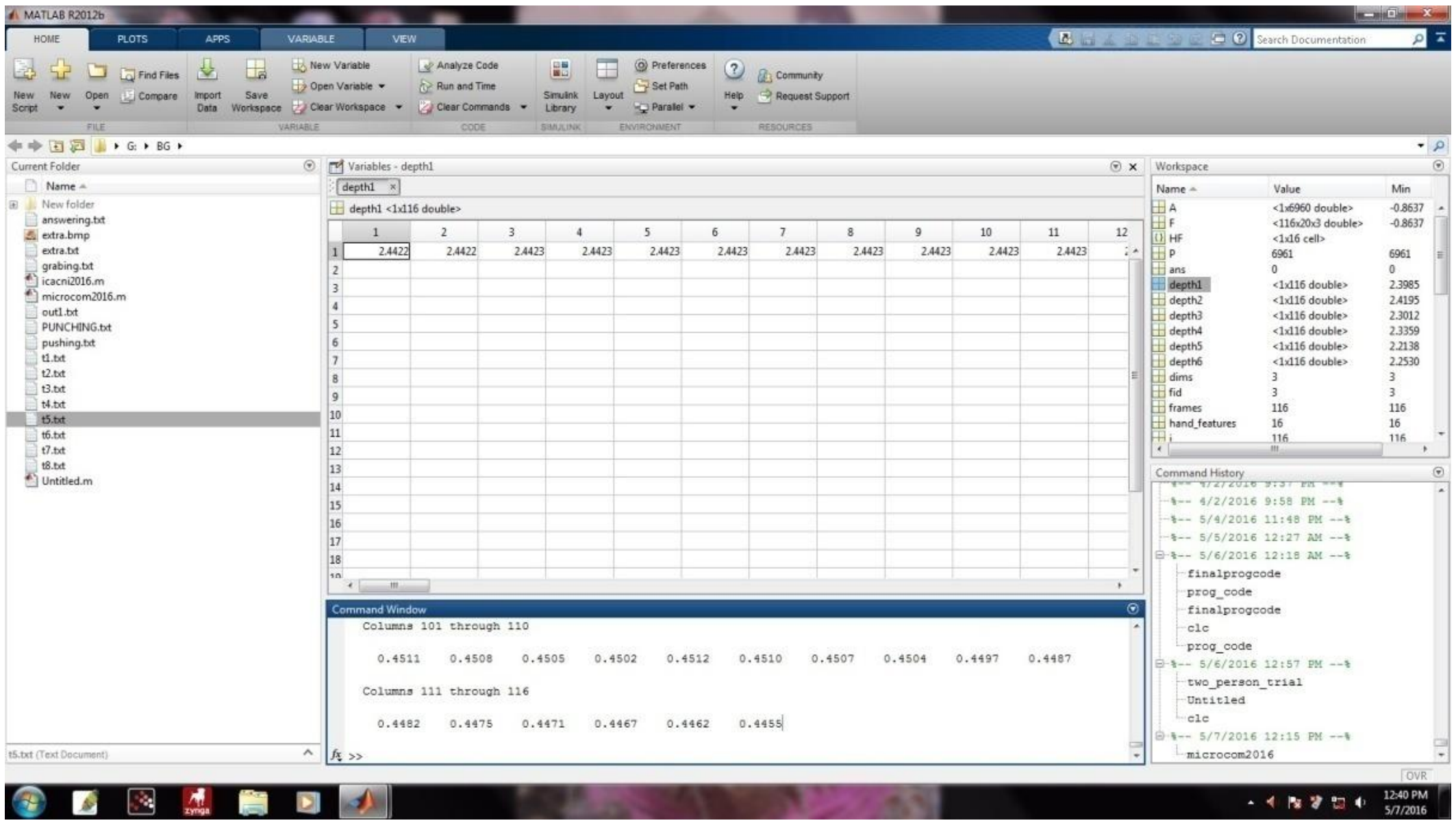

8. **SVM** code has been run to recognize the gesture.

## A.2 User Guide for Two-Person Interactions

1. Install **MATLAB R2012b**
2. Select **two_person** folder given in the CD.
3. Run **t3.txt** file in the folder using **MATLAB R2012b**.
4. Select particular iteration to observe the state of the interaction.

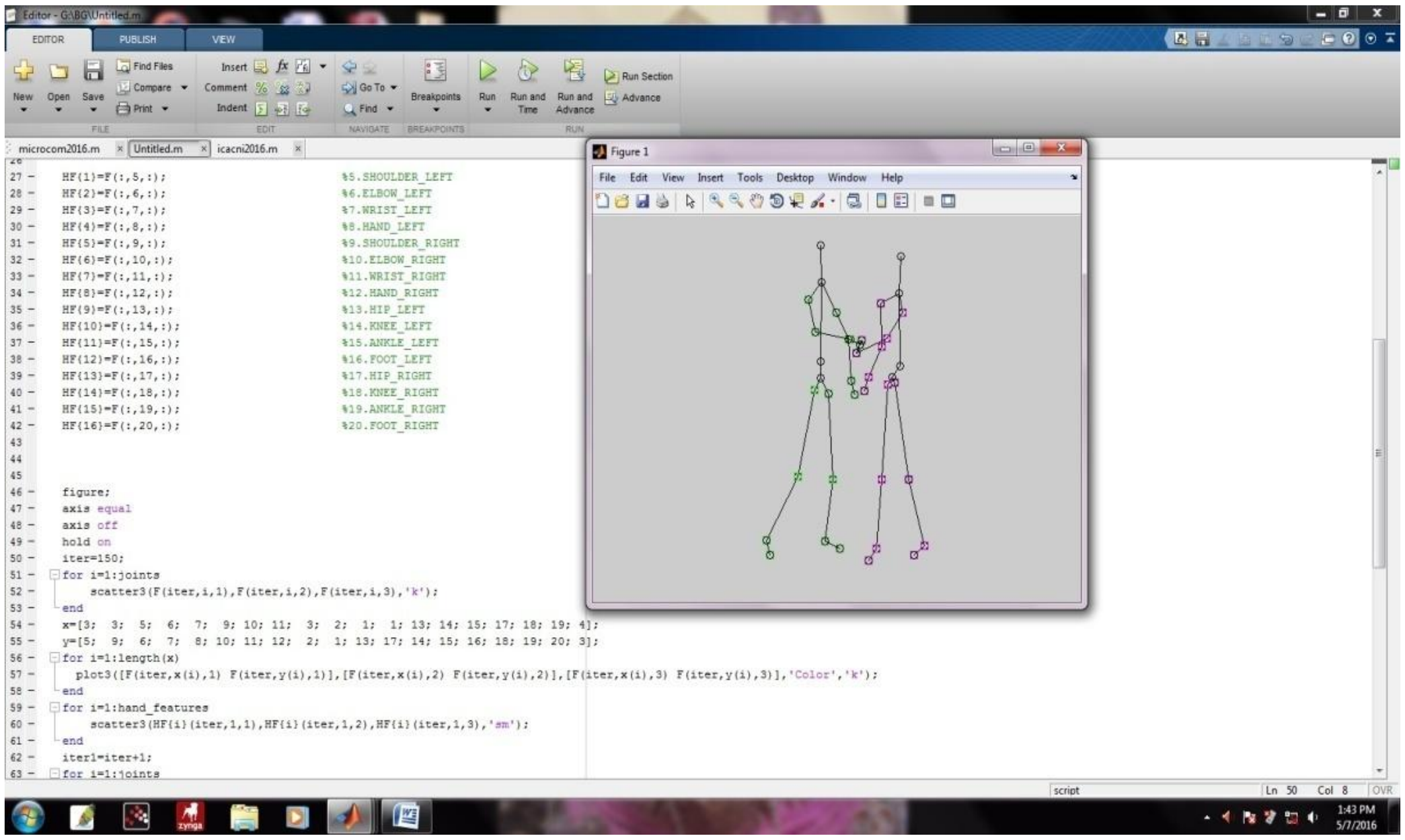

5. The required joints have been selected followed by weight adaptation formulae.
6. The direction cosines formulae have been put on to calculate the respective angles.

.

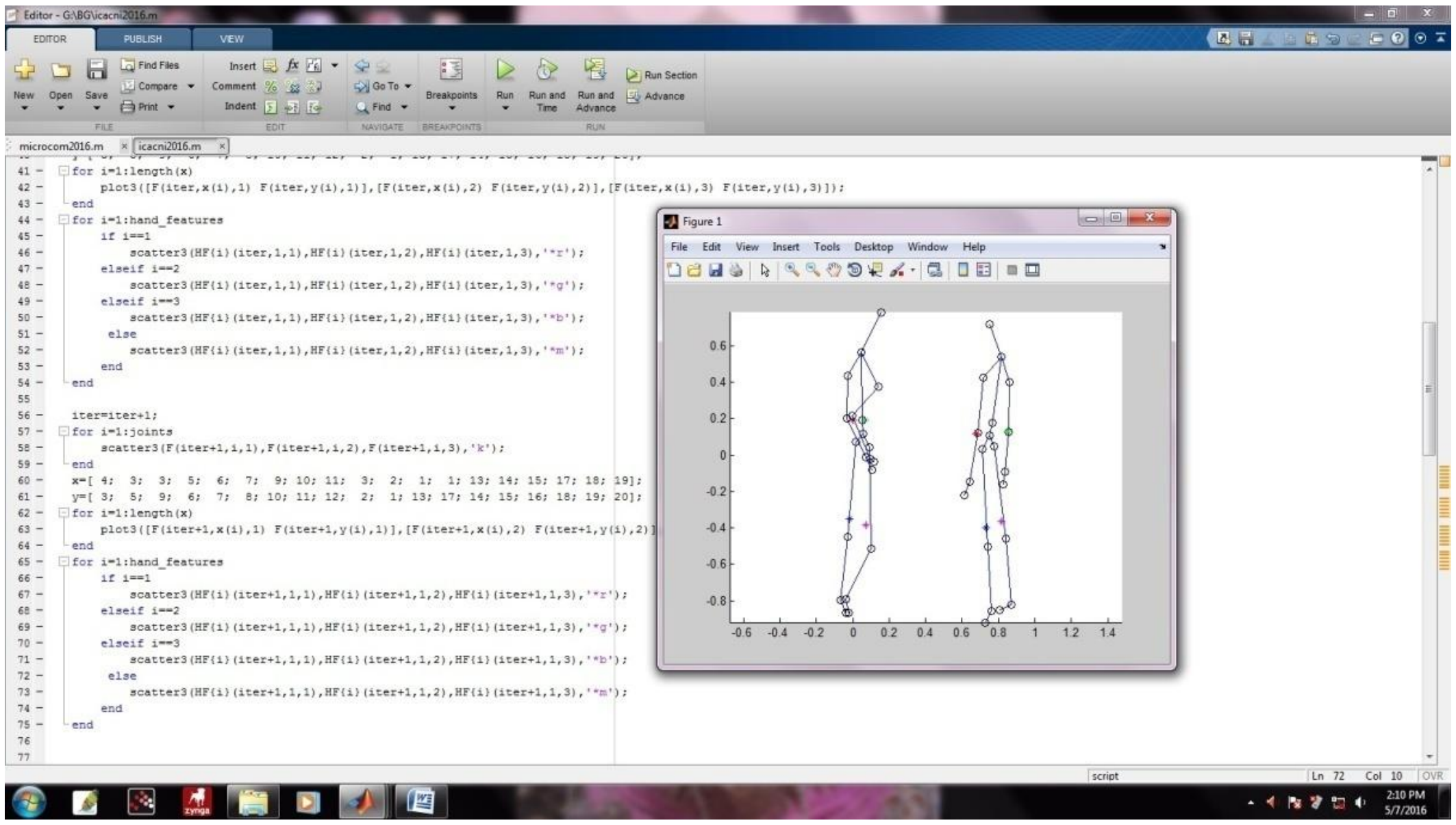

7. The command window shows the angles through the columns.

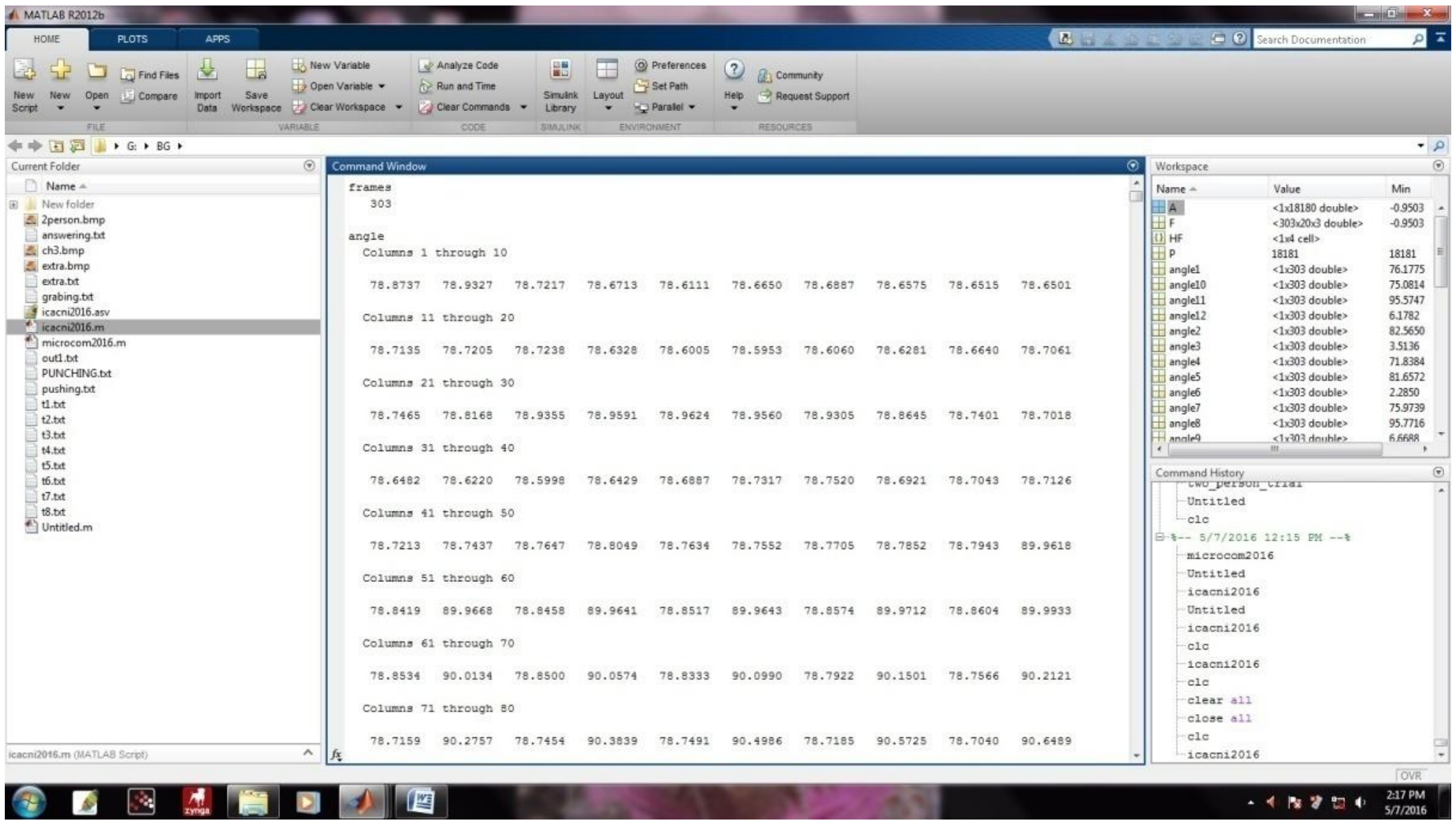

8. Run **EDT** code for two person interaction detection.

# APPENDIX B

# MATLAB CODES
# AND
# SIMULATION

*This Appendix provides the MATLAB computer codes for single person and two person hand gesture recognition. The codes are meant to be executed in the MATLAB R2012b environment with suitable text files.*

### B.1 Matlab Code for Single person hand gesture recognition

```
clc
clear all
close all

dims=3;
hand_features=16;
joints=20;

fid = fopen('pushing.txt');
A = fscanf(fid, '%g', [1 inf]);
P=1;
total=length(A);
frames=total/(dims*joints);
F=zeros(frames,joints,dims);
for i=1:frames
    for j=1:joints
        for k=1:dims
            F(i,j,k)=A(P);
            P=P+1;
        end
    end
end
fclose(fid);

HF{1}=F(:,2,:);                                 %2.SPINE
HF{2}=F(:,3,:);                                 %3.SHOULDER_CENTER
HF{3}=F(:,5,:);                                 %5.SHOULDER_LEFT
HF{4}=F(:,9,:);                                 %9.SHOULDER_RIGHT
HF{5}=F(:,6,:);                                 %6.ELBOW_LEFT
HF{6}=F(:,10,:);                                %10.ELBOW_RIGHT
HF{7}=F(:,7,:);                                 %7.WRIST_LEFT
HF{8}=F(:,11,:);                                %11.WRIST_RIGHT
HF{9}=F(:,8,:);                                 %8.HAND_LEFT
HF{10}=F(:,12,:);                                %12.HAND_RIGHT

HF{11}=(F(:,3,:)+F(:,5,:)+F(:,6,:))/3;     %centroid (SC,SL,EL)
HF{12}=(F(:,3,:)+F(:,9,:)+F(:,10,:))/3;    %centroid (SC,SR,ER)
HF{13}=(F(:,5,:)+F(:,6,:)+F(:,7,:))/3;     %centroid (SL,EL,WL)
HF{14}=(F(:,9,:)+F(:,10,:)+F(:,11,:))/3;   %centroid (SR,ER,WR)
HF{15}=(F(:,6,:)+F(:,7,:)+F(:,8,:))/3;     %centroid (EL,WR,HL)
HF{16}=(F(:,10,:)+F(:,11,:)+F(:,12,:))/3;  %centroid (ER,WR,HR)

figure;
axis equal
axis off
hold on
iter=70;
```

```
for i=1:joints
    scatter3(F(iter,i,1),F(iter,i,2),F(iter,i,3),'k');
end
x=[3;  3;  5;  6;  7;  9; 10; 11;  3;  2;  1;  1; 13; 14; 15; 17;
18; 19; 4];
y=[5;  9;  6;  7;  8; 10; 11; 12;  2;  1; 13; 17; 14; 15; 16; 18;
19; 20; 3];
for i=1:length(x)
    if i==10
        plot3([F(iter,x(i),1)      F(iter,y(i),1)],[F(iter,x(i),2)
F(iter,y(i),2)],[F(iter,x(i),3) F(iter,y(i),3)],'Color','k');
    elseif i<11
        plot3([F(iter,x(i),1)      F(iter,y(i),1)],[F(iter,x(i),2)
F(iter,y(i),2)],[F(iter,x(i),3)
F(iter,y(i),3)],'Color','k','LineWidth',2.0);
    else
        plot3([F(iter,x(i),1)      F(iter,y(i),1)],[F(iter,x(i),2)
F(iter,y(i),2)],[F(iter,x(i),3) F(iter,y(i),3)],'Color','k');
    end
end
for i=11:hand_features
    if i==11

scatter3(HF{i}(iter,1,1),HF{i}(iter,1,2),HF{i}(iter,1,3),'*r');
    elseif i==12

scatter3(HF{i}(iter,1,1),HF{i}(iter,1,2),HF{i}(iter,1,3),'*r');
    elseif i==13

scatter3(HF{i}(iter,1,1),HF{i}(iter,1,2),HF{i}(iter,1,3),'*g');
    elseif i==14

scatter3(HF{i}(iter,1,1),HF{i}(iter,1,2),HF{i}(iter,1,3),'*g');
    else

scatter3(HF{i}(iter,1,1),HF{i}(iter,1,2),HF{i}(iter,1,3),'*b');
    end
end

%%
x=[ 2; 3; 5];
y=[ 3; 5; 2];
for i=1:length(x)
       plot3([HF{x(i)}(iter,1,1)
HF{y(i)}(iter,1,1)],[HF{x(i)}(iter,1,2)
HF{y(i)}(iter,1,2)],[HF{x(i)}(iter,1,3)
HF{y(i)}(iter,1,3)],'Color','r','LineWidth',2.0,'LineStyle','-');
end
x=[ 2; 4; 6];
y=[ 4; 6; 2];
```

```
for i=1:length(x)
    plot3([HF{x(i)}(iter,1,1)
HF{y(i)}(iter,1,1)],[HF{x(i)}(iter,1,2)
HF{y(i)}(iter,1,2)],[HF{x(i)}(iter,1,3)
HF{y(i)}(iter,1,3)],'Color','r','LineWidth',2.0,'LineStyle','-');
end
%%
x=[ 3; 5; 7];
y=[ 5; 7; 3];
for i=1:length(x)
    plot3([HF{x(i)}(iter,1,1)
HF{y(i)}(iter,1,1)],[HF{x(i)}(iter,1,2)
HF{y(i)}(iter,1,2)],[HF{x(i)}(iter,1,3)
HF{y(i)}(iter,1,3)],'Color','g','LineWidth',2.0,'LineStyle','-');
end
x=[ 4; 6; 8];
y=[ 6; 8; 4];
for i=1:length(x)
   plot3([HF{x(i)}(iter,1,1)
HF{y(i)}(iter,1,1)],[HF{x(i)}(iter,1,2)
HF{y(i)}(iter,1,2)],[HF{x(i)}(iter,1,3)
HF{y(i)}(iter,1,3)],'Color','g','LineWidth',2.0,'LineStyle','-');
end
%%
x=[5; 7; 9];
y=[7; 9; 5];
for i=1:length(x)
 plot3([HF{x(i)}(iter,1,1)
HF{y(i)}(iter,1,1)],[HF{x(i)}(iter,1,2)
HF{y(i)}(iter,1,2)],[HF{x(i)}(iter,1,3)
HF{y(i)}(iter,1,3)],'Color','b','LineWidth',2.0,'LineStyle','-');
end
x=[ 6; 8; 10];
y=[ 8; 10; 6];
for i=1:length(x)
 plot3([HF{x(i)}(iter,1,1)
HF{y(i)}(iter,1,1)],[HF{x(i)}(iter,1,2)
HF{y(i)}(iter,1,2)],[HF{x(i)}(iter,1,3)
HF{y(i)}(iter,1,3)],'Color','b','LineWidth',2.0,'LineStyle','-');
end

%% display the distances
x=[ 1; 1; 1; 1; 1; 1];
y=[ 11; 12; 13; 14; 15; 16];
for i=1:length(x)
 plot3([HF{x(i)}(iter,1,1)
HF{y(i)}(iter,1,1)],[HF{x(i)}(iter,1,2)
HF{y(i)}(iter,1,2)],[HF{x(i)}(iter,1,3)
HF{y(i)}(iter,1,3)],'Color','m','LineWidth',2.0,'LineStyle','--
');
```

```
end
%% length between spine and six centroids %%
temp1=HF{11};
temp2=HF{12};
temp3=HF{13};
temp4=HF{14};
temp5=HF{15};
temp6=HF{16};
for i=1:frames
    length1(i)=[sqrt((F(i,2,1)-temp1(i,1,1))^2)+((F(i,2,2)-
temp1(i,1,2))^2)+((F(i,2,3)-temp1(i,1,3))^2)];
    depth1(i)=(F(i,2,3)+temp1(i,1,3))/2;
    length_norm1(i)=length1(i)/depth1(i);

    length2(i)=[sqrt((F(i,2,1)-temp2(i,1,1))^2)+((F(i,2,2)-
temp2(i,1,2))^2)+((F(i,2,3)-temp2(i,1,3))^2)];
    depth2(i)=(F(i,2,3)+temp2(i,1,3))/2;
    length_norm2(i)=length2(i)/depth2(i);

    length3(i)=[sqrt((F(i,2,1)-temp3(i,1,1))^2)+((F(i,2,2)-
temp3(i,1,2))^2)+((F(i,2,3)-temp3(i,1,3))^2)];
    depth3(i)=(F(i,2,3)+temp3(i,1,3))/2;
    length_norm3(i)=length3(i)/depth3(i);

    length4(i)=[sqrt((F(i,2,1)-temp4(i,1,1))^2)+((F(i,2,2)-
temp4(i,1,2))^2)+((F(i,2,3)-temp4(i,1,3))^2)];
    depth4(i)=(F(i,2,3)+temp4(i,1,3))/2;
    length_norm4(i)=length4(i)/depth4(i);

    length5(i)=[sqrt((F(i,2,1)-temp5(i,1,1))^2)+((F(i,2,2)-
temp5(i,1,2))^2)+((F(i,2,3)-temp5(i,1,3))^2)];
    depth5(i)=(F(i,2,3)+temp5(i,1,3))/2;
    length_norm5(i)=length5(i)/depth5(i);

    length6(i)=[sqrt((F(i,2,1)-temp6(i,1,1))^2)+((F(i,2,2)-
temp6(i,1,2))^2)+((F(i,2,3)-temp6(i,1,3))^2)];
    depth6(i)=(F(i,2,3)+temp6(i,1,3))/2;
    length_norm6(i)=length6(i)/depth6(i);
end

disp('length')
disp(length1)
disp(length2)
disp(length3)
disp(length4)
disp(length5)
disp(length6)
```

## B.2 Matlab Code for Two-Person Interactions

```
clc
clear all
close all

dims=3;
hand_features=4;
joints=20;

fid = fopen('t3.txt');
A = fscanf(fid, '%g', [1 inf]);
P=1;
total=length(A);
frames=total/(dims*joints);
F=zeros(frames,joints,dims);
for i=1:frames
    for j=1:joints
        for k=1:dims
            F(i,j,k)=A(P);
            P=P+1;
        end
    end
end
fclose(fid);
disp('frames')
disp(frames)

HF{1}=(0.271*F(:,5,:))+(0.449*F(:,6,:))+(0.149*F(:,7,:))+(0.131*F
(:,8,:));           % {ShoulderLeft,ElbowLeft,WristLeft,HandLeft}
HF{2}=(0.271*F(:,9,:))+(0.449*F(:,10,:))+(0.149*F(:,11,:))+(0.131
*F(:,12,:));        %
{ShoulderRight,ElbowRight,WristRight,HandRight}
HF{3}=(0.348*F(:,13,:))+(0.437*F(:,14,:))+(0.119*F(:,15,:))+(0.09
6*F(:,16,:));       % {HipLeft,KneeLeft,AnkleLeft,FootLeft}
HF{4}=(0.348*F(:,17,:))+(0.437*F(:,18,:))+(0.119*F(:,19,:))+(0.09
6*F(:,20,:));       % {HipRight,KneeRight,AnkleRight,FootRight}

figure
hold on
axis equal
iter=98;
%ieration no (modify it to display the corresponding frame no)
for i=1:joints
    scatter3(F(iter,i,1),F(iter,i,2),F(iter,i,3),'k');
end
x=[ 4;  3;  3;  5;  6;  7;  9; 10; 11;  3;  2;  1;  1; 13; 14;
15; 17; 18; 19];
```

```
y=[ 3;  5;  9;  6;  7;  8; 10; 11; 12;  2;  1; 13; 17; 14; 15;
16; 18; 19; 20];
for i=1:length(x)
    plot3([F(iter,x(i),1) F(iter,y(i),1)],[F(iter,x(i),2)
F(iter,y(i),2)],[F(iter,x(i),3) F(iter,y(i),3)]);
end
for i=1:hand_features
    if i==1

scatter3(HF{i}(iter,1,1),HF{i}(iter,1,2),HF{i}(iter,1,3),'*r');
    elseif i==2

scatter3(HF{i}(iter,1,1),HF{i}(iter,1,2),HF{i}(iter,1,3),'*g');
    elseif i==3

scatter3(HF{i}(iter,1,1),HF{i}(iter,1,2),HF{i}(iter,1,3),'*b');
     else

scatter3(HF{i}(iter,1,1),HF{i}(iter,1,2),HF{i}(iter,1,3),'*m');
    end
end

iter=iter+1;
for i=1:joints
    scatter3(F(iter+1,i,1),F(iter+1,i,2),F(iter+1,i,3),'k');
end
x=[ 4;  3;  3;  5;  6;  7;  9; 10; 11;  3;  2;  1;  1; 13; 14;
15; 17; 18; 19];
y=[ 3;  5;  9;  6;  7;  8; 10; 11; 12;  2;  1; 13; 17; 14; 15;
16; 18; 19; 20];
for i=1:length(x)
    plot3([F(iter+1,x(i),1) F(iter+1,y(i),1)],[F(iter+1,x(i),2)
F(iter+1,y(i),2)],[F(iter+1,x(i),3) F(iter+1,y(i),3)]);
end
for i=1:hand_features
    if i==1

scatter3(HF{i}(iter+1,1,1),HF{i}(iter+1,1,2),HF{i}(iter+1,1,3),'*
r');
    elseif i==2

scatter3(HF{i}(iter+1,1,1),HF{i}(iter+1,1,2),HF{i}(iter+1,1,3),'*
g');
    elseif i==3

scatter3(HF{i}(iter+1,1,1),HF{i}(iter+1,1,2),HF{i}(iter+1,1,3),'*
b');
     else
```

```
scatter3(HF{i}(iter+1,1,1),HF{i}(iter+1,1,2),HF{i}(iter+1,1,3),'*
m');
    end
end

temp1=HF{1};
temp2=HF{2};
temp3=HF{3};
temp4=HF{4};

for i=1:frames

dc1(i)=(temp1(i,1,1)/[sqrt((temp1(i,1,1))^2+(temp1(i,1,2))^2+(tem
p1(i,1,3))^2)]);
        angle1(i)=acosd(dc1(i));

dc2(i)=(temp1(i,1,2)/[sqrt((temp1(i,1,1))^2+(temp1(i,1,2))^2+(tem
p1(i,1,3))^2)]);
        angle2(i)=acosd(dc2(i));

dc3(i)=(temp1(i,1,3)/[sqrt((temp1(i,1,1))^2+(temp1(i,1,2))^2+(tem
p1(i,1,3))^2)]);
        angle3(i)=acosd(dc3(i));

dc4(i)=(temp2(i,1,1)/[sqrt((temp2(i,1,1))^2+(temp2(i,1,2))^2+(tem
p2(i,1,3))^2)]);
        angle4(i)=acosd(dc4(i));

dc5(i)=(temp2(i,1,2)/[sqrt((temp2(i,1,1))^2+(temp2(i,1,2))^2+(tem
p2(i,1,3))^2)]);
        angle5(i)=acosd(dc5(i));

dc6(i)=(temp2(i,1,3)/[sqrt((temp2(i,1,1))^2+(temp2(i,1,2))^2+(tem
p2(i,1,3))^2)]);
        angle6(i)=acosd(dc6(i));

dc7(i)=(temp3(i,1,1)/[sqrt((temp3(i,1,1))^2+(temp3(i,1,2))^2+(tem
p3(i,1,3))^2)]);
        angle7(i)=acosd(dc7(i));

dc8(i)=(temp3(i,1,2)/[sqrt((temp3(i,1,1))^2+(temp3(i,1,2))^2+(tem
p3(i,1,3))^2)]);
        angle8(i)=acosd(dc8(i));

dc9(i)=(temp3(i,1,3)/[sqrt((temp3(i,1,1))^2+(temp3(i,1,2))^2+(tem
p3(i,1,3))^2)]);
        angle9(i)=acosd(dc9(i));
```

```
dc10(i)=(temp4(i,1,1)/[sqrt((temp4(i,1,1))^2+(temp4(i,1,2))^2+(te
mp4(i,1,3))^2)]);
        angle10(i)=acosd(dc10(i));

dc11(i)=(temp4(i,1,2)/[sqrt((temp4(i,1,1))^2+(temp4(i,1,2))^2+(te
mp4(i,1,3))^2)]);
        angle11(i)=acosd(dc11(i));

dc12(i)=(temp4(i,1,3)/[sqrt((temp4(i,1,1))^2+(temp4(i,1,2))^2+(te
mp4(i,1,3))^2)]);
        angle12(i)=acosd(dc12(i));

  end

display ('angle')
disp(angle1)
disp(angle2)
disp(angle3)
disp(angle4)
disp(angle5)
disp(angle6)
disp(angle7)
disp(angle8)
disp(angle9)
disp(angle10)
disp(angle11)
disp(angle12)


%1.HIP_CENTER
%2.SPINE
%3.SHOULDER_CENTER
%4.HEAD
%5.SHOULDER_LEFT
%6.ELBOW_LEFT
%7.WRIST_LEFT
%8.HAND_LEFT
%9.SHOULDER_RIGHT
%10.ELBOW_RIGHT
%11.WRIST_RIGHT
%12.HAND_RIGHT
%13.HIP_LEFT
%14.KNEE_LEFT
%15.ANKLE_LEFT
%16.FOOT_LEFT
%17.HIP_RIGHT
%18.KNEE_RIGHT
%19.ANKLE_RIGHT
%20.FOOT_RIGHT
```

***This is the beginning!***